\documentclass[default,iicol,fleqn]{sn-jnl}% Default with double column layout

\jyear{2023}%
\usepackage[title]{appendix}
\usepackage{graphicx}
\usepackage{amssymb}
\usepackage{multicol}
\usepackage{booktabs}
\usepackage{amsmath}
\usepackage{amsfonts}
\usepackage{amsthm}
\usepackage{mathtools}
\usepackage{subfig}
\usepackage{lineno}
\hypersetup{
    colorlinks=false,
    linkcolor=blue,
    filecolor=blue,     
    citecolor=black,
    urlcolor=blue,
    pdftitle={dissertation},
    }
\usepackage{algorithm,algpseudocode,algorithmicx}
\usepackage{tabularx}
\usepackage{physics}
\usepackage{MnSymbol}
\usepackage{float}
\usepackage{array,multirow}
\usepackage{color}
\definecolor{lightgray}{gray}{0.80}

\theoremstyle{thmstyletwo}%
\newtheorem{remark}{Remark}%
\definecolor{grey1}{rgb}{0.5, 0.5, 0.5}
\definecolor{green1}{rgb}{0.4660, 0.6740, 0.1880} 
\definecolor{blue1}{rgb}{0, 0.4470, 0.7410} 
\definecolor{red1}{rgb}{0.8500, 0.3250, 0.0980}
\definecolor{yellow1}{rgb}{0.9290, 0.6940, 0.1250}
\definecolor{purple1}{rgb}{0.4940, 0.1840, 0.5560}
\definecolor{lightblue1}{rgb}{0.3010, 0.7450, 0.9330}
\definecolor{bordeaux1}{rgb}{0.6350, 0.0780, 0.1840}

\definecolor{burntorange}{rgb}{0.74902,0.341176,0}

\usepackage{color, soul}
\definecolor{blue2}{RGB}{125, 249, 255}
\DeclareRobustCommand{\reviewerI}[1]{{\sethlcolor{pink}\hl{#1}}}
\DeclareRobustCommand{\reviewerII}[1]{{\sethlcolor{yellow}\hl{#1}}}
\DeclareRobustCommand{\reviewerIII}[1]{{\sethlcolor{lightblue1}\hl{#1}}}

\soulregister\reviewerI{1}
\soulregister\reviewerII{1}
\soulregister\reviewerIII{1}

\soulregister\ref7
\soulregister\pageref7
\soulregister\eqref7
\soulregister\cite7

\makeatletter
\renewcommand*\env@matrix[1][*\c@MaxMatrixCols c]{%
  \hskip -\arraycolsep
  \let\@ifnextchar\new@ifnextchar
  \array{#1}}
\makeatother

\newcommand{\vect}[1]{\boldsymbol{#1}} 									% a vector
\newcommand{\mat}[1]{\mathbf{#1}} 											% a vector
\newcommand\abss[1]{\left\vert#1\right\vert}       % absolute value

\newcommand{\ndofs}{n_\text{dof}}																		% number of elements
\newcommand{\m}{m_\text{dof}}																	% dimension
\renewcommand{\r}{r}																	% regularity

\newcommand{\bspline}{N}

\newcommand{\phic}{\boldsymbol{\varphi}}

\newcommand{\qhat}{q}

\newcommand{\Pd}{\mathcal{P}_{\mat{d}}}
\newcommand{\Hd}{\mathcal{H}_{\mat{d}}}

\newcommand{\Pdh}{\mat{P}_{\mat{d}}}
\newcommand{\Hdh}{\mat{H}_{\mat{d}}}

\begin{document}

\title[Article Title]{Nonlinear dynamic analysis of shear- and torsion-free rods using isogeometric discretization and outlier removal}
% and robust time integration


\author*[1]{\fnm{Thi-Hoa} \sur{Nguyen}}\email{hoa.nguyen@uib.no}

\author[1]{\fnm{Bruno A.} \sur{Roccia}}\email{bruno.roccia@uib.no}

\author[2]{\fnm{Ren\'e R.} \sur{Hiemstra}}\email{hiemstra@mechanik.tu-darmstadt.de}

\author[1]{\fnm{Cristian G.} \sur{Gebhardt}}\email{cristian.gebhardt@uib.no}

\author[2]{\fnm{Dominik} \sur{Schillinger}}\email{dominik.schillinger@tu-darmstadt.de}

\affil*[1]{\orgdiv{Geophysical Institute and Bergen Offshore Wind Centre}, \orgname{University of Bergen}, \orgaddress{\country{Norway}}}

\affil[2]{\orgdiv{Institute for Mechanics, Computational Mechanics Group}, \orgname{Technical University of Darmstadt}, \orgaddress{\country{Germany}}}

%%==================================%%
%% Abstract %%
%%==================================%%

\abstract{
In this paper, we present a discrete formulation of nonlinear shear- and torsion-free rods introduced by Gebhardt and Romero in \cite{gebhardt_2021_beam} that uses isogeometric discretization and robust time integration. 
Omitting the director as an independent variable field, we reduce the number of degrees of freedom and obtain discrete solutions in multiple copies of the Euclidean space $\left(\mathbb{R}^3\right)$, which is larger than the corresponding multiple copies of the manifold $\left(\mathbb{R}^3 \cross S^2\right)$ obtained with standard Hermite finite elements. 
For implicit time integration, we choose the same integration scheme as Gebhardt and Romero in \cite{gebhardt_2021_beam} that is a hybrid form of the midpoint and the trapezoidal rules. 
In addition, we apply a recently introduced approach for outlier removal by Hiemstra et al. \cite{hiemstra_outlier_2021} that reduces high-frequency content in the response without affecting the accuracy, ensuring robustness of our nonlinear discrete formulation. 
We illustrate the efficiency of our nonlinear discrete formulation for static and transient rods under different loading conditions, demonstrating good accuracy in space, time and the frequency domain. Our numerical example coincides with a relevant application case, the simulation of mooring lines.
}

%%================================%%
%% Abstract %%
%%================================%%


\maketitle

\section{Introduction}

Nonlinear rods have a plethora of applications in science and engineering, 
for example, in the analysis of DNA molecules \cite{Benham1979,Schlick1995,Shi1994}, 
the dynamics of cables \cite{Boyer2011,Coyne1990}, 
the mechanical analysis of M\"obius bands \cite{Moore2019}, or the 
stability of elastic knots \cite{Audoly2007,Ivey1999}, among others. 
The shear-free model of rods is based on the assumption of cross-sections that remain flat and perpendicular to the tangent vector associated with the curve that describes the rod axis \cite{Giusteri2018,OReilly2017}. 
In the context of linear rods, the Euler-Bernoulli and Rayleigh models are well-established \cite{Antman1972,HAN1999}. 
For nonlinear rods, one of the most widely used models is the so-called Kirchhoff rod, which can be considered a generalization of the Rayleigh model \cite{Antman1974,Antman2005,Langer1996}. 

In general, it is not possible to formulate the governing equation of non-shearable rods through a truly unconstrained variational statement, particularly in dynamics problems, due to the non-integrable nature of vanishing shear deformations \cite{Giusteri2018,OReilly2017}. 
In \cite{Romero2020}, Romero and Gebhardt developed an unconstrained variational formulation for this type of rod, but rely on certain simplification hypotheses. 
Recently, in \cite{gebhardt_2021_beam}, Gebhardt and Romero introduced 
a new unconstrained structural model for nonlinear initially straight rods that do not exhibit shear and torsion. 
This model provides a variational formulation for shear- and torsion-free rods and is a special case of the static and dynamic variational principles for nonlinear Kirchhoff rods developed in \cite{Romero2020}. Moreover, 
it can be considered as the non-shearable counterpart of the torsion-free beam model introduced in \cite{Romero2014}. 
The main advantage of the nonlinear rod formulation \cite{gebhardt_2021_beam} is that it is an unconstrained variational statement which can be employed for both static and dynamic problems. It is in its simplest representation and does not include any non-integrable constraints such as the one enforcing non-twisting conditions \cite{gebhardt_2021_beam}. This relies on the two approaches employed when deriving this rod formulation in kinematic and energetic settings, particularly to model the shear- and torsion-free behavior: (i) the construction of a configuration space that completely prevents shear, and (ii) discarding the torsion contribution to potential and kinetic energy.

We are aware of the cable formulation introduced in \cite{Raknes2013} which, given the same material law and initial straight geometry, essentially models similar slender beam-like structures as the rod formulation introduced in \cite{gebhardt_2021_beam}, which are represented by their middle curve and are shear- and torsion-free. 
Nevertheless, these two formulations are two different formulations due to the following two essential differences: 
(i) the rod formulation \cite{gebhardt_2021_beam} does not depend on the definition or choice of the Frenet-Serret triad or any triad, while this is essential for the one introduced in \cite{Raknes2013}, particularly, for its strain measure; and 
(ii) the rod formulation \cite{gebhardt_2021_beam} considers the contribution of the rod director to the kinetic energy, which is not the case when using the cable formulation \cite{Raknes2013}. 
Moreover, to achieve the shear- and torsion-free behavior, these formulations employ different approaches: while the formulation \cite{Raknes2013} is based on the cable kinematics, which are equivalent to the Euler-Bernoulli assumptions, the rod formulation \cite{gebhardt_2021_beam} models this behavior via constructing appropriate configuration spaces. 
Such configuration spaces also require the straight initial geometry of the rods, while the cable formulation in \cite{Raknes2013} is valid for arbitrary initial cable geometry.

Among the several approaches to solve the governing equations of non-shearable rods, we can mention classical nodal and isogeometric finite elements \cite{Boyer2004,Greco2014,Maurin2018,Meier2014,Zhao2012,Raknes2013}. 
One of the key advantages of isogeometric finite elements is the higher-order smoothness of spline basis functions, which naturally fulfills the $C^1$ continuity required by the rod formulation. 
As a consequence, they have broad applications in the analysis of beam and shell structures, see e.g.  \cite{Alaydin2021,Benson_shell_2010,Benson_shell_2013,Borkovic2022,Echter_shell_2013,Kiendl_shell_2009,Oesterle2022,Oesterle_shell_2017}. 
For the recently developed nonlinear rod formulation in \cite{gebhardt_2021_beam}, the spatial discretization scheme applied so far is the one based on nodal finite elements. It relies on cubic Hermite functions to represent the discrete rod configuration, which in turn is decomposed into nodal positions and nodal directors. 
We refer to this scheme hereinafter as the standard discretization scheme. 
The discrete solutions obtained lie 
in the manifold $\left(\mathbb{R}^3 \cross S^2\right)^n$, where $n$ is the number of discrete nodes. 
This discretization scheme establishes the first attempt to numerically solve the shear- and torsion-free Kirchhoff rod.

In this paper, 
we investigate an alternative spatial discretization scheme in the context of isogeometric analysis (IGA) for the rod formulation of \cite{gebhardt_2021_beam}. 
In particular, we discretize the rod configuration in terms of the position of control points, without considering the director as an independent variable field. 
Hence, the number of degrees of freedom can be reduced and the discrete solution lies in multiple copies of the Euclidean space $\mathbb{R}^3$ which is a larger space than the corresponding multiple copies of the manifold $\left(\mathbb{R}^3 \cross S^2\right)$ of the standard scheme. 
We utilize the higher-order smoothness of spline functions that naturally fulfill the $C^1$ continuity required by the rod formulation (and beyond). 
We illustrate, via static benchmarks of two- and three-dimensional cantilever rods, that isogeometric discretizations and the standard scheme  
achieve a comparable level of accuracy. 
Moreover, we show for a geometrically nonlinear cantilever rod bent to a circle that the convergence behavior is comparable to an optimal convergence in the $H^2$ semi-norm as linear fourth-order problems, but shows a smaller rate in the $H^1$ semi-norm and the $L^2$ norm. 
We also show via this example that decreasing the continuity of spline basis functions generally does not affect the accuracy and convergence, but reduces the convergence rate in $H^2$ semi-norm when using odd polynomial degrees.

For time integration in our dynamic computations, we employ the same implicit integration scheme as \cite{gebhardt_2021_beam}, which is a hybrid combination of the midpoint and trapezoidal rules. 
This type of implicit scheme has been shown to achieve second-order accuracy, approximately preserve the energy, and exactly preserves the linear and angular momentum \cite{gebhardt_implicit_2020}, \cite{guo_time_int_2022,Wen_time_int_2022}. 
We show, via dynamic benchmarks of two- and three-dimensional rods, that the isogeometric discretization scheme using B-splines with $C^1$ continuity or higher is less robust than the standard one. 
We improve its robustness via the strong approach of outlier removal introduced in \cite{hiemstra_outlier_2021}. 
We illustrate, via an example of an unconstrained rod subjected to out-of-plane vanishing forces, 
that the mass term associated to the inertia is irregular. Hence, the configuration-dependent mass matrix of the studied formulation behaves irregularly and, therefore, cannot be simplified to a constant matrix. 
Finally, we test our rod formulation for the nonlinear behavior of swinging rods under conservative, non-conservative, and pulsating forces. 
Our results indicate that our discrete isogeometric scheme is an efficient tool for such nonlinear computations.

The structure of the paper is as follows: 
In Section \ref{sec-formulation}, we briefly review the nonlinear rod formulation. 
In addition, we derive the external forces induced by a surrounding flow, considered in our numerical examples. 
In Section \ref{sec-discretization-time}, we discuss discretization in space with isogeometric finite elements, the resulting semi-discrete formulation, and differences when compared to the standard discretization scheme based on Hermite functions. 
We also briefly recap the implicit time integration scheme that is applied in our transient computations. 
In Section \ref{sec-robust}, we numerically demonstrate the robustness of isogeometric discretizations for two- and three-dimensional benchmarks and improve it via a strong approach for outlier removal. 
In Section \ref{sec-swinging-rod}, we apply the isogeometric nonlinear rod formulation to a swinging rubber rod subjected to different loading conditions, which can be considered as a relevant application case for the simulation of mooring lines. 
In Section \ref{sec-summary}, we summarize our results and draw conclusions.

\section{Nonlinear shear- and torsion-free rods}\label{sec-formulation}

In this section, we briefly review the formulation of nonlinear shear- and torsion-free rods in a continuous setting introduced in \cite{gebhardt_2021_beam}. 
We then describe and derive the external forces induced by a surrounding flow that are considered in the numerical examples of this work. 
We start with a brief recap of required fundamental equations and definitions in differential geometry that are later utilized for the rod formulation.

%--------------------------------------------------------------------
\subsection{Preliminaries}

Consider an arbitrary regular one-parameter curve $\phic \,=\, \phic(s)$ in the ambient space $\mathbb{R}^3$, where $s \,\in\, [0,L]$ is the arc-length coordinate. 
Since $\phic$ is regular, its first derivative with respect to $s$, denoted as $\phic^\prime$, is non-zero, i.e. $\phic^\prime \,\neq\, \mat{0}$. 
The Frenet-Serret moving frame associated with the curve $\phic$ is then: 
\begin{align}\label{eq-triad}
\hspace{-0.2cm}
    \mat{d} \, := \, \frac{\phic^\prime}{\abss{\phic^\prime}} \, , \quad 
    \mat{t} \, := \, \frac{\phic^{\prime \prime}}{\abss{\phic^{\prime \prime}}} \, , \quad 
    \mat{b} \, := \, \frac{\phic^\prime \, \times \phic^{\prime \prime}}{\abss{\phic^\prime \, \times \phic^{\prime \prime}}} \, ,
\end{align}
where $(\cdot)^\prime$ denotes the first derivative with respect to the arc-length $s$, i.e. $(\cdot)^\prime = \partial (\cdot) / \partial \, s$, and $\abss{\cdot} \,:\, \mathbb{R}^3 \,\to\, \mathbb{R}_{\geq\,0}$ denotes the Euclidean vector norm.
We refer to $\mat{d}$ as the director of the curve $\phic$. 
We note that $\mat{t}$, and thus also $\mat{b}$, is ill-defined at points where 
$\abss{\phic^{\prime \prime}} = 0$, while the director $\mat{d}$ is well-defined everywhere along the curve $\phic$.

The director $\mat{d}$ lives in the unit sphere $S^2 \,:=$ \\
$\left\{ \, \mat{d} \,\in\, \mathbb{R}^3 \, \vert\, \mat{d} \,\cdot\, \mat{d} \,=\, 1 \, \right\}$ that is a nonlinear, smooth, compact, two-dimensional manifold \cite{Eisenberg1979APO,Romero2017}. 
The tangent bundle associated with $S^2$ is also a manifold, which is given by
$T S^2 \,:=$ 
$\left\{ \, (\mat{d},\mat{c}) \,\in\, S^2 \,\times\,\mathbb{R}^3 \,, \mat{d} \,\cdot\, \mat{c} \,=\, 0 \, \right\}$. 
We recall that the covariant derivative of a smooth vector field $\mat{v}\,:\,S^2 \,\to\, T S^2$ along a vector field $\mat{w}\,:\,S^2 \,\to\, T S^2$ is a vector field in $T S^2$ evaluated at $\mat{d}$, given by:
\begin{align}\label{eq-covar-deriv}
    \nabla_{\mat{w}} \, \mat{v} \,:=\, \left(\, \mat{I} \,-\, \mat{d} \,\otimes\, \mat{d} \,\right) \, D\,\mat{v} \,\cdot\, \mat{w},
\end{align}

\noindent
where $\mat{I}$ denotes the identity matrix, and $D\,\mat{v}$ the derivative of $\mat{v}$. 
The covariant derivative $\nabla_{\mat{w}} \, \mat{v}$ is 
the projection of $D\,\mat{v}$ in the direction of $\mat{w}$ onto the tangent plane at $\mat{d}$ \cite{Eisenberg1979APO,Romero2017}.

For the rod formulation \cite{gebhardt_2021_beam} considered in this work, 
we are particularly interested in the covariant derivative of $\phic^\prime$ in the direction of $\mat{d}^\prime$. Applying \eqref{eq-covar-deriv}, this covariant derivative takes the following form \cite{gebhardt_2021_beam}:
\begin{align}\label{eq-covar-deriv-rod}
    \nabla_{\mat{d}^\prime} \, \phic^\prime \,=\, \underbrace{\left(\,\mat{I} - \mat{d} \, \otimes \, \mat{d}\, \right)}_{\Pd} \, \phic^{\prime\prime} \,,
\end{align}

\noindent
where $\mat{d}^\prime$ is computed by taking the derivative of \eqref{eq-triad}, i.e., $\mat{d}^\prime \,=\, \frac{1}{\abss{\phic^\prime}} \, \Pd \, \phic^{\prime\prime}$. 
We refer to $\Pd$ as the orthogonal projection operator.

%--------------------------------------------------------------------
\subsection{Strong and weak forms}

Let the curve $\phic$ now be the configuration of Kirchhoff rods, dependent on the arc-length $s$ and time $t$, $\phic\,=\, \phic(s,\,t)$, $(s,\,t) \,\in\, [0,\,L] \, \times \, [0,\,T]$, that are initially straight, shear-, torsion-free, and transversely isotropic \cite{gebhardt_2021_beam}. 
Next, let us Consider the following set for the rod configurations: 
\begin{equation}\label{eq-manifold}
    \begin{split}
        & \mathcal{D} := \left\{ \phic \, \in \, \left[C^2 (0, \,L)\right]^3, \;
        \abss{\phic^\prime} \, > \, 0, \; \right. \\
        & \qquad \qquad \left. \phic(0,\,t) = \mathbf{0}, \;
        \phic^\prime \, (0,\,t) \, = \, \mat{E}_3 \right\} \, ,
    \end{split}
\end{equation}
where $C^2(0, \,L)$ is the space of $C^2$ continuous functions on $(0, \,L)$, 
$\mat{E}_i$, $i=1,2,3$, are the canonical Cartesian basis of $\mathbb{R}^3$. 
For simplicity, 
we adopt here the clamped boundary condition at $s=0$.

We recall, from \cite{gebhardt_2021_beam}, 
the strong form of the equations of motion governing the space-time evolution for the Kirchhoff rod:
\begin{equation}\label{s-eom}
    \begin{split}
        & \mat{n}^\prime + \left(\, \frac{1}{\abss{\phic^\prime}} \, \mat{d} \,\times\, \nabla_{\mat{d}^\prime} \, \mat{m} \, \right)^\prime \, = \, \\
        & \qquad \qquad A_\rho\,\ddot{\phic} \,+\, \left(\, \frac{1}{\abss{\phic^\prime}} \, \mat{d} \,\times\, I_\rho \, \nabla_{\dot{\mat{d}}} \, \dot{\mat{d}} \right)^\prime \,-\, \mat{f}^{\text{ext}} \,,
    \end{split}    
\end{equation}

\noindent
where $\mat{n}$ and $\mat{m}$ are the stress measures, defined as: 
\begin{align}\label{eq-stress}
    \mat{n} \,=\, EA \, \boldsymbol{\epsilon} \,, \qquad \mat{m} \,=\, EI\, \boldsymbol{\kappa} \,,
\end{align}

\noindent
respectively, which are conjugated with the following strain measures:
\begin{align}\label{eq-strain}
    \boldsymbol{\epsilon} \,:=\, \phic^\prime \,-\, \mat{d} \,, \qquad \boldsymbol{\kappa} \,:=\, \mat{d} \,\times\,\mat{d}^\prime \,.
\end{align}

Here, $A_\rho$ and $I_\rho$ are the mass per unit length and the inertia density, respectively, i.e. $A_\rho \,=\, \rho\,A$ and $I_\rho \,=\, \rho\,I$, where $\rho$ is the mass density, $A$ the cross-section area and $I$ the moment of inertia of the rod. 
$\mat{f}^{\text{ext}}$ is the external generalized forces, and  
the dot notation in the superscript 
denotes the derivative with respect to time $t$, i.e. $\dot{(\cdot)} \,=\, \partial (\cdot) / \partial \, t$. 
We note that since the director $\mat{d}$ is well-defined along the rod $\phic \,\in\, \mathcal{D}$ (see also \eqref{eq-manifold}), as discussed in the previous subsection, the strain measures \eqref{eq-strain} are also well-defined at every point of the rod.

At time $t=0$, we require the following initial conditions: 
\begin{subequations}
    \begin{align}
        & \phic \,=\, \phic_0 \quad & \text{on } (s,\,t) \,\in\, [0,\,L] \, \times \, [0] \,, \\ 
        & \dot{\phic} \,=\, \mat{v}_0 \quad & \text{on } (s,\,t) \,\in\, [0,\,L] \, \times \, [0] \,.
    \end{align}
\end{subequations}

Additionally, we require at all times the following boundary conditions; for instance, clamped-free ends:
\begin{subequations}
    \begin{align}
        \hspace{-0.2cm} \text{on } & (s,\,t) \,\in\, [0] \, \times \, [0,\,T]: \,,  \nonumber \\
        & \phic \,=\, \mat{0}\,, \qquad 
        \phic^\prime \,=\, \mat{E}_3 \,, \\
        \hspace{-0.2cm} \text{on } & (s,\,t) \,\in\, [L] \, \times \, [0,\,T]: \,, \nonumber \\
        \hspace{-0.2cm} & \mat{n} \,+\, \frac{1}{\abss{\phic^\prime}} \, \mat{d} \,\times\, \left(\, \nabla_{\mat{d}^\prime} \, \mat{m} \,-\, I_\rho \, \nabla_{\dot{\mat{d}}} \, \dot{\mat{d}} \,\right) \,=\, \mat{0} \,, \\
        & \frac{1}{\abss{\phic^\prime}} \, \mat{d} \,\times\, \mat{m} \,=\, \mat{0} \,.
    \end{align}
\end{subequations}

According to \cite{gebhardt_2021_beam}, the weak form corresponding to \eqref{s-eom} is then: 
\begin{equation}\label{w-eom}
    \begin{split}
        \int_0^S \, & \delta\phic \,\cdot\, \left( 
            \mathcal{M}\left(\phic^\prime\right)\,\hat{\nabla}_{\dot{\phic}} \, \dot{\phic} \,+\, \right. \\
            & \qquad \left. \mathcal{B}\left(\phic^\prime,\,\phic^{\prime\prime}\right)^T \, \boldsymbol{\sigma} \,-\, \mat{f}^{\text{ext}} \,\right) \, \mathrm{d} \, s \,=\, 0\,,
    \end{split}    
\end{equation}

\noindent
where the mass operator, $\mathcal{M}$, and the linearized strain operator, $\mathcal{B}$, are given by:
\begin{align}
    \hspace{-0.2cm} & \mathcal{M} \,=\, \mathcal{M}\left(\phic^\prime\right) \, :=\, A_\rho \, \mat{I} \,+\, \nonumber \\
    & \qquad \qquad \qquad \qquad (\cdot)^{\prime\,T} \, I_\rho \, \frac{1}{\abss{\phic^\prime}^2} \, \Pd \, (\cdot)^\prime \, \label{mass-op}\\
    \hspace{-0.2cm} & \mathcal{B} \,=\, \mathcal{B}\left(\phic^\prime,\,\phic^{\prime\prime}\right) \,:=\, \nonumber \\
    & \quad 
    \begin{bmatrix}
        \mat{I} \,-\, \frac{1}{\abss{\phic^\prime}} \, \Pd  & \mat{0} \\
        -\frac{1}{\abss{\phic^\prime}^2} \, \left[\phic^{\prime\prime}\right]_\times \, \Hd     & \frac{1}{\abss{\phic^\prime}} \, \left[\mat{d}\right]_\times 
    \end{bmatrix} \; \begin{bmatrix}
        (\cdot)^\prime \\ (\cdot)^{\prime\prime}
    \end{bmatrix} \,. \label{B-op}
\end{align}
Here, 
$\boldsymbol{\sigma} \,:=\, [\mat{n} \quad \mat{m}]^T$, $\Hd$ is the Householder operator\footnote{The Householder operator is also known as a Householder reflection or elementary reflector.}, $\Hd \,:=\, \mat{I} \,-\, 2 \, \mat{d} \,\otimes\, \mat{d}$, and $[\mat{a}]_\times$ denotes the skew-symmetric matrix of a vector $\mat{a} \,=\, \left[a_1 \quad a_2 \quad a_3\right]^T$, i.e.:
\begin{align}
    [\mat{a}]_\times \,=\, \begin{bmatrix}
        0 & -a_3 & a_2 \\ a_3 & 0 & -a_1 \\ -a_2 & a_1 & 0
    \end{bmatrix} \,. \nonumber
\end{align}
The field covariant derivative $\hat{\nabla}_{(\cdot)} \, ( \cdot)$ is the extension of the covariant derivative \eqref{eq-covar-deriv}.

\section{Isogeometric discrete rod model and implicit time integration}\label{sec-discretization-time}

In this section, 
we first discuss an alternative spatial discretization scheme of the rod formulation \eqref{w-eom}, reviewed in the previous section. 
We employ isogeometric discretizations, which utilize the higher-order continuity of smooth spline functions fulfilling the $C^1$-continuity required by the considered rod formulation. 
We show that this alternative discretization scheme yields different semi-discrete formulations, reduces the number of degrees of freedom, and leads to a larger solution space than the standard one based on nodal finite elements using cubic Hermite functions. 
We then numerically demonstrate, for a geometrically nonlinear rod bent to a circle, that the isogeometric discretizations show optimal convergence in $H^2$ semi-norm and a smaller convergence rate in $H^1$ semi-norm and $L^2$ norm. 
In addition, we briefly review the implicit time integration scheme employed in this work. 
We close this section with a discussion of applying the strong approach of outlier removal \cite{hiemstra_outlier_2021} to improve the robustness of isogeometric discretizations.

%--------------------------------------------------------------------
\subsection{Spatial discretizations}\label{sec-discretization}

The rod formulation \eqref{w-eom} requires discretizations of at least $C^1$-continuity. 
To fulfill this, the standard discretization scheme based on nodal finite element employs cubic Hermite functions and discretizes both the nodal spatial position and nodal director as two variable fields \cite{gebhardt_2021_beam}. 
In this work, we want to utilize the higher-order continuity of smooth spline functions that allow lower polynomial degree and one can omit the director as a variable field.
Thus, we discretize the rod configuration, $\phic(s,\,t) \,\in\, \mathcal{D}$, and its variation, $\delta \phic(s,\,t)$, by a weighted finite sum of $\m$ B-splines, $\bspline_i \,(s)$, with continuity $C^\r$ and polynomial degree $p$ \cite{Piegl1996,Schumaker2007}, where $r$ is the continuity order, $1 \,\leq\, r \,\leq\, p-1$, as follows: 
\begin{subequations}\label{eq-discretize}
    \begin{align}
        & \phic(s,\,t) \,\approx\, \phic_h \,(s,\,t) \,=\,\nonumber\\
        & \qquad \qquad \sum_i^{\m} \, \bspline_i \, (s) \, \mat{x}_i\,(t) \,=\, \mat{\bspline} \, \mat{\qhat} \,, \\
        & \delta \phic(s) \,\approx\, \delta \phic_h (s) \,=\, \nonumber \\
        & \qquad \qquad \sum_i^{\m} \, \bspline_i \, (s) \, \delta \mat{x}_i \,=\, \mat{\bspline} \, \delta \mat{\qhat} \,.
    \end{align}    
\end{subequations}

\noindent
Here, 
$\phic_h\,=\,\phic_h\,(s,\,t) \in \mathbb{R}^3$ denotes the discrete rod configuration in space, 
$\mat{x}_i \, \in \mathbb{R}^3$ is the time-dependent position of the $i^{\text{th}}$ control point, 
$\mat{\qhat}\,=\,\mat{\qhat}(t) \in (\mathbb{R}^3)^{\m}$ is the vector of unknown time-dependent coefficients, 
and $\delta \phic_h(s)$, $\delta \mat{x}_i$, and $\delta \mat{\qhat}$ their variations, respectively. 
The discrete director $\mat{d}$ and strain/stress measures follow directly from their definitions in \eqref{eq-triad} and \eqref{eq-stress}-\eqref{eq-strain}, respectively.

Introducing \eqref{eq-discretize} into the variational formulation \eqref{w-eom}, we obtain the following semi-discrete formulation: 
\begin{equation}\label{semi-deom}
    \begin{split}
        & \text{Find } \mat{\qhat}(t) \,\in\, \mathbb{R}^{3\,\m}, t\,\in\,[0,\,T], \text{ such that}: \\
        & \int_0^S \, \delta \mat{\qhat} \, \cdot \, \left(\, \mat{M} (\mat{\qhat}) \, \nabla_{\dot{\mat{\qhat}}} \, \dot{\mat{\qhat}} \,+\, \mat{B} (\mat{\qhat})^T \, \boldsymbol{\sigma}_h \right.  \\
        & \qquad \quad \left. -\, \mat{\bspline}^T \, \mat{f}^{\text{ext}} \,\right) \, \mathrm{d} \, s \,=\, \mat{0} \quad \forall \, \delta \mat{\qhat} \,\in\, \mathbb{R}^{3\,\m} \,.
    \end{split}
\end{equation}

\noindent
Here, the mass matrix $\mat{M}$ and the matrix $\mat{B}$, resulting from the operators \eqref{mass-op} and \eqref{B-op}, respectively, are:
    \begin{align}
        & \mat{M} \,=\, \mat{M} (\mat{\qhat}) \,=\, \underbrace{A_\rho \, \mat{\bspline}^T \, \mat{I} \, \mat{\bspline}}_{\mat{M}_1} \,+\nonumber \\
        & \qquad \qquad \underbrace{I_\rho \, \frac{1}{\abss{\phic_h^\prime}^2} \, \left(\mat{\bspline}^\prime\right)^T \, \Pdh \, \mat{\bspline}^\prime}_{\mat{M}_2} \label{mass-matrix}\\
        & \mat{B} \,=\, \mat{B}(\mat{\qhat}) \,:=\, \nonumber \\
        & \begin{bmatrix}
            \mat{I} \,-\, \frac{1}{\abss{\phic_h^\prime}} \, \Pdh  & \mat{0} \\
            -\frac{1}{\abss{\phic_h^\prime}^2} \, \left[\phic_h^{\prime\prime}\right]_\times \, \Hdh     & \frac{1}{\abss{\phic_h^\prime}} \, \left[\mat{d}_h\right]_\times 
        \end{bmatrix} \; \begin{bmatrix}
            \mat{\bspline}^\prime \\ \mat{\bspline}^{\prime\prime}
        \end{bmatrix} \, , \label{B-matrix}
    \end{align}
    
\noindent
with $\Pdh \,=\, \mat{I} \,-\, \mat{d}_h \,\otimes\, \mat{d}_h$, and $\Hdh \,=\, \mat{I} \,-\, 2\,\mat{d}_h \,\otimes\, \mat{d}_h$. The discrete stress measures $\boldsymbol{\sigma}_h$ are:
\begin{align}
    \boldsymbol{\sigma}_h \,=\, \begin{bmatrix}
        \mat{n}_h \\ \mat{m}_h
    \end{bmatrix} \,& =\, \begin{bmatrix}
        EA \, \boldsymbol{\epsilon}_h \\ EI \, \boldsymbol{\kappa}_h
    \end{bmatrix} \nonumber \\
    & = \begin{bmatrix}
        EA \, \left(\, \phic_h^\prime \,-\, \mat{d}_h \,\right) \\ EI \, \mat{d}_h \,\times\, \mat{d}_h^\prime
    \end{bmatrix} \,.
\end{align}
The term $\mat{M} (\mat{\qhat}) \, \nabla_{\dot{\mat{\qhat}}} \, \dot{\mat{\qhat}}$ in \eqref{semi-deom}, derived in \cite{gebhardt_2021_beam}, takes the following form:
\begin{align}
    \hspace{-0.7cm} & \mat{M} (\mat{\qhat}) \, \nabla_{\dot{\mat{\qhat}}} \, \dot{\mat{\qhat}} \,=\,
    \mat{M} \, \ddot{\mat{\qhat}} \,-\, \nonumber \\ \hspace{-0.7cm} & 
    2\,I_\rho \, \frac{1}{\abss{\phic_h^\prime}^2} \, \left(\mat{\bspline}^\prime\right)^T \, \left[\, \frac{1}{\abss{\phic_h^\prime}} \, \left(\mat{d}_h \, \cdot \, \dot{\phic}_h^\prime\right) \Pdh \,+\, \mat{d}_h \,\odot\, \mat{d}_h \,\right] \, \dot{\mat{\qhat}} \nonumber \\
    \hspace{-0.7cm} & \,+\, 2\,I_\rho \, \frac{1}{\abss{\phic_h^\prime}^3} \, \left(\mat{\bspline}^\prime\right)^T \, \left[\, \Pdh \,\odot\, \left( \dot{\phic}_h^\prime \,\otimes\, \mat{d}_h \right) \,\right] \, \dot{\phic}_h^\prime \,,
\end{align}

\noindent
where $\odot$ denotes the symmetric product between two vectors $\mat{a}_1$, $\mat{a}_2$, or two second-order tensors $\mat{A}_1$, $\mat{A}_2$, that is:
\begin{subequations}
    \begin{align}
        & \mat{a}_1 \,\odot\, \mat{a}_2 \,=\, \frac{1}{2} \, \left(\,\mat{a}_1 \,\otimes \, \mat{a}_2 \,+\, \mat{a}_2 \,\otimes \, \mat{a}_1 \, \right) \,, \\ 
        & \mat{A}_1 \,\odot\, \mat{A}_2 \,=\, \frac{1}{2} \, \left(\,\mat{A}_1 \, \mat{A}_2 \,+\, \mat{A}_2^T \, \mat{A}_1^T \, \right) \,.
    \end{align}
\end{subequations}

\begin{remark}
    The semi-discrete formulation \eqref{semi-deom} includes neither rotational degrees of freedom nor nodal directors. For cases involving prescribed directions such as clamped boundaries, one can employ for example the bending strip method \cite{Kiendl2010} or the Nitsche's method \cite{Guo2015}. In this work, we strongly enforce the prescribed directions as built-in constraints in the trial and test spaces via the extraction operator (see Section \ref{sec:outlier-removal} and Algorithm \ref{alg-int-outlier-removal}).
\end{remark}

%--------------------------------------------------------------------
\subsection{Isogeometric versus classical nodal finite elements}\label{sec-static-benchmark}

The isogeometric discretization scheme and the standard one based on nodal finite element using cubic Hermite functions both are based on the isoparametric concept. 
When the former applies cubic $C^1$ B-splines, the basis functions of these two schemes span the same function space. 
However, they belong to two different classes of finite element methods. 
While the standard scheme employs classical nodal finite elements, the isogeometric scheme is in the context of isogeometric analysis, where we deal with control points instead of element nodes \cite{hughes_isogeometric_2005, Cottrell:09.1}. 
Moreover, the former requires two variable fields that are the nodal spatial position and the nodal director \cite{gebhardt_2021_beam}, while the latter only considers the positions of the control points as a variable field.
Hence, the isogeometric scheme can reduce the number of degrees of freedom (dofs). 
In particular, a discretization with $n_e$ elements using $C^r$ B-splines of degree $p$ leads to $3\left[n_e \, (p-r) + r + 1\right]$ dofs\footnote{The number of basis functions is $m_{k}-p-1$, where $m_{k}$ is the number of knots in the knot vector. We assume that the $C^r$ B-splines of degree $p$ are defined on an open knot vector with interior knots repeated $(p-r)$-times.}. 
Applying B-splines with maximum continuity $C^{p-1}$ leads to $3(n_e + p)$ dofs, 
while the standard scheme leads to $5(n_e+1)$ dofs. 
For example, employing B-splines of minimum required polynomial degree of $p=2$, that are $C^1$ continuous, leads to a smaller number of dofs of $3(n_e+2)$ for the same number of elements. 
Using cubic $C^1$ B-splines that are in the same function space as cubic Hermite functions, however, leads to more dofs of $6(n_e+1)$ for the same number of elements. 
Due to the different variable fields of these two schemes, they lead to different solution spaces. 
Particularly, using the standard scheme results in a discrete solution in the manifold $\phic_h \in \left(\mathbb{R}^3 \cross S^2\right)^{\m}$ since the director belongs to a unit sphere $S^2$. 
This necessary means that the standard scheme also preserves the manifold structure of the continuous rod. 
Isogeometric discretizations, however, lead to a solution in the Euclidean space $\phic_h \in \left(\mathbb{R}^3\right)^{\ndofs}$, which is a larger space but does not ensure the underlying manifold structure of the rod.

\begin{remark}
    We note that in general, spline basis functions have supports up to $p+1$ elements \cite{hughes_isogeometric_2005, Cottrell:09.1}, where $p$ is their polynomial degree, which is larger than the support of basis functions in classical finite elements. However, this does not lead to a larger bandwidth of matrices obtained from the two discretization schemes \cite[p. 92-97]{Cottrell:09.1}. 
    Hence, given the same number of degrees of freedom, these two schemes require similar computational cost. For a detailed comparison between the two discretization schemes, we refer to \cite[p. 92-97]{Cottrell:09.1}.
\end{remark}

\begin{figure*}[ht!]
    \centering
    \includegraphics[width=1\textwidth]{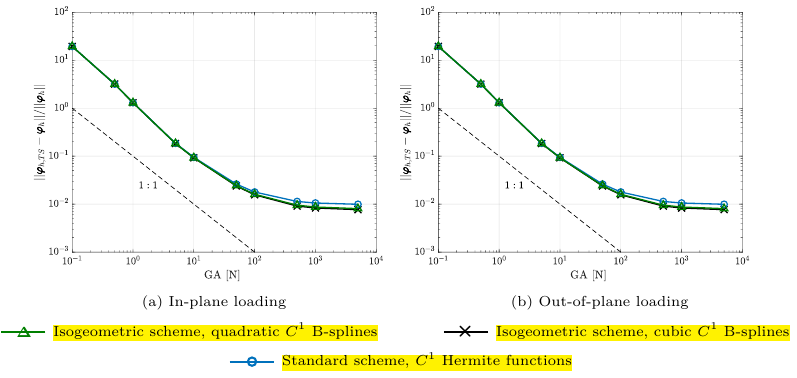}
   
    \caption{Convergence of the relative error between the geometrically exact beam and the nonlinear rod model \cite{gebhardt_2021_beam}, computed with different discretizations on a mesh of $40$ elements, obtained at the last load step.}
    \label{fig-static-TSconverge}
\end{figure*}

We now numerically demonstrate, for two- and three-dimensional static benchmarks from\cite[Sec.~5.1]{gebhardt_2021_beam}, that the isogeometric scheme, with a possibly 
smaller number of dofs, and the standard one approximately achieve the same accuracy. 
We 
consider an initially straight, transversely isotropic, clamped rod of $40$ m with an axial stiffness of $EA \,=\, 100$ N, bending stiffness of $EI \,=\, 200$ Nm$^2$, and 
subjected to an in-plane and out-of-plane loading. 
We compare the response obtained with isogeometric discretizations using quadratic and cubic $C^1$ B-splines, with classical nodal finite elements using cubic $C^1$ Hermite functions, and with a geometrically exact beam model including shear and torsion deformations using linear $C^0$ Lagrange polynomials. 
The results of the last two approaches are provided by \cite{gebhardt_2021_beam}. 
For the first two approaches, we discretize the rod with the same number of $40$ uniform elements with the force divided into $55$ uniform load steps. 
The number of dofs of the isogeometric scheme using quadratic, cubic B-splines, and the standard scheme is $126$, $246$, and $205$, respectively. We have chosen a tolerance of $10^{-10}$ for the Newton-Raphson scheme, which required up to $6$ iterations in all cases. 
We obtained visually indistinguishable snapshots when compared to the standard scheme (see \cite[Fig.~1,4]{gebhardt_2021_beam}) in all cases. 
In Fig.~\ref{fig-static-TSconverge}, we compare the difference between the geometrically exact beam formulation and the isogeometric beam formulation with the one between the former and the standard Hermite scheme. 
To this end, we plot the relative $L^2$ error between the deformed rod obtained with the geometrically exact beam and the one obtained with the isogeometric (black and green curves), and the standard scheme (blue), as a function of the shear stiffness $GA$ employed in the geometrically exact beam model. 
In \cite{gebhardt_2021_beam}, the authors employed this convergence study to point out that the geometrically exact beam model converges to the model using their rod formulation when $GA$ increases, 
since the normal directors tend to become tangent to the deformed rod axis with increasing $GA$. 
We adopt this study to illustrate that the isogeometric scheme results in the same behavior as the standard one, as observed in Fig.~\ref{fig-static-TSconverge} for both the in-plane and out-of-plane loading cases, irrespective of the spline basis employed in the isogeometric scheme. 
We note that the constant error level between the geometrically exact beam and the nonlinear rod of \cite{gebhardt_2021_beam} at large values of $GA$ is due to shear and torsion deformations considered in the former but not in the latter.

\begin{figure*}[ht!]
    \centering
    \def\svgwidth{0.9\textwidth}
    \input{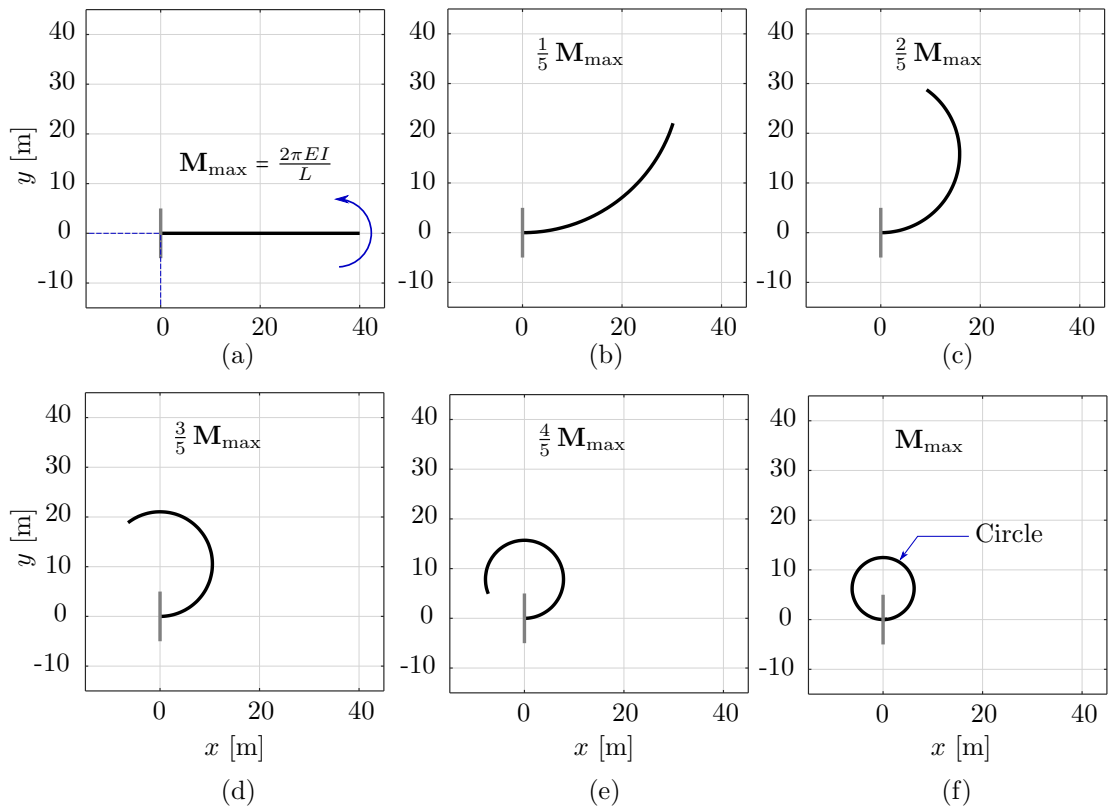}
    \vspace{0.2cm}
    \caption{Deformed configurations of a clamped rod bent to a circle at different load steps, computed with quadratic $C^1$ B-splines ($p=2$) and a mesh of $40$ elements.}
    \label{fig-circle-snapshot}
\end{figure*}

\begin{figure*}[ht!]
    \centering
    \def\svgwidth{0.94\textwidth}
    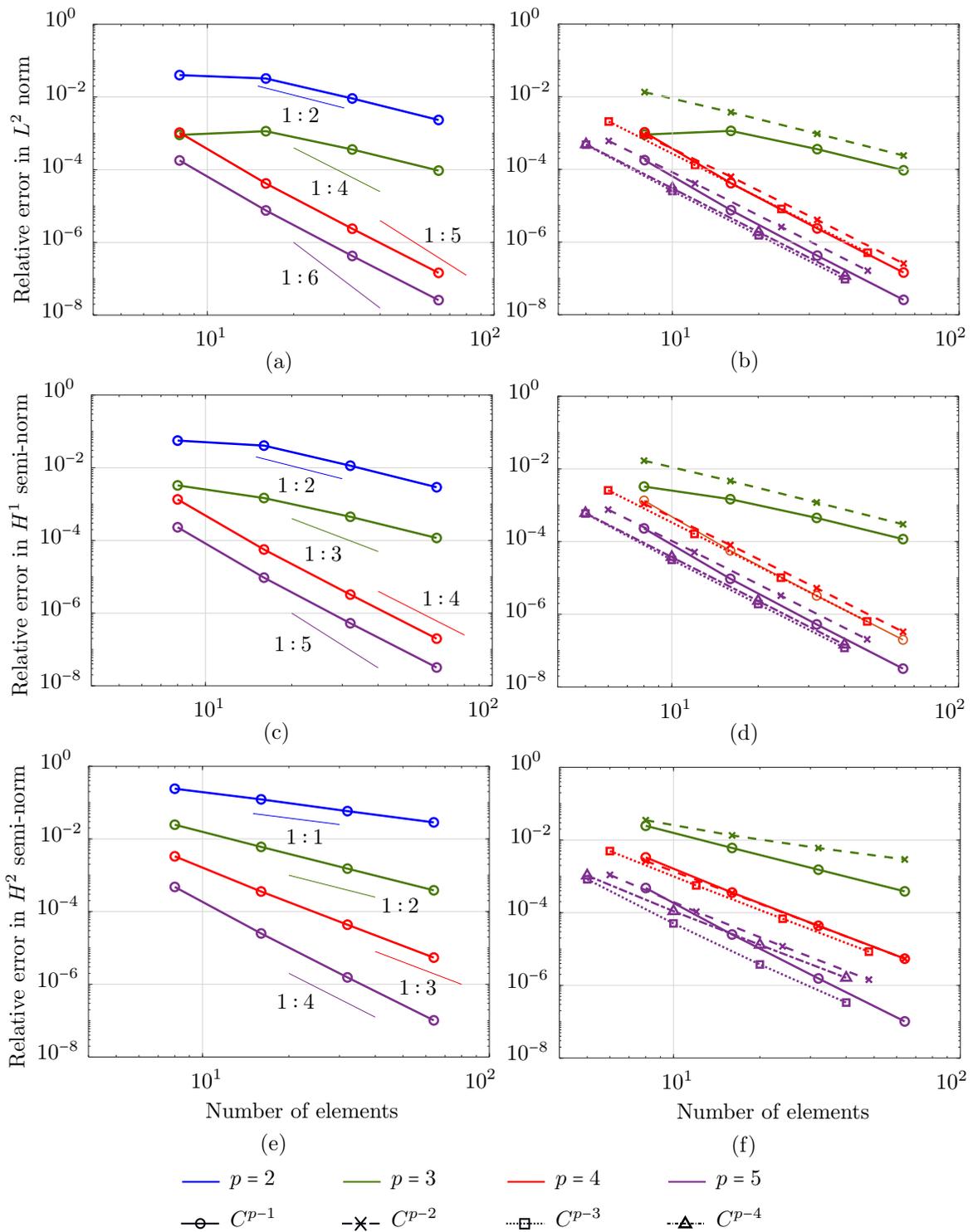
    \vspace{0.2cm}
    \caption{Convergence of relative errors of the clamped rod bent to a circle computed with $C^{p-1}$ B-splines of different degrees $p$ (left column) and with $C^r$ B-splines, $1\,\leq\,r\,\leq\,p-1$ (right column). The reference rate shown in figures on the right column is the convergence rate of linear fourth-order problems \cite{tagliabue_error_2014}}.
    \label{fig-circle-converge}
\end{figure*}

To study the convergence behavior of the isogeometric scheme with mesh refinement, we consider a geometrically nonlinear benchmark of planar roll-up. 
Fig. \ref{fig-circle-snapshot}a illustrates the initial rod subjected to a bending moment $M_{\text{max}}\,=\,\frac{2\,EI\,\pi}{L}$ Nm at its free end, where $L$ is the rod initial length. 
We choose the same material parameters and value of $L$ as in the static benchmarks above. 
We illustrate the deformed rod in a sequence of six load steps, obtained with the isogeometric scheme using quadratic $C^1$ B-splines ($p=2$) in Fig. \ref{fig-circle-snapshot}. 
We observe that, as expected, the deformed rod closes a circle in the last load step. 
In the left column of Fig. \ref{fig-circle-converge}, we plot the convergence of the relative error in $L^2$ norm, $H^1$, and $H^2$ semi-norm, obtained with $C^{p-1}$ B-splines of different polynomial degrees $p$, compared to the exact circle in the last load step. 
As a reference, we included the convergence rate of linear fourth-order problems using isogeometric discretizations based on \cite{tagliabue_error_2014}. 
We observe the same optimal convergence behavior obtained in $H^2$ semi-norm as the linear case for all degrees (see Fig. \ref{fig-circle-converge}e). 
The error in the $H^1$ semi-norm (see Fig. \ref{fig-circle-converge}c), however, converges with the same rate of linear cases only when using even degrees, but with one order lower when using odd degrees. 
Focusing on the error in the $L^2$ norm, we see that the convergence rate is smaller for all $p\,\geq\,3$ (see Fig. \ref{fig-circle-converge}a). 
Furthermore, cubic $C^2$ B-splines illustrate the same convergence rate as quadratic, and quintic functions the same as quartic ones. 
We note that this is not the first time that the different convergence behavior between even and odd degrees has been observed. Such behavior is well-known in isogeometric collocation methods \cite{auricchio_collocation_2010,kiendl_collocation_2015}, however, to our best knowledge, is not yet established for nonlinear problems.

\begin{remark}
    For fourth-order problems, one can mathematically prove the optimal convergence merely in the $H^2$ norm, not in $H^1$ or $L^2$ norms (see e.g. \cite[Eq.~3.13]{tagliabue_error_2014}). Convergence can be proven in $H^1$ and $L^2$ norms, generally using the Lax-Milgram lemma and the Poincar\'e inequality, where the latter leads to additional scaling factors associated with the mesh size, i.e. the proof is not the one of the optimal convergence according to the Lax-Milgram lemma. This necessary means that one may not expect optimal convergence in $H^1$ and $L^2$ norms for fourth-order problems.
\end{remark}

In the second part of the convergence study, 
we investigate the impact of the smoothness of spline basis functions on the convergence of isogeometric discretizations.  
We compute the relative error obtained with $C^r$ B-splines of different polynomial degrees $p\,\geq\,3$, where $1\,\leq\,r\,\leq\,p-1$, and illustrate this 
on the right column of Fig. \ref{fig-circle-converge}. 
Different markers correspond to different continuous orders $r$. 
We can see that the continuity of B-splines does not affect the convergence rate in both the $L^2$ norm and the $H^1$ semi-norm for this planar roll-up example. 
Focusing on the error in the $H^2$ semi-norm, reducing the continuity of B-spline basis functions decreases the convergence rate in cases of odd degrees (see the green and purple curves in Fig. \ref{fig-circle-converge}f), 
however, does not affect this rate in the case of even degree ($p=4$, red curve in Fig. \ref{fig-circle-converge}f). 
We note that the impact of reducing or increasing continuity of spline basis functions is not the same when studying different benchmarks and cases, see e.g. \cite{Cottrell2007,puzyrev_spectral_2018}. 
Furthermore, based on empirical results, we observe that increasing or decreasing the number of quadrature points also does not affect the convergence behavior of the discretizations employed in this example. 
Further numerical investigations of a linear simply supported beam subjected to sinusoidal loads, using the nonlinear rod formulation \cite{gebhardt_2021_beam}, 
shows the same convergence behavior as in the linear cases for all error norms, polynomial degrees, and continuity. 
Hence, we suggest that the convergence behavior in the planar roll-up example results from the nonlinear behavior captured by the employed rod formulation. 
A mathematical error estimate for the considered rod formulation \cite{gebhardt_2021_beam} is outside the scope of this work and is postponed to future work.

%--------------------------------------------------------------------
\subsection{Time integration scheme}\label{sec-time-int}

For time integration of our numerical examples in subsequent sections, we apply the same implicit scheme as in \cite{gebhardt_2021_beam}, which is 
a hybrid combination of the midpoint and trapezoidal rules. 
This implicit scheme 
achieves second-order accuracy, approximately preserves energy, and exactly preserves the linear and angular momenta \cite{gebhardt_2021_beam,gebhardt_implicit_2020}, 
which we also verify via an example of an elastic pendulum in Appendix \ref{sec-appendix1}. 
We note that the employed integration scheme 
is based on the one introduced in \cite{gebhardt_implicit_2020}. 
In this work, to observe the occurrence and investigate the effects of all contents of the response, including the spurious high-frequency contents (see also discussions in the next section), when using isogeometric discretizations, 
we choose to eliminate the dissipation terms of the original scheme. 
We note that this choice is also to serve the comparison purpose with the results obtained with the standard nodal finite elements employed in \cite{gebhardt_2021_beam}. 
Moreover, in \cite{gebhardt_implicit_2020}, the authors have discussed and shown that the original scheme \cite{gebhardt_implicit_2020} is one possible realization of 
the well-established Energy-Dissipative-Momentum-Conserving method \cite{Armero2001,Armero2001a,Armero2003,Romero2002}.

\begin{remark}
    The mass matrix \eqref{mass-matrix} of the considered rod formulation \cite{gebhardt_2021_beam} is configuration-dependent due to its second counterpart $\mat{M}_2$. Thus, an explicit time integration scheme would not be applicable. 
\end{remark}

Consider the semi-discrete formulation \eqref{semi-deom} in space evaluated at time instant $t_{n\,+\,\frac{1}{2}}\in[t_n,t_{n+1}]$:
\begin{equation}\label{eq0}
    \begin{split}
        & \int_0^S \, \delta \mat{\qhat} \, \cdot \, \left(\, \mat{M} (\mat{\qhat}) \, \nabla_{\dot{\mat{\qhat}}} \, \dot{\mat{\qhat}} \,+\, \mat{B} (\mat{\qhat})^T \, \boldsymbol{\sigma}_h \right.  \\
        & \qquad \qquad \qquad \left. -\, \mat{\bspline}^T \, \mat{f}^{\text{ext}} \,\right)_{n+\frac{1}{2}} \, \mathrm{d} \, s \,=\, \mat{0} \,.
    \end{split}    
\end{equation}

\noindent
We approximate the inertial term $\left(\,\mat{M} (\mat{\qhat}) \, \nabla_{\dot{\mat{\qhat}}} \, \dot{\mat{\qhat}}\,\right)_{n+\frac{1}{2}}$ using an extended version of the midpoint rule as follows:
\begin{align}\label{eqs}
    & \left(\,\mat{M} (\mat{\qhat}) \, \nabla_{\dot{\mat{\qhat}}} \, \dot{\mat{\qhat}}\,\right)_{n+\frac{1}{2}} \,\approx\, \nonumber \\
    & \qquad \frac{\mat{M}\left(\mat{\qhat}_{n+1}\right) \, \dot{\mat{\qhat}}_{n+1} \,-\, \mat{M}\left(\mat{\qhat}_{n}\right) \, \dot{\mat{\qhat}}_{n}}{\Delta \, t} \; +\, \\
    \hspace{-0.2cm} & \left\{\, 2\,I_\rho \, \frac{1}{\abss{\phic_h^\prime}^3} \, \left(\mat{\bspline}^\prime\right)^T \, \left[\, \Pdh \,\odot\, \left( \dot{\phic}_h^\prime \,\otimes\, \mat{d}_h \right) \,\right] \, \dot{\phic}_h^\prime \,\right\}_{n+\frac{1}{2}} \,,  \nonumber
\end{align}

\noindent
where $\Delta t$ is the time step. 
The internal term \\
$\left(\,\mat{B} (\mat{\qhat})^T \, \boldsymbol{\sigma}_h\,\right)_{n+\frac{1}{2}}$ is approximated by using the trapezoidal rule as follows:
\begin{equation}\label{eqs2}
    \begin{split}
        & \left(\,\mat{B} (\mat{\qhat})^T \, \boldsymbol{\sigma}_h\,\right)_{n+\frac{1}{2}} \,\approx \\
        & \qquad \frac{\mat{B} \left(\mat{\qhat}_{n+1}\right)^T \, \boldsymbol{\sigma}_{h,n+1} \,+\, \mat{B} \left(\mat{\qhat}_n\right)^T \, \boldsymbol{\sigma}_{h,n}}{2}\,.
    \end{split}    
\end{equation}

\noindent
We also approximate $\mat{\qhat}_{n+\frac{1}{2}}$ and $\dot{\mat{\qhat}}_{n+\frac{1}{2}}$ using the trapezoidal and midpoint rules, respectively, as follows:
\begin{align}\label{eqm}
    \hspace{-0.5cm} \mat{\qhat}_{n+\frac{1}{2}} \,\approx\, \frac{\mat{\qhat}_{n+1} \,+\, \mat{\qhat}_n}{2} \,, \quad 
    \dot{\mat{\qhat}}_{n+\frac{1}{2}} \,\approx\, \frac{\mat{\qhat}_{n+1} \,-\, \mat{\qhat}_n}{\Delta \, t} \,,
\end{align}
and $\dot{\mat{\qhat}}_{n+1}$ as:
\begin{align}\label{eqe}
    \dot{\mat{\qhat}}_{n+1} \,\approx\, \frac{2}{\Delta \, t} \, \left(\, \mat{\qhat}_{n+1} \,-\, \mat{\qhat}_n \,\right) \,-\, \dot{\mat{\qhat}}_n \,.
\end{align}

\noindent
Introducing the approximations \eqref{eqs}-\eqref{eqe} into \eqref{eq0} leads to a system of discrete nonlinear equations in space and time:
\begin{align}\label{eq-nrph}
    \mat{g} \, (\mat{\qhat}_{n+1}) \,=\, \mat{0} \, ,
\end{align}

\noindent
which can be normalized and solved using, for instance, the Newton-Raphson method, for $\mat{\qhat}_{n+1}$. 
$\dot{\mat{\qhat}}_{n+1}$ can be then obtained using \eqref{eqe}. 
The configuration-independent external forces are evaluated at time instant $t_{n\,+\,\frac{1}{2}}$, while the configuration-dependent forces induced by a surrounding flow, discussed in Appendix \ref{sec-flow-force}, are approximated using the midpoint rule along with the approximations \eqref{eqm}-\eqref{eqe}. 
We note that solving \eqref{eq-nrph} using the Newton-Raphson method requires each term in \eqref{eqs}-\eqref{eqs2} and force terms in cases of configuration-dependent forces to be linearized. 
We derive this and their resulting counterparts to the tangent stiffness matrix in Appendix \ref{sec-appendix2}.

%--------------------------------------------------------------------
\subsection{Outlier removal}\label{sec:outlier-removal}

\begin{algorithm*}
\textbf{Input}:
	$\mat{\qhat}_0$,
	$\dot{\mat{\qhat}}_0$ (initial conditions)\\
\textbf{Output}: $\mat{\qhat}\,(t)$, $\dot{\mat{\qhat}}\,(t)$
\begin{algorithmic}[1]
    \State $\mat{C} \,=\, \mathcal{C}\left(p,\,n_{ele},\, \text{continuity } C^r,\, \text{boundary conditions} \right)$     \Comment{Extraction operator of \cite{hiemstra_outlier_2021}}
    \State $n\,=\,0$        \Comment{index of the time step}
	\For{$t$ in $\Delta\,t\,:\,\Delta\,t\,:\,T$}
	\State $\mat{\qhat}\,(t)\,=\, \mat{\qhat}_n$, $\dot{\mat{\qhat}}\,(t)\,=\, \dot{\mat{\qhat}}_n$
    \State $\mat{\qhat}_{n+1}\,=\, \mat{\qhat}_n$   \Comment{inital guess for Newton-Raphson scheme}
	\State $\Delta\,\mat{\qhat}_{n+1} \,=\, 1.0$    \Comment{initialize} 
	\While{$\Delta\,\mat{\qhat}_{n+1} \,\geq\, 10^{-10}$}
	\State $\mat{r} \,=\, -\,\mat{g}\,\left(\mat{\qhat}_{n+1}\right)$   \Comment{residual vector, see also Eq. \eqref{eq-nrph}}
    \State $\mat{K} \,=\, \frac{\partial \, \mat{g}}{\partial \, \mat{\qhat}_{n+1}}$    \Comment{Tangent stiffness matrix, see also Appendix \ref{sec-appendix2}} 
    \State $\mat{r} \,=\, \mat{C}^T \, \mat{r}$, $\mat{K} \,=\, \mat{C}^T \, \mat{K} \, \mat{C}$    \Comment{Removing outliers}
    \State $\Delta\,\mat{\qhat}_{n+1} \,=\, \mat{K}^{-1}\,\mat{r}$
    \State $\mat{\qhat}_{n+1} \,+=\, \Delta\,\mat{\qhat}_{n+1}$
    \State $\dot{\mat{\qhat}}_{n+1} \,\approx\, \frac{2}{\Delta \, t} \, \left(\, \mat{\qhat}_{n+1} \,-\, \mat{\qhat}_n \,\right) \,-\, \dot{\mat{\qhat}}_n$    \Comment{Eq. \eqref{eqe}}
	\EndWhile
    \State $n \,+=\, 1$
	\EndFor
\caption{Implicit time integration scheme employing the strong approach of outlier removal of \cite{hiemstra_outlier_2021}.}\label{alg-int-outlier-removal}
\end{algorithmic}
\end{algorithm*}

Our empirical results, discussed in the next section, indicate that 
the excessive high-frequency contents of the response obtained with isogeometric discretizations lead to unstable computations. 
Based on this observation, we propose to improve the robustness of these discretizations via 
the strong approach of outlier removal introduced in \cite{hiemstra_outlier_2021}, which entirely and merely removes the spurious outlier modes corresponding to the highest frequencies. 
The motivation to remove the outliers is based on the fact that in nonlinear dynamic analysis, the modes are coupled. 
Thus, when the outliers are excited at a time instance, other high-frequency modes might be excited at that time instance and later ones as well. 
We note that the modes mentioned here and throughout this work in the context of nonlinear analysis are the ones corresponding to a linearized problem at a time instance. 
Moreover, comparing our results with the one obtained with nodal finite elements employed in \cite{gebhardt_2021_beam}, we want to emphasize that 
the occurrence of outliers is purely due to the choice of the isogeometric discretization scheme, 
not the employed rod formulation \cite{gebhardt_2021_beam} or the chosen time integration scheme.

To remove outliers, 
the fundamental idea 
of \cite{hiemstra_outlier_2021} 
is to strongly enforce sufficient higher-order continuity at the boundary in terms of additional boundary conditions. 
This idea originates from the observations made with a set of second- and fourth-order eigenvalue problems, where outliers occur due to the lower-order continuity at the boundary of the discrete mode shapes. 
In \cite{hiemstra_outlier_2021}, the authors 
strongly enforce these additional constraints by constructing a new constrained subspace of the original space of B-splines basis functions. 
They computed the so-called extraction operator $\mathcal{C}$ that linearly combines the original basis functions, such that the resulting basis satisfies the enforced constraints. 
One can compute the operator $\mathcal{C}$ by 
finding a basis for the null space of a matrix including merely boundary constraints (see \cite[Eq.~(25a)]{hiemstra_outlier_2021}). 
$\mathcal{C}$ is computed for the complete space of spline basis functions but only modifies functions associated with the boundary conditions. 
It has the structure of a block diagonal matrix consisting of block matrices at each constrained boundary and the identity matrix (see \cite[Eq.~(26)]{hiemstra_outlier_2021}). 
$\mathcal{C}$ preserves the important properties of B-splines, such as non-negativity and minimum local support.

In this work, we apply the extraction operator $\mathcal{C}$ to both the right- and left-hand sides of the semi-discrete formulation \eqref{semi-deom} by multiplying with the matrix $\mat{C}$ at each time step, where $\mat{C}$ is the matrix expression of $\mathcal{C}$ \cite{hiemstra_outlier_2021}. 
Since $\mathcal{C}$ only depends on the chosen spline space and the boundary conditions, $\mat{C}$ remains constant and can be computed once before the time integration procedure. 
It is associated with the property of the chosen spline space, does not depend on the choice of the time integration scheme, and thus can be applied together with other schemes than the one chosen in this work. 
$\mat{C}$ inherits the structure of $\mathcal{C}$, i.e. $\mat{C}$ is a block diagonal matrix. 
In general, $\mat{C}$ is not a square matrix but a sparse matrix. 
We illustrate the implicit time integration scheme employing the outlier removal approach \cite{hiemstra_outlier_2021} in Algorithm \ref{alg-int-outlier-removal}. 
We note that in our computations, together with the additional boundary conditions for outlier removal, we also strongly enforce the essential boundary conditions via $\mat{C}$. 
We apply $\mat{C}$ to the global stiffness matrix $\mat{K}$ before solving the matrix equation at each time step. 
Since $\mat{C}$ is a sparse matrix, the product $\mat{C}^T \mat{K} \, \mat{C}$ is also a sparse matrix. 
For the technical details of the computation of $\mat{C}$, we refer to \cite{hiemstra_outlier_2021}. 
We note that there exist other approaches to remove outliers, either strongly \cite{Manni2021} or weakly \cite{deng_outlier_2021,horger_penalty_2019}. For an overview of these approaches, we refer to \cite{hiemstra_outlier_2021,Nguyen_outlier_2022} and references therein.

An alternative line of approaches that also tackle spurious high-frequency modes such as outliers is to employ time integration schemes with high-frequency dissipation such as the generalized-$\alpha$ method \cite{Raknes2013,chung1993}, the HHT-$\alpha$ method \cite{Hilber1977}, the $\rho_{\infty}$-Bathe method \cite{Noh2019}. 
Such schemes are robust and have found their wide applications in structural dynamics and fluid mechanics communities. 
In general, integration schemes with higher-frequency dissipation damp out the spurious high-frequency responses by introducing numerical damping, which can be regulated via a dissipation parameter. 
In this work, since we want to observe the occurrence and investigate the effect of outlier modes on the performance of isogeometric discretizations, we chose another time integration scheme that is the hybrid combination of the midpoint and trapezoidal rules and does not damp out the high-frequency modes. 
Furthermore, for the comparison purpose between the isogeometric discretization scheme and the standard scheme using nodal finite elements employed in \cite{gebhardt_2021_beam}, we employed the same time integration scheme as in \cite{gebhardt_2021_beam}. 
We note that using schemes with high-frequency dissipation and choosing a zero dissipation parameter, i.e. zero numerical damping, also allows the occurrence of outliers. 
For the comparison between two discretization schemes, however, the integration scheme employed in \cite{gebhardt_2021_beam} is the simplest choice for our work.

\section{Robustness of isogeometric discretizations}\label{sec-robust}

In this section, we numerically demonstrate, for two- and three-dimensional benchmarks, that the isogeometric discretization scheme using cubic $C^1$ or $C^2$ splines is less robust than the standard one based on nodal finite elements using cubic Hermite functions. 
We then discuss important factors, such as the high-frequency contents of the response and round-off errors due to floating-point arithmetic, which may negatively affect the robustness of the employed discretization scheme. 
We show that employing the strong approach of outlier removal \cite{hiemstra_outlier_2021}, discussed in the previous section, improves the robustness of isogeometric schemes. 
In addition, we discuss the influence of the configuration-dependent mass matrix \eqref{mass-matrix} on the accuracy of the response.

%--------------------------------------------------------------------
\subsection{Two- and three-dimensional benchmarks}\label{sec-dyn-benchmark}

\begin{figure*}[ht!]
    \centering
    \def\svgwidth{1\textwidth}
    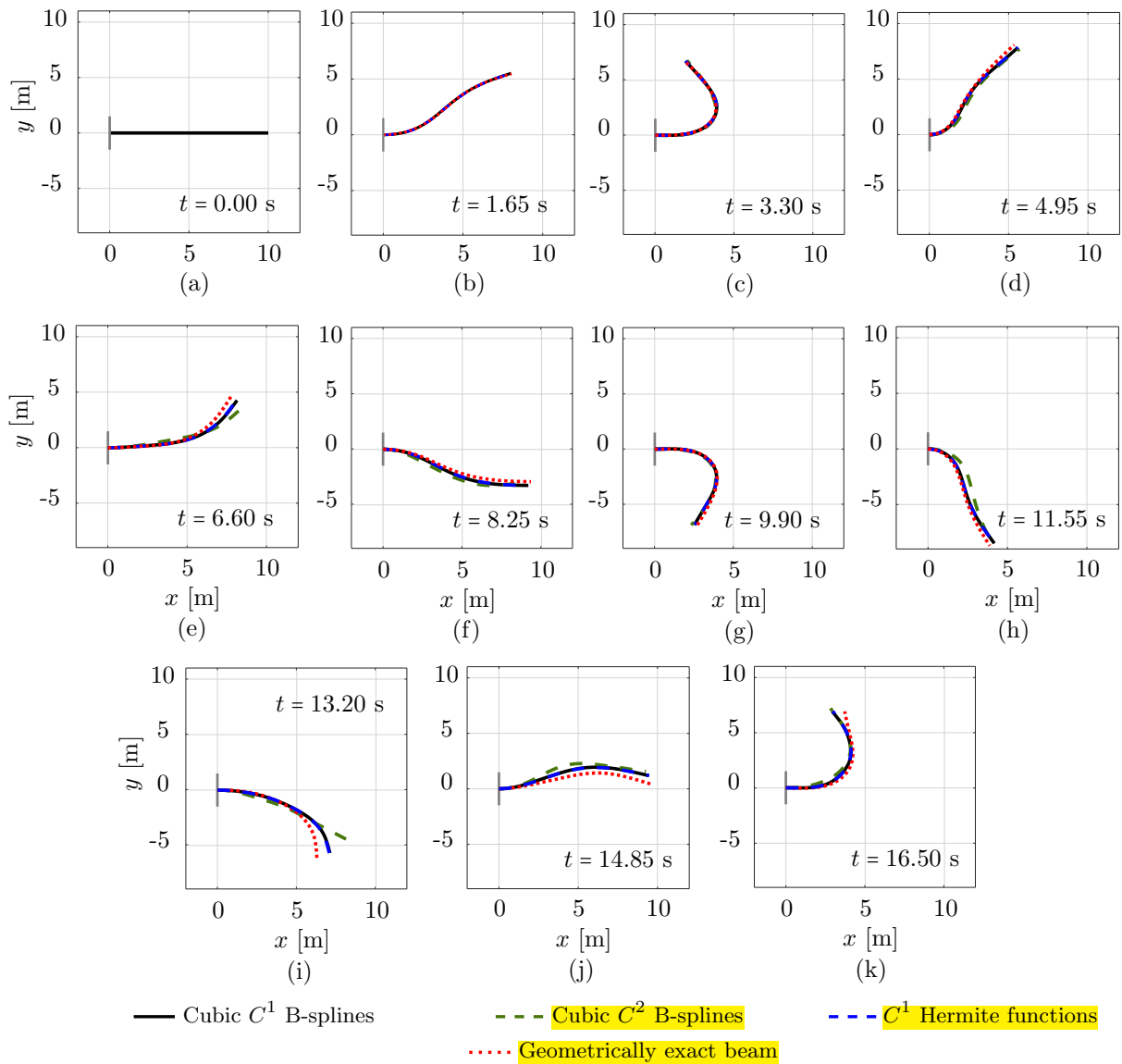
    \vspace{0.0cm}
    \caption{Deformed configurations of a clamped rod subjected to an in-plane loading at different time steps, computed with different discretizations and a geometrically exact beam model.}
    \label{fig-dyn-2d-snapshots}
\end{figure*}

\begin{figure*}[ht!]
    \centering
    \def\svgwidth{0.73\textwidth}
    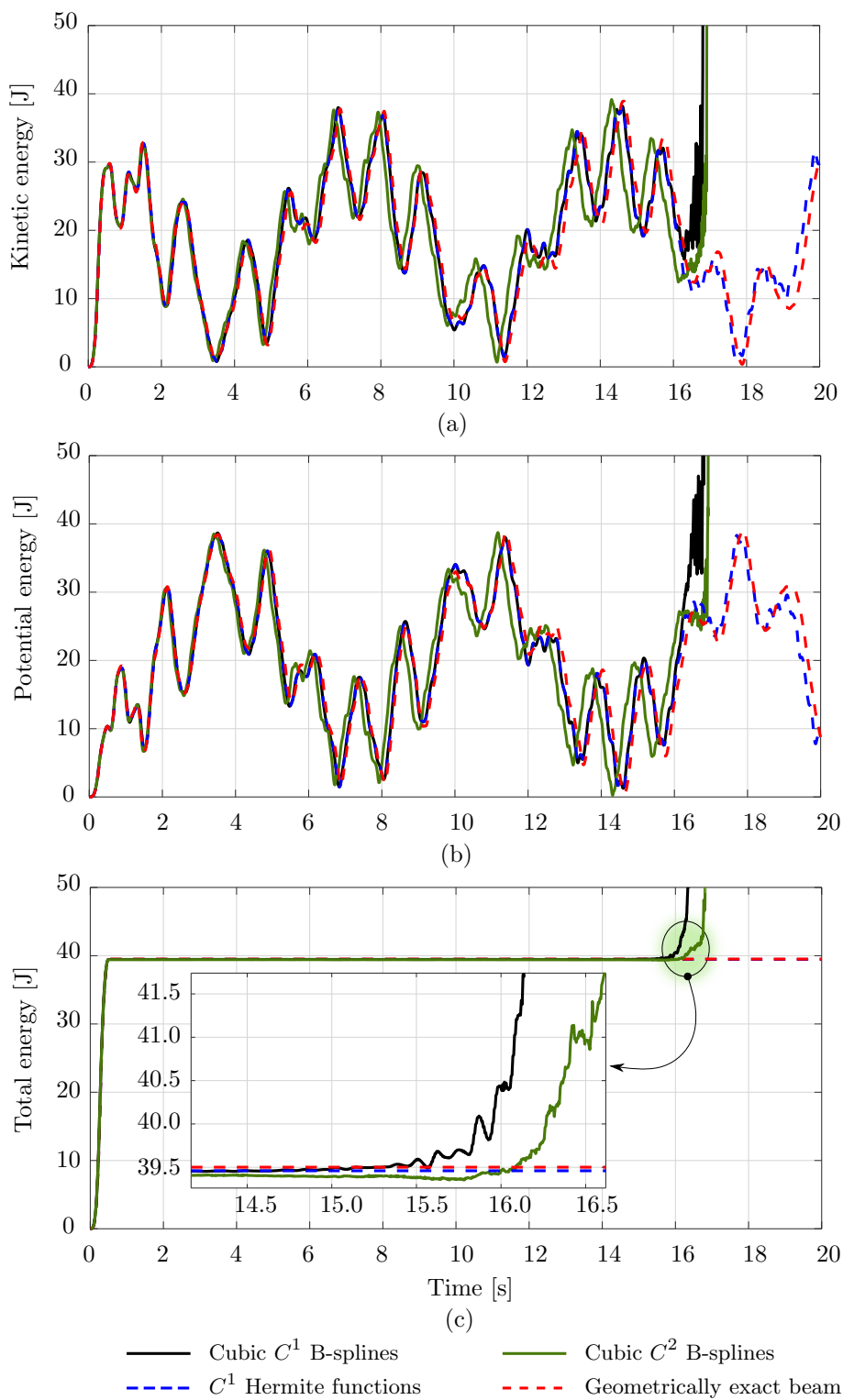
    \vspace{0.2cm}
    \caption{The energy of a clamped rod subjected to a vanishing in-plane loading, computed with different discretizations and a geometrically exact beam model.}
    \label{fig-dyn-2d-energy1}
\end{figure*}

\begin{figure*}[ht!]
    \centering
    \def\svgwidth{1\textwidth}
    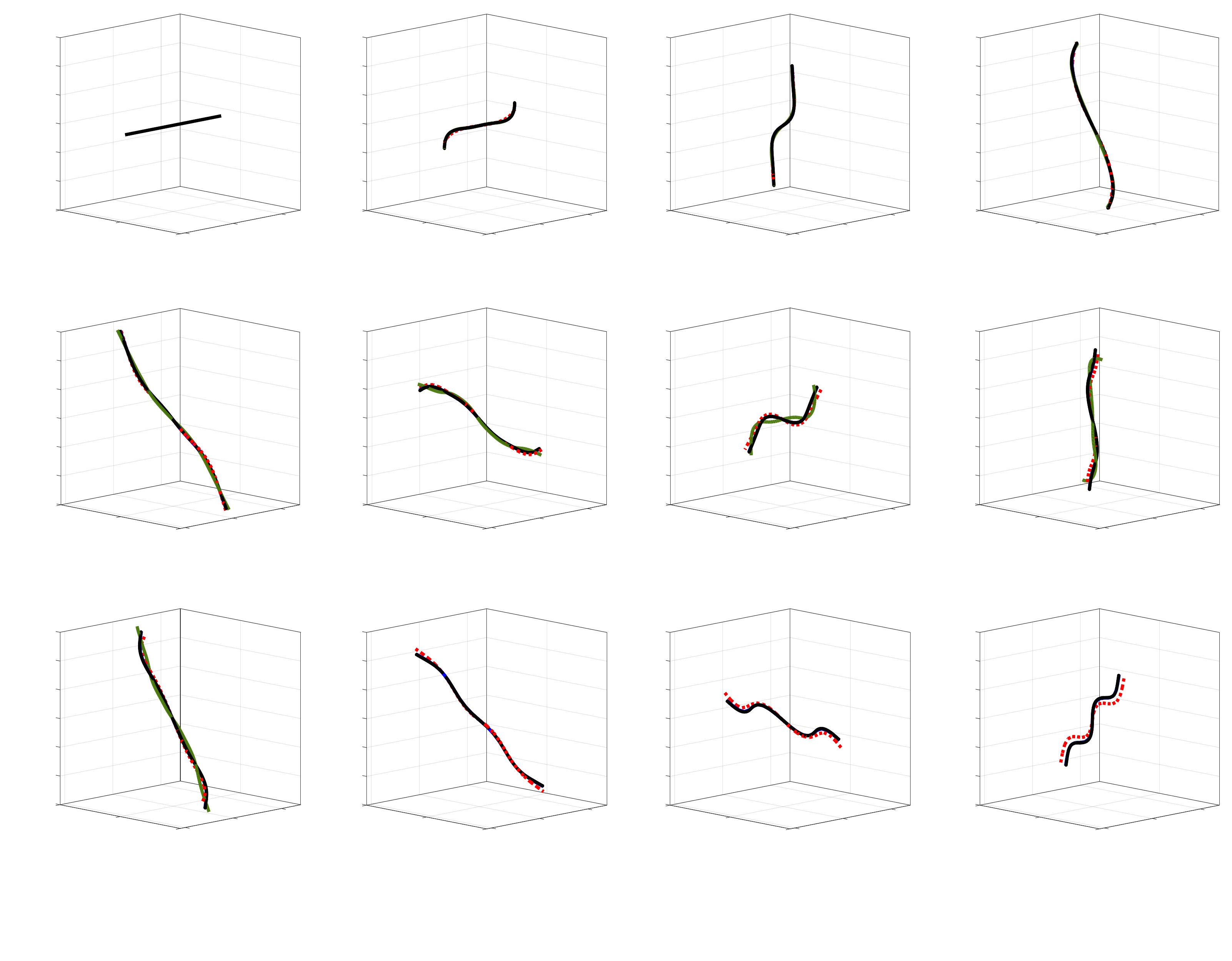
    \vspace{0.0cm}
    \caption{Deformed configurations of an unconstrained rod subjected to a vanishing out-of-plane loading at different time steps, computed with different discretizations and a geometrically exact beam model.}
    \label{fig-dyn-3d-snapshots}
\end{figure*}

We consider the two dynamic benchmarks of \cite[Sec.~5.2]{gebhardt_2021_beam} and compare the responses obtained with isogeometric discretizations, using cubic $C^1$ and $C^2$ B-splines, against the standard scheme. 
In all cases, the rod is uniformly discretized into $20$ elements, which leads to $126$, $69$, and $105$ dofs using these three approaches, respectively. 
We also include the solution obtained with the geometrically exact beam model including shear and torsion using linear $C^0$ Lagrange polynomials as a reference, which is provided by \cite{gebhardt_2021_beam}.

The first example consists of an initially straight, transversely isotropic, clamped rob subjected to the following in-plane vanishing load at its free end:
\begin{align}\label{eq-vanish-F}
    \hspace{-0.7cm} \mat{F}(t) \,=\, \begin{cases}
        \frac{t}{0.5\,t_c} \, & \mat{F}_c \,, \quad 0\,\leq\,t\,\leq\,0.5\,t_c \,, \\
        \frac{2}{t_c}\, \left(\,t_c\,-\,t\,\right) \, & \mat{F}_c \,, \quad 0.5\,t_c\,<\,t\,\leq\,t_c \,, \\
        0 \, & \mat{F}_c \,, \quad t \,>\, t_c \,,
    \end{cases}
\end{align}
where $t_c$ and $\mat{F}_c$ are chosen to be $t_c\,=\,0.5$ s and $\mat{F}_c \,=\, (0,\, 30,\, 0)$ N, respectively. 
Thus, the rod deforms in the $xy$-plane. 
The rod has an initial length of $10$ with an initial director of $(1,\;0,\;0)$, a circular cross-section with a diameter of $0.01$ m, Young's modulus $E\,=\,2\cdot10^{11} $ N/m$^2$, and mass density $\rho\,=\,7900$ kg/m$^3$. 
We choose the same time step of $\Delta\,t \,=\, 0.005$ s, a simulation time of $20$ s, and a tolerance of $10^{-10}$ for the Newton-Raphson scheme, as done in \cite{gebhardt_2021_beam}. 
Fig. \ref{fig-dyn-2d-snapshots} illustrates the deformed configurations in a sequence of eleven load steps, obtained with cubic $C^1$ (black curves) and $C^2$ (green dashed curves) B-splines, 
with the standard scheme using cubic $C^1$ Hermite functions (blue dashed curves), 
and with the geometrically exact beam model using linear $C^0$ Lagrange polynomials (red dotted curves). 
We note that for this example, after $16.5$ s, the computations using isogeometric discretizations become unstable and the Newton scheme does not converge anymore, while the one using the standard scheme remained stable during the computation time of $20$ s. 
Nevertheless, we observe that during the first $16.5$ s the isogeometric scheme using cubic $C^1$ splines and the standard one result in virtually identical responses, since their basis functions span the same space and have the same approximation power. 
Comparing these results to the one obtained with the geometrically exact beam, we see that the difference between these two models increases progressively in time. 
This may result from \reviewerII{different employed time integration schemes for these models}, as discussed in \cite{gebhardt_2021_beam}. 
Focusing on the responses obtained with cubic $C^2$ B-splines (green dashed curves), we can see that these differ from those obtained with cubic $C^1$ B-splines at certain time steps.

\begin{remark}
    For this benchmark, computations using the isogeometric scheme with quadratic $C^1$ B-splines become unstable already after $3.5$ s. Thus, in this work, we did not apply quadratic B-splines for any dynamic benchmark.
\end{remark}

To gain better insights, we illustrate the time history of the kinetic, potential, and total energy resulting from the aforementioned approaches in Fig. \ref{fig-dyn-2d-energy1}. 
We observe that, starting around $4$ s, there is a phase shift between the responses obtained with cubic $C^2$ and $C^1$ B-splines (Fig. \ref{fig-dyn-2d-energy1}a,b). 
In particular, 
using cubic $C^2$ B-splines leads to a smaller phase than $C^1$ B-splines, which consequently leads to different responses (black and green curves) observed in Fig. \ref{fig-dyn-2d-snapshots}. 
We note that for the case of cubic spline functions, our convergence study of the planar roll-up in Section \ref{sec-static-benchmark} (see green curves in Fig.~\ref{fig-circle-converge}b,d,f), shows a slightly higher error level when 
using cubic $C^2$ than $C^1$ B-splines. 
This decrease in the approximation power of cubic splines with higher continuity for this nonlinear rod may relate to the different phases and responses observed in Fig. \ref{fig-dyn-2d-snapshots}. 
Focusing on the responses around $16$ s, before the energy, obtained with isogeometric discretizations, shoots up, indicating unstable computations, we see that high-frequency modes are excited and lead to fluctuations in the response. 
Due to a smaller phase, we expect that this generally occurs earlier when using cubic $C^2$ B-splines than $C^1$ splines. This necessary means that increasing the continuity of spline basis functions reduces the robustness of the corresponding discretizations. 
Nevertheless, before the computations become unstable, the isogeometric scheme approximately preserves 
the same total energy as the standard one and the geometrically exact beam model (Fig. \ref{fig-dyn-2d-energy1}c). 
We conclude that it is less robust than the standard one based on nodal finite elements \cite{gebhardt_2021_beam}. 
Using discretizations with
cubic splines of higher continuity reduces the phase of the responses, which may 
reduce their robustness and may relate to their reduced approximation power in this case of nonlinear rods \cite{gebhardt_2021_beam}.

\begin{remark}
    We note that a weaker continuity than the $C^1$ continuity, the geometric $G^1$ continuity, can be applied for Kirchhoff rods (see e.g. \cite{Greco2014,Greco2016}) and obtained with isogeometric discretizations, which may achieve better robustness than those using $C^1$ continuity or higher. Enforcing the $G^1$ continuity is equivalent to enforcing the continuity of the unit director at the element interface. For more details of this condition and an overview of different approaches for its enforcement, we refer for instance to \cite{Greco2014,Greco2016} and references therein. Studying the performance of isogeometric discretizations with $G^1$ continuity for the nonlinear rod \cite{gebhardt_2021_beam} and comparing them with those of $C^1$ continuity or higher is out of the scope of this work and thus is considered for future work.
\end{remark}

\begin{figure*}[ht!]
    \centering
    \def\svgwidth{0.70\textwidth}
    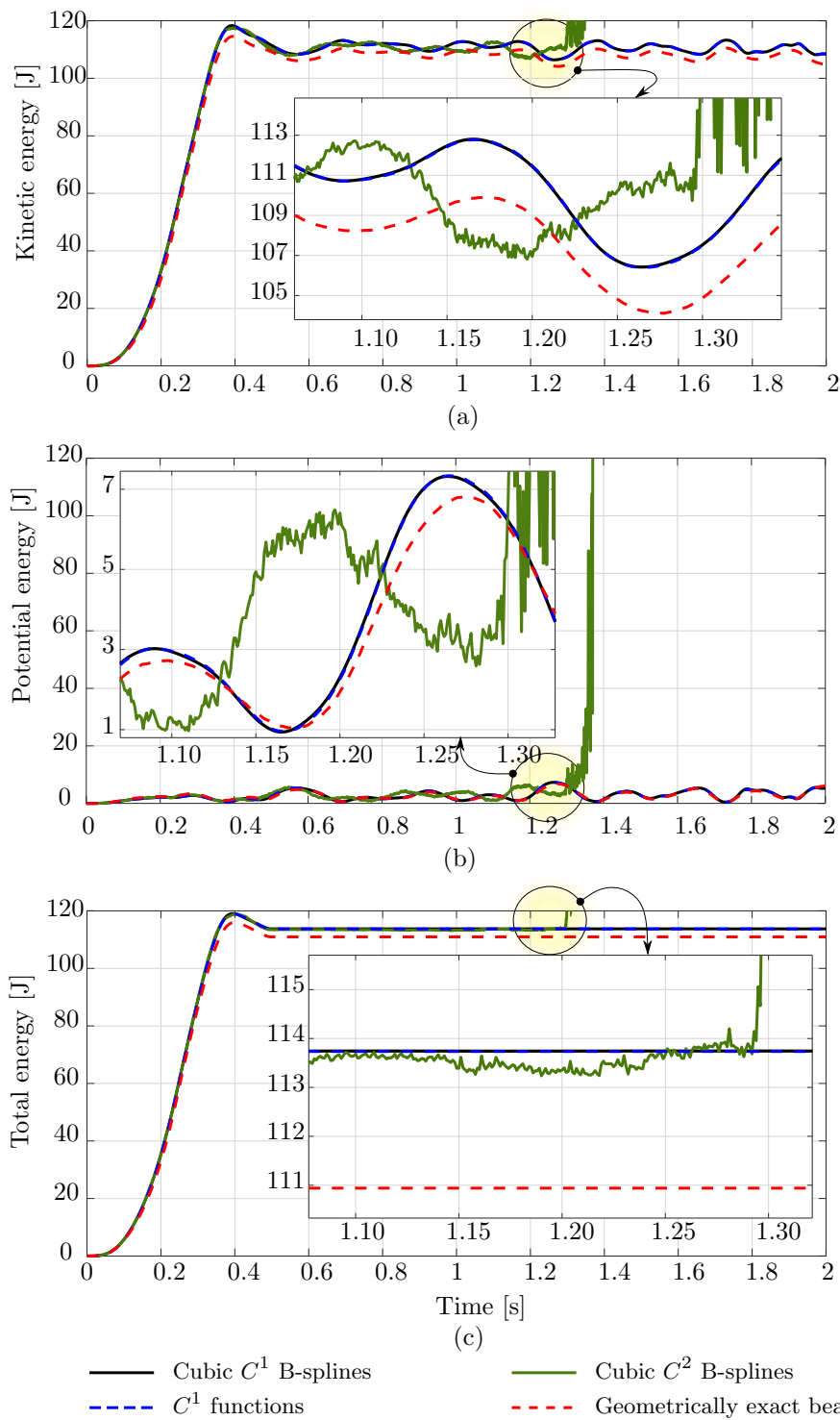
    \vspace{0.2cm}
    \caption{The energy of an unconstrained rod subjected to a vanishing out-of-plane loading, computed with different discretizations and a geometrically exact beam model.}
    \label{fig-dyn-3d-energy}
\end{figure*}

\begin{figure*}[ht!]
    \centering
    \def\svgwidth{0.70\textwidth}
    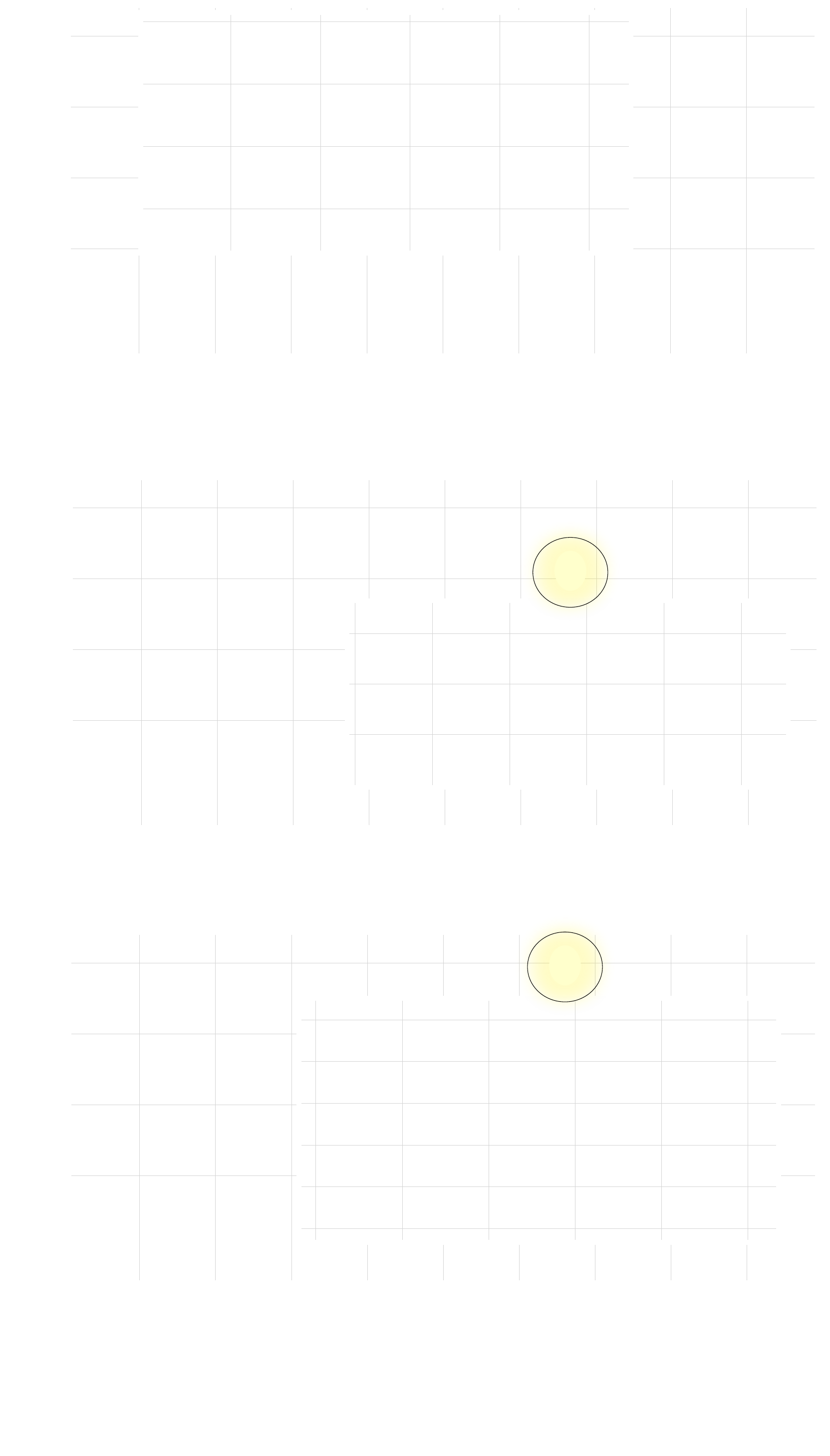
    \vspace{0.2cm}
    \caption{Three components of the angular momentum of an unconstrained rod subjected to a vanishing out-of-plane loading, computed with different discretizations and a geometrically exact beam model.}
    \label{fig-dyn-3d-angMom}
\end{figure*}

The second benchmark is an unconstrained rod of the same length and material as the clamped rod studied above. 
The rod has a smaller cross-section with a diameter of $0.005$ m and is 
subjected to the vanishing load \eqref{eq-vanish-F}, with 
$t_c=0.5$ s and $\mat{F}_c$ given by: 
\begin{align}
    & \mat{F}_c \,=\, (-30,\, -30,\, 0)\,\text{N} \quad \text{at } s\,=\,0 \,, \nonumber \\
    & \mat{F}_c \,=\, (30,\, 30,\, 0)\,\text{N} \quad \text{at } s\,=\,L \,, \nonumber \\
    & \mat{F}_c \,=\, (0,\, 0,\, -24)\,\text{N} \quad \text{at } s\,=\,L/20 \,, \nonumber \\
    & \mat{F}_c \,=\, (0,\, 0,\, 24)\,\text{N} \quad \text{at } s\,=\,19\,L/20 \,. \nonumber
\end{align}
 
\noindent
Consequently, the rod deforms freely in three-dimensional space. 
For this benchmark, we also choose the same time step $\Delta\,t \,=\, 0.001$ s, a simulation time of $2$ s, and a tolerance of $10^{-10}$ for the Newton-Raphson scheme, as done in \cite{gebhardt_2021_beam}. 
Fig. \ref{fig-dyn-3d-snapshots} illustrates the deformed configurations in a sequence of twelve load steps obtained with the aforementioned approaches. 
For this example, after $1.35$ s the computation using cubic $C^2$ B-splines becomes unstable and the Newton scheme does not converge anymore, while that using cubic $C^1$ B-splines remains stable during the simulation time of $2$ s. 
We have similar observations as in the case of the clamped rod above: both the isogeometric scheme using cubic $C^1$ B-splines and the standard one show the same accuracy, while the former using $C^2$ B-splines leads to distinct responses. 
This is also illustrated for the energy in Fig. \ref{fig-dyn-3d-energy} and the three components of the angular momentum in Fig. \ref{fig-dyn-3d-angMom}. 
We also observe a phase shift 
in the case of cubic $C^2$ B-splines, before the high-frequency modes affect the response and the computation becomes unstable. 
We see that, for this benchmark, using smoother splines of $C^2$ also leads to less robust computation than using $C^1$, as discussed in the first benchmark of a clamped rod.

%--------------------------------------------------------------------
\subsection{Robustness improvement with outlier removal}\label{sec:robust_outlier_removal}

\begin{figure*}[ht!]
    \centering
    \def\svgwidth{0.65\textwidth}
    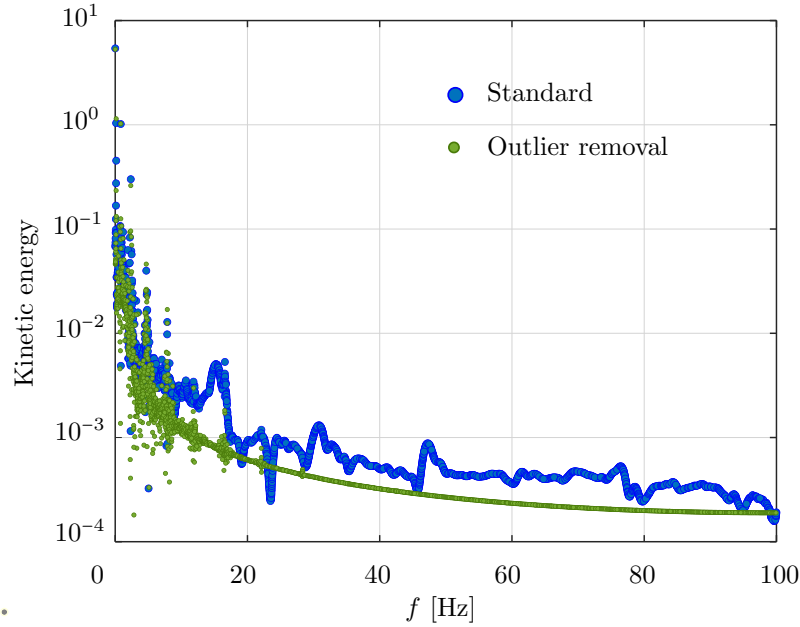
    \vspace{0.2cm}
    \caption{Fast Fourier transformation of the kinetic energy for the clamped rod in Fig.~\ref{fig-dyn-2d-snapshots}, subjected to a half of the load during a longer time period of $100$ s.}
    \label{fig-fft-energy}
\end{figure*}

The results discussed in the previous subsection indicate that excited high-frequency contents in the response reduce the robustness of isogeometric discretizations. 
To gain better insights, we perform the Fast Fourier Transformation (FFT) of the kinetic energy in the case of the clamped rod studied in the previous subsection, subjected to half of the load during a longer simulation time, $T=100$ s. 
To avoid noises in the transformed signal, we consider the energy as long as it is smaller than a threshold of $40$ J. 
Fig. \ref{fig-fft-energy} illustrates the FFT of the kinetic energy obtained with cubic $C^1$ B-splines (blue circles), where we can observe that not only one, but almost all high-frequency modes are excited.

\begin{table*}[ht]
	\centering
	{\small
	\begin{tabularx}{0.9\linewidth}{ >{\centering\arraybackslash\hsize=.12\hsize}X | >{\centering\arraybackslash\hsize=.29\hsize}X >{\centering\arraybackslash\hsize=.29\hsize}X >{\centering\arraybackslash\hsize=.29\hsize}X }
	\toprule
    \multirow{2}{*}{$n_{ele}$}   &   \multirow{2}{*}{cubic Hermite functions}     & cubic $C^1$ B-splines,     & cubic $C^1$ B-splines, \\
     & & standard & outlier removal \\
    \midrule
    \multicolumn{4}{c}{$\Delta\,t\,=\,0.0025$ s} \\
    \midrule
    $2$  & $1.000000000000000$ &   $1.000000000000002$ &  $1.000000000000001$ \\
    $4$  & $1.000000000000002$ &   $1.000000000000001$ &  $0.999999999999972$ \\
    $8$  & $0.999999999999988$ &   $1.000000000000044$ &  $1.000000000000038$ \\
    $16$ & $0.999999999999940$ &   $1.000000000000002$ &  $0.999999999999941$ \\
    $32$ & $0.999999999999979$ &   $0.999999999999240$ &  $1.000000000000137$ \\
	\midrule
    \multicolumn{4}{c}{$\Delta\,t\,=\,0.005$ s} \\
    \midrule
    $2$  & $0.999999999999999$ &   $0.999999999999999$ &  $1.000000000000000$ \\
    $4$  & $1.000000000000006$ &   $0.999999999999998$ &  $0.999999999999997$ \\
    $8$  & $1.000000000000005$ &   $0.999999999999981$ &  $0.999999999999984$ \\
    $16$ & $1.000000000000013$ &   $1.000000000000125$ &  $0.999999999999771$ \\
    $32$ & $0.999999999999995$ &   $1.000000000000421$ &  $1.000000000000139$ \\
    \midrule
    \multicolumn{4}{c}{$\Delta\,t\,=\,0.01$ s} \\
    \midrule
    $2$  & $0.999999999999997$ &   $0.999999999999992$ &  $0.999999999999997$ \\
    $4$  & $1.000000000000004$ &   $1.000000000000003$ &  $1.000000000000003$ \\
    $8$  & $0.999999999999998$ &   $1.000000000000004$ &  $0.999999999999982$ \\
    $16$ & $0.999999999999999$ &   $1.000000000000045$ &  $0.999999999999932$ \\
    $32$ & $1.000000000000020$ &   $0.999999999999663$ &  $1.000000000000187$ \\
	\bottomrule
\end{tabularx}}
\caption{Determinant of $\mat{\tilde{A}}$ in Eq. \eqref{eq-matrixA}.
}\label{tab-detA}
\end{table*}

\begin{figure*}[ht!]
    \centering
    \def\svgwidth{0.85\textwidth}
    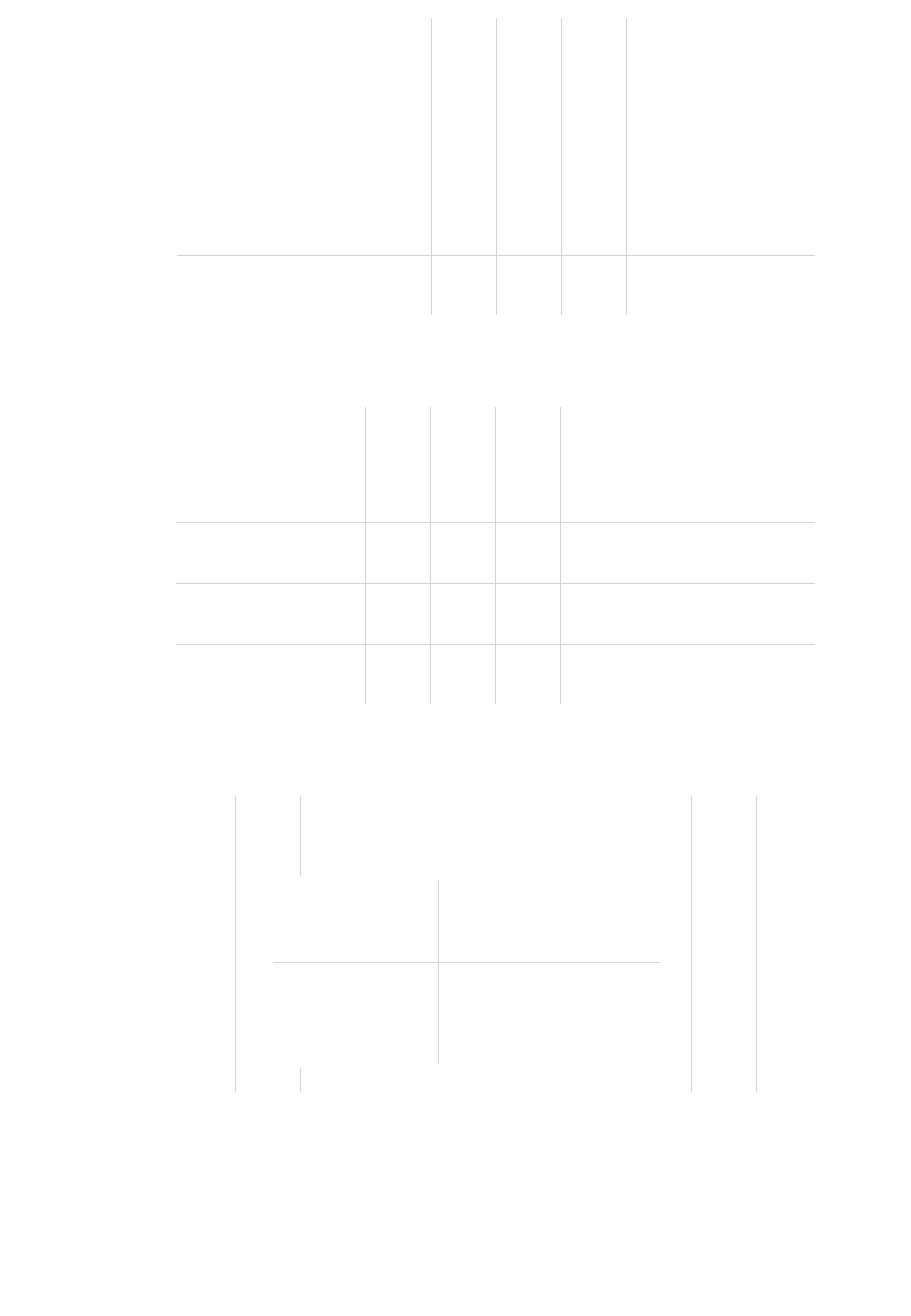
    \vspace{0.2cm}
    \caption{The energy of the clamped rod subjected to a vanishing in-plane loading in Fig.~\ref{fig-dyn-2d-snapshots}, computed with different basis functions, time steps, with and without outlier removal.}
    \label{fig-dyn-2d-energy2}
\end{figure*}

As discussed in Section~\ref{sec:outlier-removal}, we employ the strong approach of outlier removal \cite{hiemstra_outlier_2021} to remove the high-frequency contents in the response, such that the computations become more robust. 
In Fig. \ref{fig-fft-energy}, we include the kinetic energy (green circles) with the outliers removed. 
We observe that almost all high-frequency contents are removed, as expected.  
Fig. \ref{fig-dyn-2d-energy2} shows the kinetic, potential, and total energy of the clamped rod studied in the previous subsection. 
Here, we present a comparison among the energy obtained with isogeometric discretizations using cubic $C^1$ and $C^2$ B-splines, employing outlier removal, (black and green curves, respectively), the energy obtained with the standard scheme \cite{gebhardt_2021_beam} (blue dashed curve), and the geometrically exact beam model (red dashed curve). 
We observe that, as expected, the outlier removal approach improves the robustness of the isogeometric scheme. 
\reviewerII{We note and recall the conclusions in \cite{hiemstra_outlier_2021} that the employed outlier removal approach does not affect the accuracy of the original discretizations and hence results in the same deformed configurations.} 
We see that in the case using cubic $C^2$ B-splines (green curves), other remaining high-frequency modes are excited, as illustrated in the inset figure in Fig. \ref{fig-dyn-2d-energy2}c. 
Hence, we expect that in this case, the computation will become unstable at a later time, which might occur in the case using cubic $C^1$ B-splines as well. 
Moreover, we can see that the phase shift between cubic $C^1$ and $C^2$ B-splines is not affected by the outlier removal approach. 
We thus assume that this phase shift results from the different continuity of the spline basis functions, which may relate to their different approximation power, as discussed in the previous section. 
To investigate the effect of the time step on the robustness, we also include the results obtained with cubic $C^1$ B-splines using a smaller time step of $\Delta\,t \,=\, 0.0025$ s (purple curve) in Fig. \ref{fig-dyn-2d-energy2}. 
We see that decreasing the time step also improves the robustness of the discretization scheme. 
Focusing on the accuracy of the results, we observe that neither employing the outlier removal approach nor decreasing the time step, negatively affects the accuracy of the responses. 
We conclude that the robustness of the isogeometric scheme can be improved using the strong approach of outlier removal or by decreasing the time step, without compromising the obtained accuracy.

\begin{remark}
    \reviewerII{For the considered rods, we also studied the effect of increasing the polynomial degree $p$, either of $C^1$ or $C^{p-1}$ B-splines, on the responses, combining with the outlier removal approach \cite{hiemstra_outlier_2021}. We observe that irrespective of constant or increasing continuity, increasing $p$ approximately leads to the same deformed configurations and energy responses before the computations become unstable. 
    We also saw that in general, increasing $p$, however, results in responses containing more high-frequency contents, i.e. less robust computations at later time.}
\end{remark}

To gain a better understanding of how outlier modes and time step size affect the robustness of the discretization scheme, we consider the free vibration of a linear one-dimensional fourth-order problem such as an unconstrained Euler-Bernoulli beam. The semi-discrete equations of motion, in matrix form, are:
\begin{align}\label{eq-lin-beam}
    \mat{M}\, \ddot{\mat{u}} \,+\, \mat{K}\,\mat{u} \,=\, \mat{0} \,,
\end{align}

\noindent
where $\mat{M}$, $\mat{K}$, and $\mat{u}$ are the global mass matrix, global stiffness matrix, and the unknown displacement vector of the control points, respectively. 
Employing the implicit time integration scheme discussed in Section \ref{sec-time-int}, the inertial and internal elastic terms are approximated at the time instant $t_{n+\frac{1}{2}}$ as follows:
\begin{align}
    & \mat{M}\, \ddot{\mat{u}}_{n+\frac{1}{2}} \,\approx\, \frac{\mat{M}\, \dot{\mat{u}}_{n+1}\,-\, \mat{M}\, \dot{\mat{u}}_{n}}{\Delta\,t}\,, \\
    & \mat{K}\,\mat{u}_{n+\frac{1}{2}} \,\approx\, \frac{\mat{K}\,\mat{u}_{n+1} \,+\, \mat{K}\,\mat{u}_{n}}{2}\,.
\end{align}

\noindent
Inserting these approximations into \eqref{eq-lin-beam} and applying \eqref{eqe} to approximate $\dot{\mat{u}}_{n+1}$, we obtain the following system of equations:
\begin{align}
    \underbrace{\begin{bmatrix}
        \Delta\,t \, \mat{K} & 2\,\mat{M} \\
        -2\,\mat{I} & \Delta\,t \, \mat{I}
    \end{bmatrix}}_{\mat{A}_L} \, & \begin{bmatrix}
        \mat{u}_{n+1} \\ \dot{\mat{u}}_{n+1}
    \end{bmatrix} \,=\, \underbrace{\begin{bmatrix}
        -\Delta\,t \, \mat{K} & 2\,\mat{M} \\
        -2\,\mat{I} & -\Delta\,t \, \mat{I}
    \end{bmatrix}}_{\mat{A}_R} \, \begin{bmatrix}
        \mat{u}_{n} \\ \dot{\mat{u}}_{n}
    \end{bmatrix} \nonumber \\
    & \begin{bmatrix}
        \mat{u}_{n+1} \\ \dot{\mat{u}}_{n+1}
    \end{bmatrix} \,=\, \underbrace{\mat{A}_L^{-1} \, \mat{A}_R}_{\tilde{\mat{A}}} \, \begin{bmatrix}
        \mat{u}_{n} \\ \dot{\mat{u}}_{n}
    \end{bmatrix} \, . 
\end{align}

In linear cases, the stiffness and mass matrices, and thus $\tilde{\mat{A}}$, do not depend on the configuration, and remain constant during the time integration. 
Consequently, we can compute the response at time step $t_{n+1}$ in terms of the initial conditions as follows:
\begin{align}\label{eq-matrixA}
    \hspace{-0.5cm}\begin{bmatrix}
        \mat{u}_{n+1} \\ \dot{\mat{u}}_{n+1}
    \end{bmatrix} \,=\, \underbrace{\tilde{\mat{A}} \, \ldots \, \tilde{\mat{A}}}_{(n+1) \text{ times}} \, \begin{bmatrix}
        \mat{u}_{0} \\ \dot{\mat{u}}_{0}
    \end{bmatrix} \,=\, \tilde{\mat{A}}^{n+1} \, \begin{bmatrix}
        \mat{u}_{0} \\ \dot{\mat{u}}_{0}
    \end{bmatrix} \, .
\end{align}

\noindent
The requirement for \eqref{eq-matrixA} to have an unique solution is that the matrix $\tilde{\mat{A}}$ is a convergent matrix \cite{burden_NumAnalysis}. 
This necessarily requires $\abss{\det\left( \tilde{\mat{A}} \right)} \, < \, 1$. Table \ref{tab-detA} illustrates the determinant of $\tilde{\mat{A}}$ computed with cubic Hermite functions, and cubic $C^1$ B-splines with and without outlier removal, using different time steps and meshes. 
We observe that the determinant is either smaller or larger than $1.0$ 
with a tolerance in the range of $\left[10^{-13},\,10^{-15}\right]$, 
i.e. machine accuracy. 
Such a problem can be attributed, in principle, to round-off errors due to floating-point arithmetic, where the numerical deviation from $1.0$ highly depends on the time step, mesh size, and basis functions, i.e. on the time integration and spatial discretization schemes. 
We can also see that neither using cubic Hermite functions, employing outlier removal, nor reducing the time step, guarantees that $\abss{\det\left( \tilde{\mat{A}} \right)} \, < \, 1$. 
Hence, there is no indication that one discretization scheme ensures the existence of the solution at an arbitrary time step $t_{n+1}$ for \eqref{eq-matrixA}, and the other schemes do not. 
For nonlinear problems, since $\tilde{\mat{A}}$ is configuration-dependent, it is not trivial to identify parameters and calibrate them in order to ensure $\abss{\det\left( \tilde{\mat{A}} \right)} \, < \, 1$. 
In addition, there might be other factors that are decisive for the existence of the solution at $t_{n+1}$, and thus for the robustness of the discretization scheme. 
This requires further study and investigation, which we postpone to future work.

%--------------------------------------------------------------------
\subsection{On the influence of the configuration-dependent mass matrix}

\begin{figure*}[ht!]
    \centering
    \def\svgwidth{0.60\textwidth}
    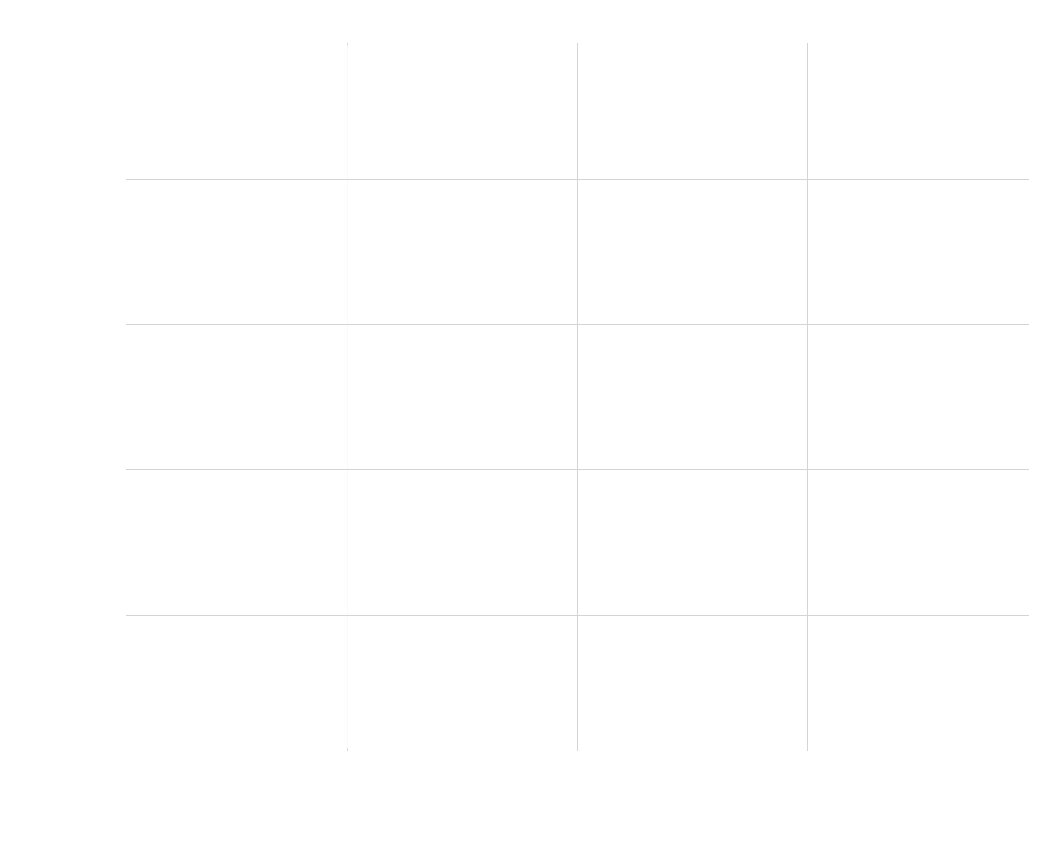
    \vspace{0.2cm}
    \caption{The relative change of the deformed configuration of an unconstrained rod computed with different scaling factors $\alpha$.}
    \label{fig-mass_perturbation}
\end{figure*}

We now briefly discuss the behavior of the configuration-dependent mass matrix \eqref{mass-matrix}, and whether this counterpart can be omitted. 
We recall that the mass matrix \eqref{mass-matrix}, $\mat{M}$, consists of two counterparts, $\mat{M}_1$ and the rotational counterpart $\mat{M}_2$. 
While $\mat{M}_1$ does not depend on the current rod configuration and remains constant, $\mat{M}_2$ is configuration-dependent. 
In most of the common nonlinear dynamic analyses, $\mat{M}_2$ is considered a small perturbation and thus is neglected, which results in a constant configuration-independent mass matrix.

To investigate the effect of $\mat{M}_2$ on the system response and whether $\mat{M}_2$ behaves as a regular perturbation, we consider the unconstrained rod benchmark studied in Section \ref{sec-dyn-benchmark}, 
computed with the isogeometric discretization scheme using cubic $C^1$ B-splines. 
In particular, we scale $\mat{M}_2$ by a factor $\alpha \,\in\, [0,\,1]$. The mass matrix \eqref{mass-matrix} then becomes:
\begin{align}
    \mat{M} \,=\, \mat{M}_1 \,+\, \alpha \, \mat{M}_2 \,. \label{mass-matrix2}
\end{align}

\noindent
Fig. \ref{fig-mass_perturbation} illustrates the relative change in the $L^2$ norm of the deformed rod, $\phic_h$, during the simulation time of $2$ s. 
Different colors correspond to different values of $\alpha$. 
We observe that this relative change increases in time and with increasing $\alpha$. 
This observation implies that, for the studied rod, $\mat{M}_2$ does not behave as a regular perturbation, and thus should not be neglected. 
Mathematical proof and analysis of this counterpart of the mass matrix is out of the scope of this paper  
and is postponed to future works.

\section{Application to a swinging rubber rod}\label{sec-swinging-rod}

In this section, we apply the nonlinear rod formulation \cite{gebhardt_2021_beam}, discretized with isogeometric discretizations, to a swinging rubber rod, which can represent for instance a sort of mooring lines or cables. Such structures are essential components, for instance, of offshore wind turbines, pumping kites, oil and gas platforms, etc. Hence, our simulation is a part and intermediate step towards the aero-hydro-elastic simulation framework accounting for such applications. 
We consider conservative and non-conservative forces such as gravity, forces induced by surrounding wind or water, and pulsating forces. 
We illustrate that our rod formulation is able to represent the rod nonlinear behavior that is a combination of 
elastic vibrations and rigid body oscillations around a static position and shape, which deforms differently at different force frequencies. 
Based on our results in the previous section, we spatially discretize the swinging rod with cubic $C^1$ B-splines ($p=3$) and improve its robustness with the strong approach of outlier removal. 
We start with benchmarking our approach via an example of a swinging rod under gravitational loading \cite{gebhardt_rod_2019,masud_rod_2000}.

%--------------------------------------------------------------------
\subsection{Benchmarking}

We consider an initially straight rod of length $L=1.0$ m, with a circular cross-section, a radius of $5 \,\cdot\, 10^{-3}$ m, Young's modulus $E\,=\,5\cdot10^{6}$ N/m$^2$, and mass density $\rho \,=\, 1100$ kg/m$^3$. 
The rod is subjected merely to 
gravitational loading \cite{gebhardt_rod_2019}, which we simulate with a direction of $(0,\;-1,\;0)$ while the initial director of the rod is $(1,\;0,\;0)$. 
Thus, the rod deforms in the $xy$-plane. 
We choose a time step of $\Delta\,t \,=\, 0.01$ s, a simulation time of $2.4$ s, and a tolerance of $10^{-10}$ for the Newton-Raphson scheme.

\begin{figure*}[ht!]
    \centering
    \def\svgwidth{0.85\textwidth}
    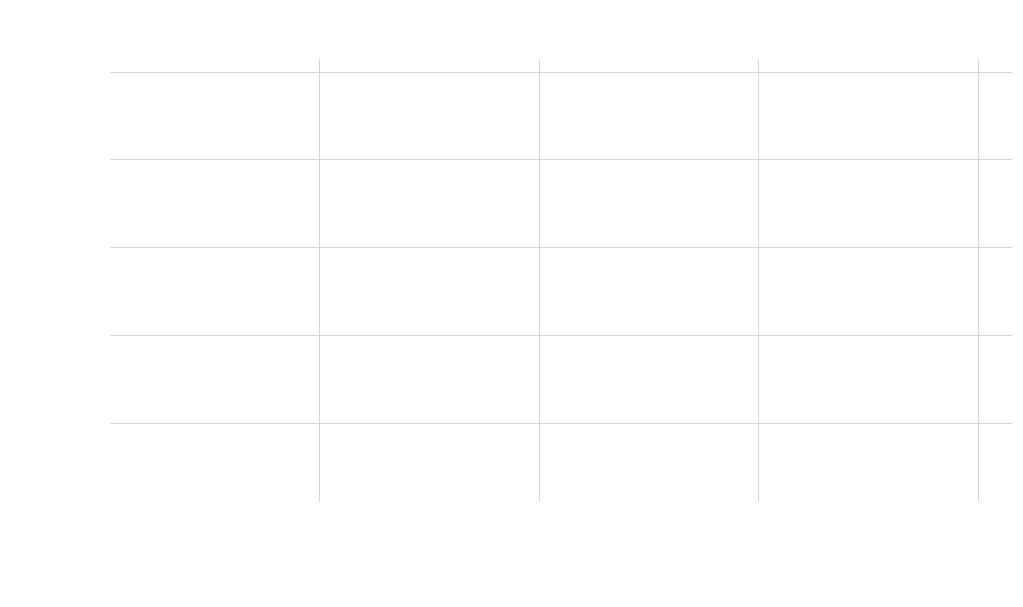
    \vspace{0.2cm}
    \caption{Deformed configurations of a swinging rubber rod due to gravity, computed with cubic $C^1$ B-splines ($p=3$) and outliers removed.}
    \label{fig-swinging_rod_snapshot}
\end{figure*}

\begin{figure*}[ht!]
    \centering
    \def\svgwidth{0.75\textwidth}
    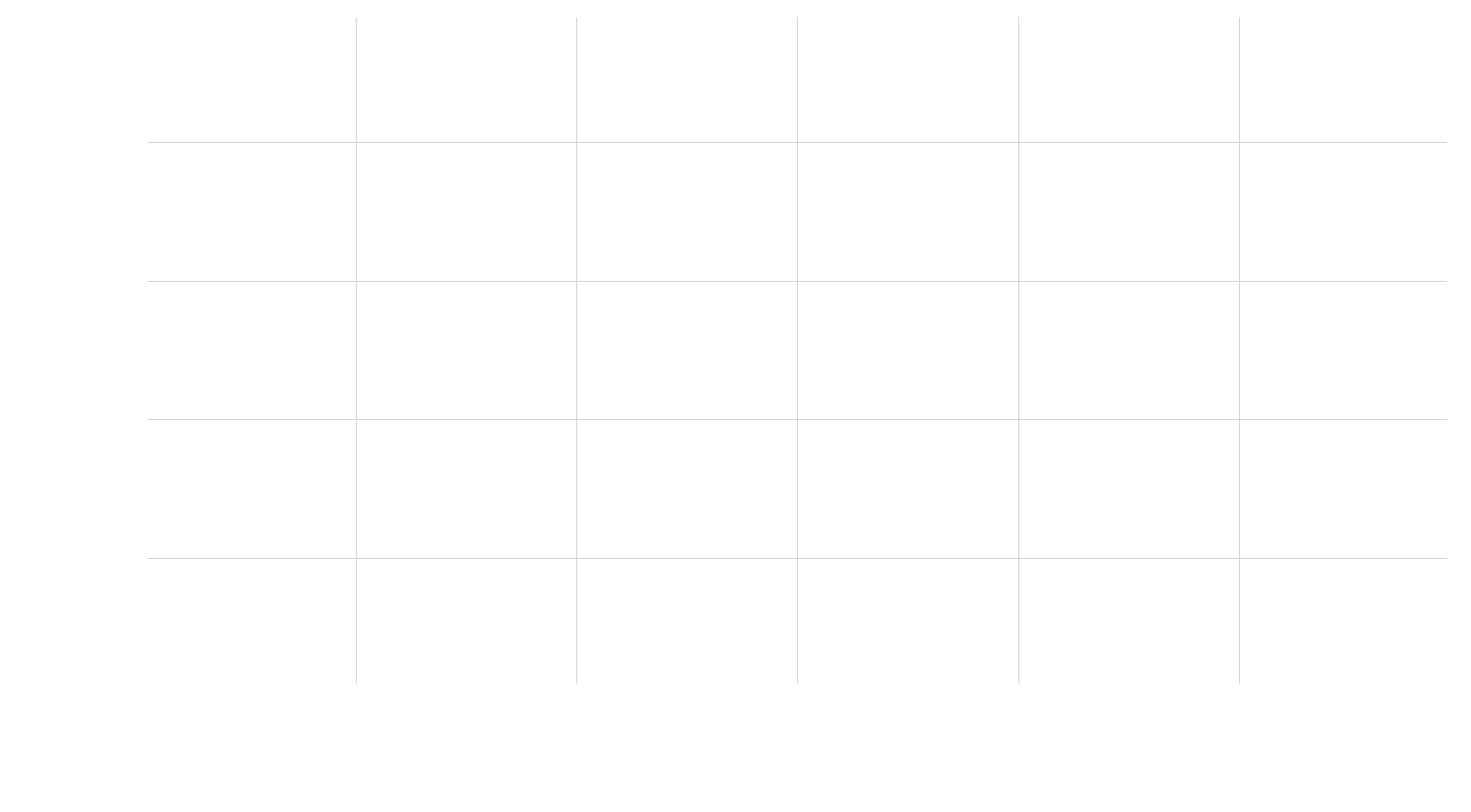
    \vspace{0.2cm}
    \caption{The tip displacements of the swinging rubber rod in Fig. \ref{fig-swinging_rod_snapshot}.}
    \label{fig-swinging_rod_tip_displ}
\end{figure*}

\begin{figure*}[ht!]
    \centering
    \def\svgwidth{0.77\textwidth}
    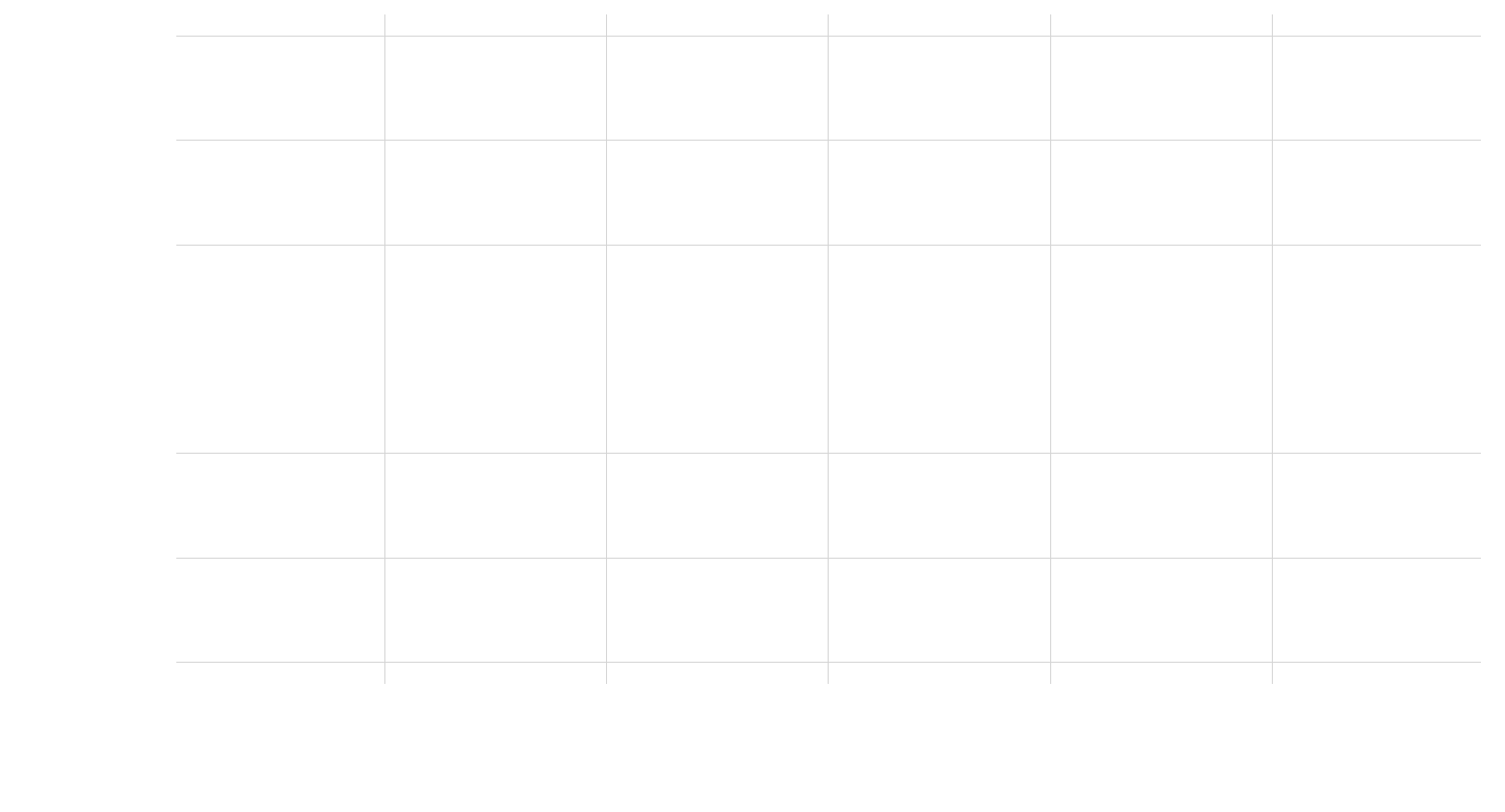
    \vspace{0.2cm}
    \caption{The energy of the swinging rubber rod in Fig. \ref{fig-swinging_rod_snapshot}.}
    \label{fig-swinging_rod_energy}
\end{figure*}

Fig. \ref{fig-swinging_rod_snapshot} illustrates the deformed configuration of the studied rod at twelve time steps, 
and Fig. \ref{fig-swinging_rod_tip_displ} shows the time evolution of the horizontal ($u_x(t)$ in blue) and the vertical displacement ($u_y(t)$ in red) at the rod free-end. 
We observe in Fig. \ref{fig-swinging_rod_snapshot} that 
the swinging rod behaves similarly to an elastic pendulum, and thus has a stable equilibrium configuration when it is aligned with the $y-$axis and its free-end is located at $(x,y,z)=(0,-1,0)$ m. 
Due to the highly nonlinear nature captured by the current formulation, the rod exhibits large elastic rotations and displacements in time. 
Comparing these results with those of \cite[Fig.~8,9]{gebhardt_rod_2019}, we observe an apparent difference after $2.0$ s, for instance, the deformed configuration at $t=2.4$ s clearly distinguishes from that in \cite[Fig.~8]{gebhardt_rod_2019}. 
We further compare the energy obtained with our approach, illustrated in Fig. \ref{fig-swinging_rod_energy}, with that in \cite[Fig.~10]{gebhardt_rod_2019}. 
We see that the total energy and its counterparts show approximately the same time history as the reference, with an exception of around $2.4$ s. 
While the reference and our results both consist of high-frequency contents around this time, the reference energy does not jump to a large value as ours. 
This is consistent with different deformed configurations after $2.0$ s observed in Fig. \ref{fig-swinging_rod_snapshot}. 
We conclude that despite the absence of outliers in our computation, our approach using cubic $C^1$ B-splines is less robust than the scheme using classical nodal finite elements with cubic Hermite functions applied in \cite{gebhardt_rod_2019}. 
Nevertheless, it leads to the same behavior for the swinging rod, given that the computation is stable.

%--------------------------------------------------------------------
\subsection{Dynamic response to wind forces}

\begin{figure}[ht!]
    \centering
    \def\svgwidth{0.6\columnwidth}
%% Accompanies image file 'Wind_profile_swinging_rod.pdf' (pdf, eps, ps)
\begingroup%
  \makeatletter%
  \providecommand\color[2][]{%
    \errmessage{(Inkscape) Color is used for the text in Inkscape, but the package 'color.sty' is not loaded}%
    \renewcommand\color[2][]{}%
  }%
  \providecommand\transparent[1]{%
    \errmessage{(Inkscape) Transparency is used (non-zero) for the text in Inkscape, but the package 'transparent.sty' is not loaded}%
    \renewcommand\transparent[1]{}%
  }%
  \providecommand\rotatebox[2]{#2}%
  \newcommand*\fsize{\dimexpr\f@size pt\relax}%
  \newcommand*\lineheight[1]{\fontsize{\fsize}{#1\fsize}\selectfont}%
  \ifx\svgwidth\undefined%
    \setlength{\unitlength}{210bp}%
    \ifx\svgscale\undefined%
      \relax%
    \else%
      \setlength{\unitlength}{\unitlength * \real{\svgscale}}%
    \fi%
  \else%
    \setlength{\unitlength}{\svgwidth}%
  \fi%
  \global\let\svgwidth\undefined%
  \global\let\svgscale\undefined%
  \makeatother%
  \begin{picture}(1,1.78571429)%
    \lineheight{1}%
    \setlength\tabcolsep{0pt}%
    \put(0,0){\includegraphics[width=\unitlength,page=1]{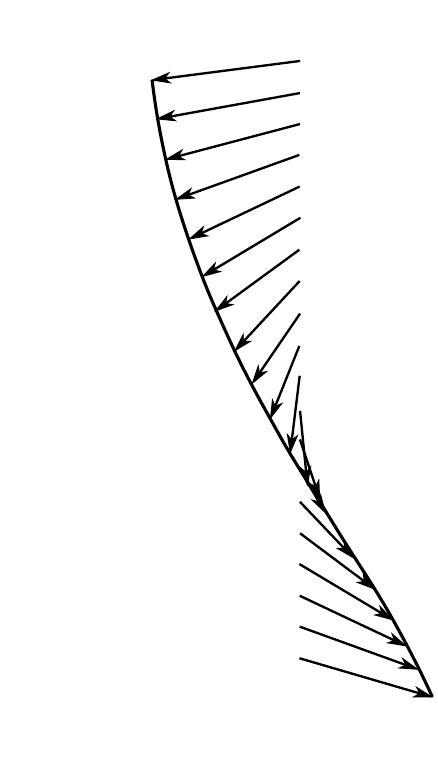}}%
    \put(0.70997935,1.4593084){\color[rgb]{0,0,0}\makebox(0,0)[lt]{\lineheight{1.25}\smash{\begin{tabular}[t]{l}$-\mathbf{E}_3$\end{tabular}}}}%
    \put(0.47853501,1.73112855){\color[rgb]{0,0,0}\makebox(0,0)[lt]{\lineheight{1.25}\smash{\begin{tabular}[t]{l}$\mathbf{E}_1$\end{tabular}}}}%
    \put(0.44719462,1.47140176){\color[rgb]{0,0,0}\makebox(0,0)[lt]{\lineheight{1.25}\smash{\begin{tabular}[t]{l}$\mathbf{E}_2$\end{tabular}}}}%
    \put(0.00593046,0.85558682){\color[rgb]{0,0,0}\makebox(0,0)[lt]{\lineheight{1.25}\smash{\begin{tabular}[t]{l}$\mathbf{v}_{wind}(z)$\\$= v_0\,\mathbf{d}_{wind}(z)$\end{tabular}}}}%
    \put(0,0){\includegraphics[width=\unitlength,page=2]{Wind_profile_swinging_rod.pdf}}%
    \put(0.44501618,1.57111167){\color[rgb]{0,0,0}\makebox(0,0)[lt]{\lineheight{1.25}\smash{\begin{tabular}[t]{l}$\beta_0$\end{tabular}}}}%
  \end{picture}%
\endgroup%

    \caption{A wind profile with constant amplitude and rotating direction around the $z$-axis.}
    \label{fig-wind_profile}
\end{figure}

\begin{figure*}[ht!]
    \centering
    \def\svgwidth{0.7\textwidth}
    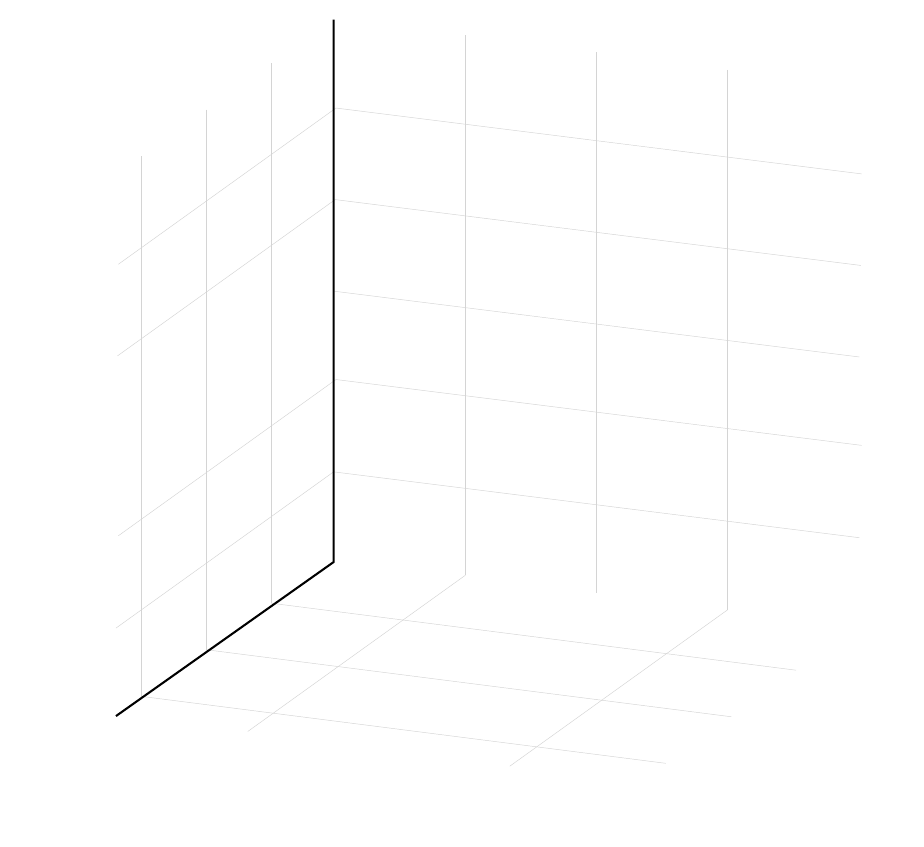
    \vspace{0.2cm}
    \caption{Deformed configurations of a swinging rubber rod due to gravity and the wind profile in Fig. \ref{fig-wind_profile}, computed with cubic $C^1$ B-splines ($p=3$) and outliers removed.}
    \label{fig-swinging_rod_snapshot3d}
\end{figure*}

\begin{figure*}[ht!]
    \centering
    \def\svgwidth{0.75\textwidth}
    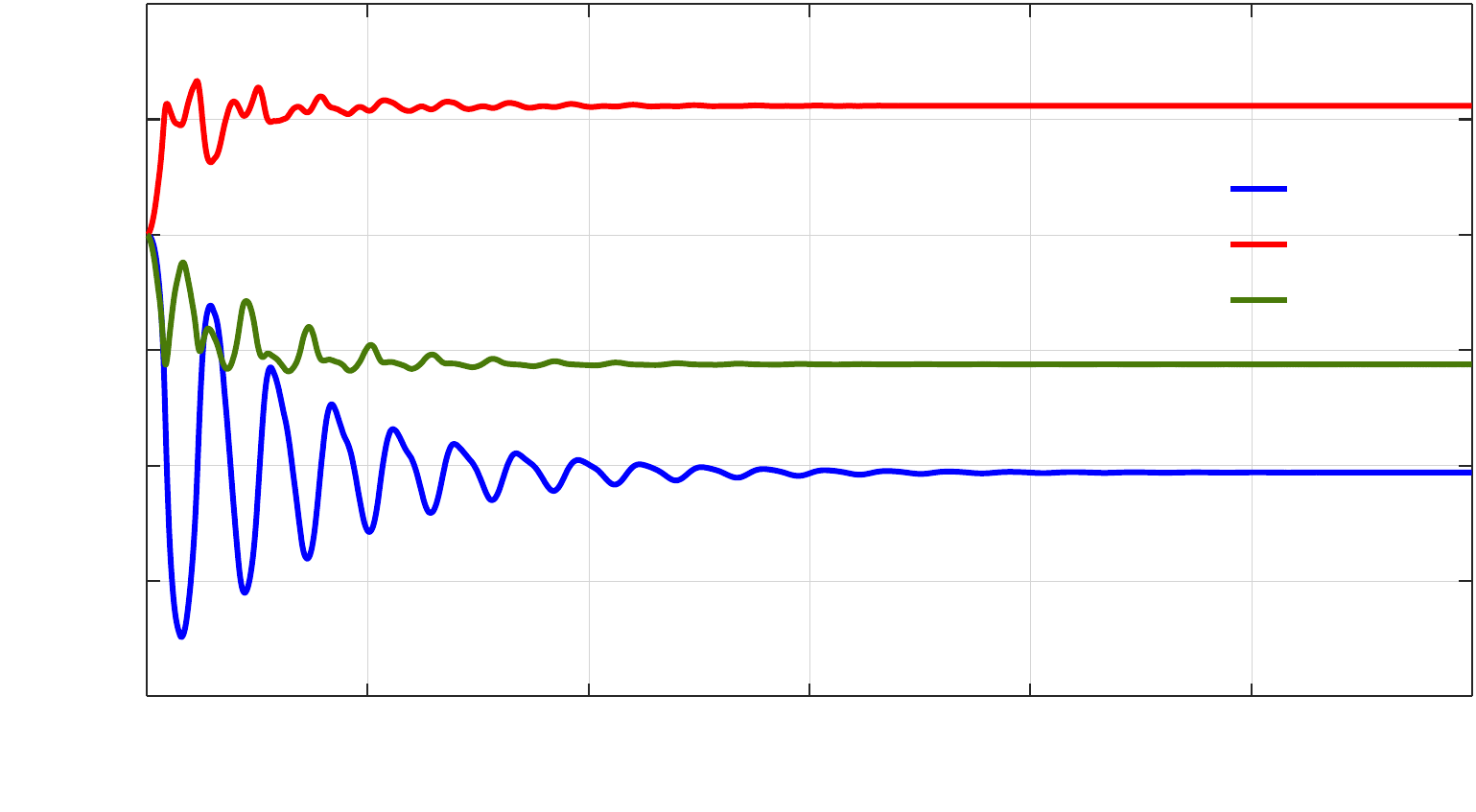
    \vspace{0.2cm}
    \caption{The tip displacements of the swinging rubber rod in Fig. \ref{fig-swinging_rod_snapshot3d}.}
    \label{fig-swinging_rod_tip_displ3d}
\end{figure*}

\begin{figure*}[ht!]
    \centering
    \def\svgwidth{0.75\textwidth}
    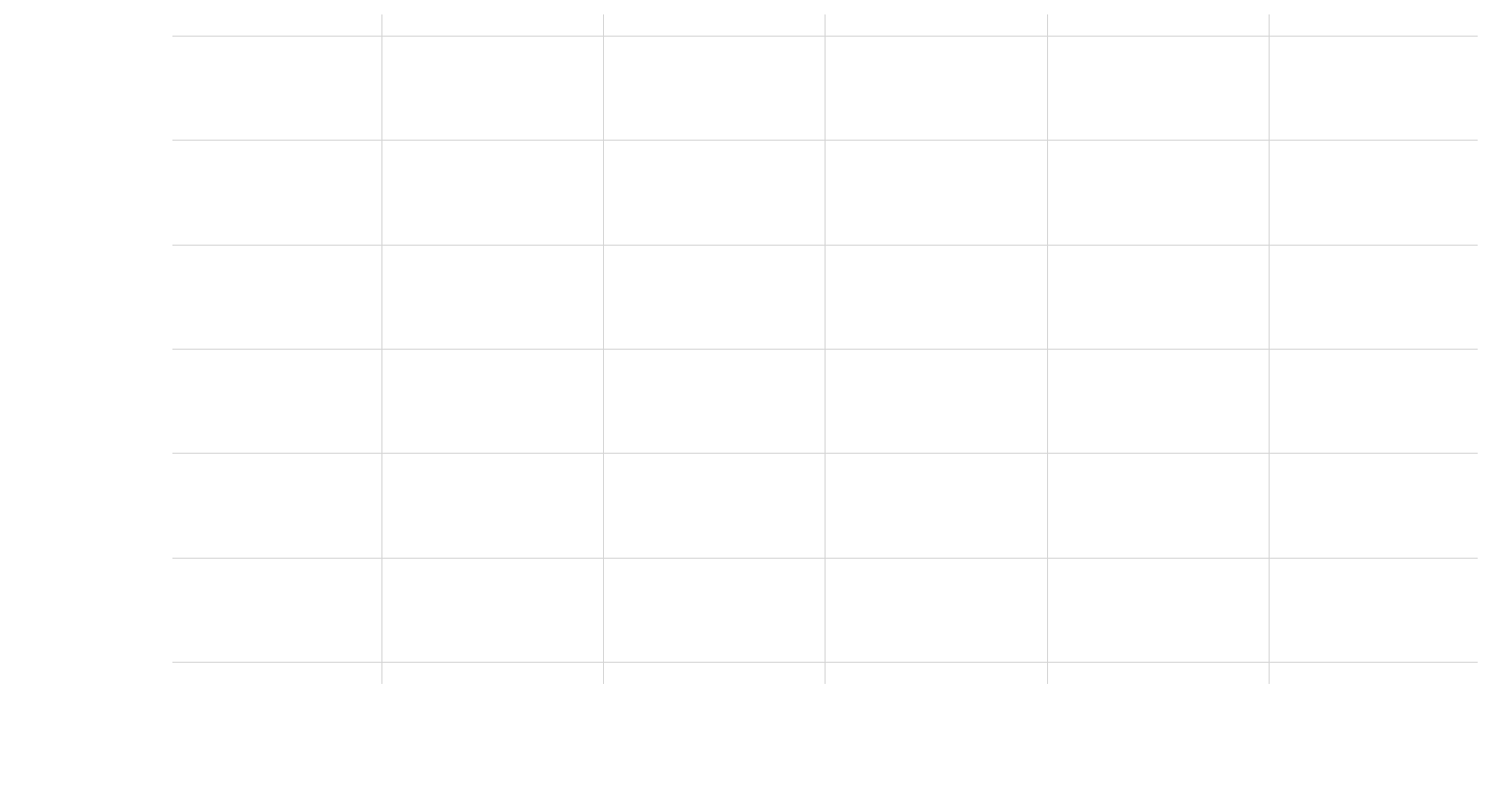
    \vspace{0.2cm}
    \caption{The energy of the swinging rubber rod in Fig. \ref{fig-swinging_rod_snapshot3d}.}
    \label{fig-swinging_rod_energy3d}
\end{figure*}

We now consider the same rod as in the previous benchmark, subjected to an additional wind field with a wind profile illustrated in Fig. \ref{fig-wind_profile}. 
The wind velocity has constant values but changing direction along the $z-$axis, that is: 
\begin{equation}\label{windprofile1}
    \begin{split}
        & \mat{v}_{wind} \, (z) \,=\, v_0 \,\mat{d}_{wind}\,(z) \,=\,  \\
        & v_0 \, \left[ \mat{E}_1 \, \cos\left(\beta_0 - 2\,\beta_0 \,\frac{z}{L}\right) \,+\, \right. \\
        & \qquad \left. \mat{E}_2 \, \sin\left(\beta_0 - 2\,\beta_0 \,\frac{z}{L}\right) \right] \,,
    \end{split}    
\end{equation}

\noindent
where $v_0$ is the constant value of the wind velocity, 
$\mat{d}_{wind}$ is the director of the wind velocity, 
$\beta_0$ is the angle of $\mat{d}_{wind}$ with respect to the $y-$axis at $z=0$, 
and $L$ is the initial length of the rod. 
For this example, we choose $v_0 = 10$ m/s, $\beta_0 = 45^{\circ}$, a simulation time of $30$ s, and an initial angle of the rod with respect to the $x-$axis of $-15^{\circ}$, that is an initial rod director of $\mat{d}=(\cos(\pi/12),\;0,\;-\sin(\pi/12))$. 
We apply the same discretization as in the previous benchmark. 
Fig. \ref{fig-swinging_rod_snapshot3d} illustrates the motion sequence of the swinging rod under the considered loading. Due to the three-dimensional wind profile, the rod shows out-of-plane deformations. 
Focusing on the configurations at $t=8.75$ s and $t=10$ s, which are no longer visually distinguishable, we assume that the rod has approximately reached a steady state configuration. 
This behavior is also illustrated in the time history of the rod tip displacement, showed in Fig. \ref{fig-swinging_rod_tip_displ3d}, and of the energy of the system, showed in Fig. \ref{fig-swinging_rod_energy3d}. 
We observe that after about $15$ s, all displacement components remain approximately constant, which is consistent with approximately zero kinetic energy at this time, i.e. the rod reaches an equilibrium configuration. 
Comparing this with the behavior of the rod in the previous benchmark without wind force, 
we can see that the wind force dampens the motion of the rod to an equilibrium configuration. 
This damping characteristic, generally of aerodynamic forces, can be seen when it is approximated by a Taylor expansion as follows:
\begin{equation}\label{flowexpand}
    \mat{F}_f = \left.\frac{\partial\mat{F}_f}{\partial\mat{q}}\right\vert_{\mat{q}=\mat{0}}\mat{q} + \left.\frac{\partial\mat{F}_f}{\partial\dot{\mat{q}}}\right\vert_{\dot{\mat{q}}=\mat{0}}\dot{\mat{q}} + 
    \left.\frac{\partial\mat{F}_f}{\partial\ddot{\mat{q}}}\right\vert_{\ddot{\mat{q}}=\mat{0}}\ddot{\mat{q}} + ...,
\end{equation}

\noindent
where the first term gives rise to the lift/thrust forces, the second term contains the so-called damping forces, and the third term is related to the added mass forces. 
An aerodynamic damping matrix can then be defined as $\mat{D}_{\text{fluid}}=-\left.\frac{\partial\mat{F}_f}{\partial\dot{\mat{q}}}\right\vert_{\dot{\mat{q}}=\mat{0}}$, which is a function of the free-stream velocity, among other parameters. 
The aerodynamic damping strongly depends on 
the magnitude of the free-stream velocity $V_{\infty}=\abss{\mat{V}_{\infty}}$. 
When $V_{\infty}$ is below a critical velocity $V_{\infty}^C$ (subcritical condition), the damping is positive and the surrounding flow absorbs the energy of the structure. 
When $V_{\infty}=V_{\infty}^C$ (critical condition), the damping is zero and then the surrounding flow does not absorb or supply the structure energy. 
When $V_{\infty}>V_{\infty}^C$ (supercritical condition), the damping becomes negative and the surrounding flow supplies energy to the structure, i.e. favoring the emergence of fluid-structure interaction instabilities such as aeroelastic flutter. 
For the swinging rod studied here, we are in a subcritical condition since the rod oscillation is damped out over time. 
Focusing on the robustness of the employed discretization scheme, we observe that the wind force also dampens out the high-frequency modes, and thus improves its robustness. 
We conclude that, given the subcritical condition, employing damping forces is another approach to improve the robustness, particularly  of the isogeometric discretization scheme studied in this paper.

%--------------------------------------------------------------------
\subsection{Dynamic response to pulsating forces}

\begin{figure*}[ht!]
    \centering
    \def\svgwidth{0.99\textwidth}
    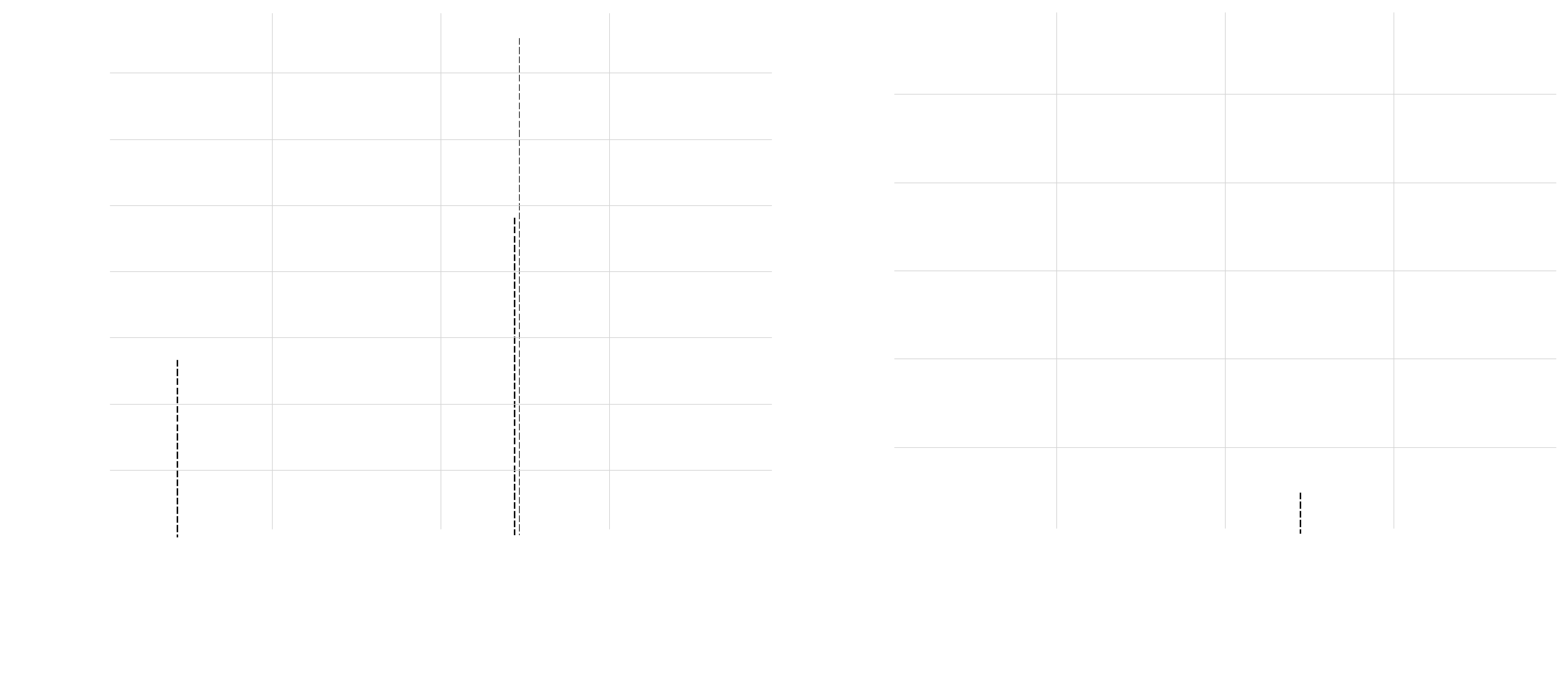
    \caption{The mean value and the amplitude of the horizontal tip displacement $u_x$ of an aluminum swinging rod at steady state, computed with cubic $C^1$ B-splines ($p=3$) and outliers removed.}
    \label{fig-alu-rod-displ-wF}
\end{figure*}

\begin{figure*}[ht!]
    \centering
    \def\svgwidth{0.75\textwidth}
    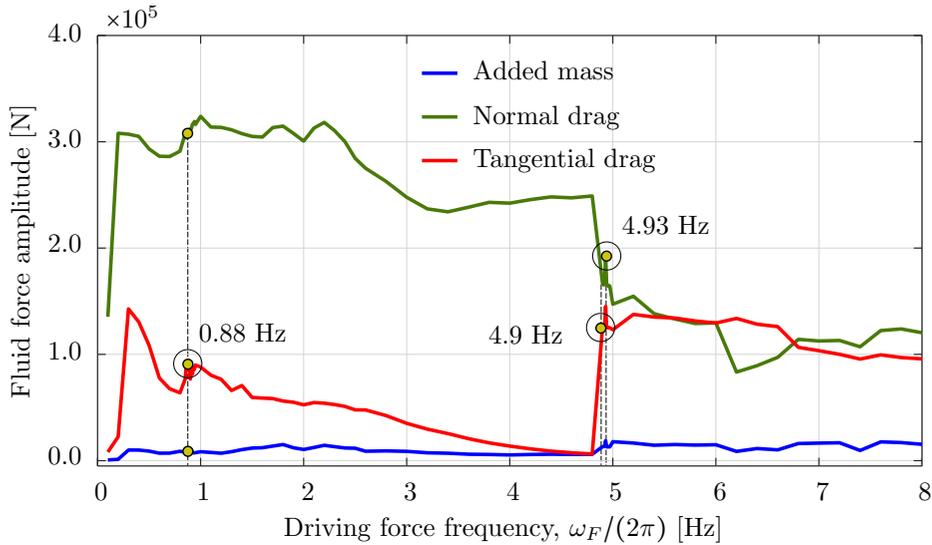
    \vspace{0.2cm}
    \caption{The amplitude of three force components induced by surrounding water at rest, integrated over the aluminium swinging rod at steady state.}
    \label{fig-alu_rod-fluid-force-wF}
\end{figure*}

\begin{figure*}[ht!]
    \centering
    \def\svgwidth{0.99\textwidth}
    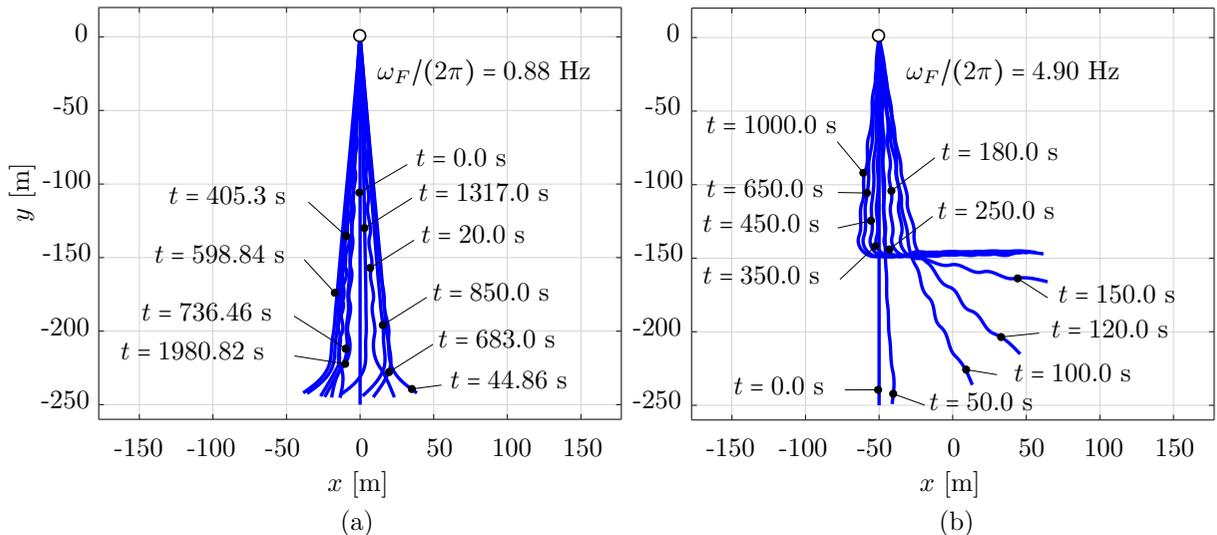
    \caption{Deformed configurations of an aluminum swinging rod at force frequencies of $0.88$ Hz and $4.90$ Hz, computed with cubic $C^1$ B-splines ($p=3$) and outliers removed.}
    \label{fig-alu-rod-snap-shots}
\end{figure*}

\begin{figure*}[ht!]
    \centering
    \def\svgwidth{0.75\textwidth}
    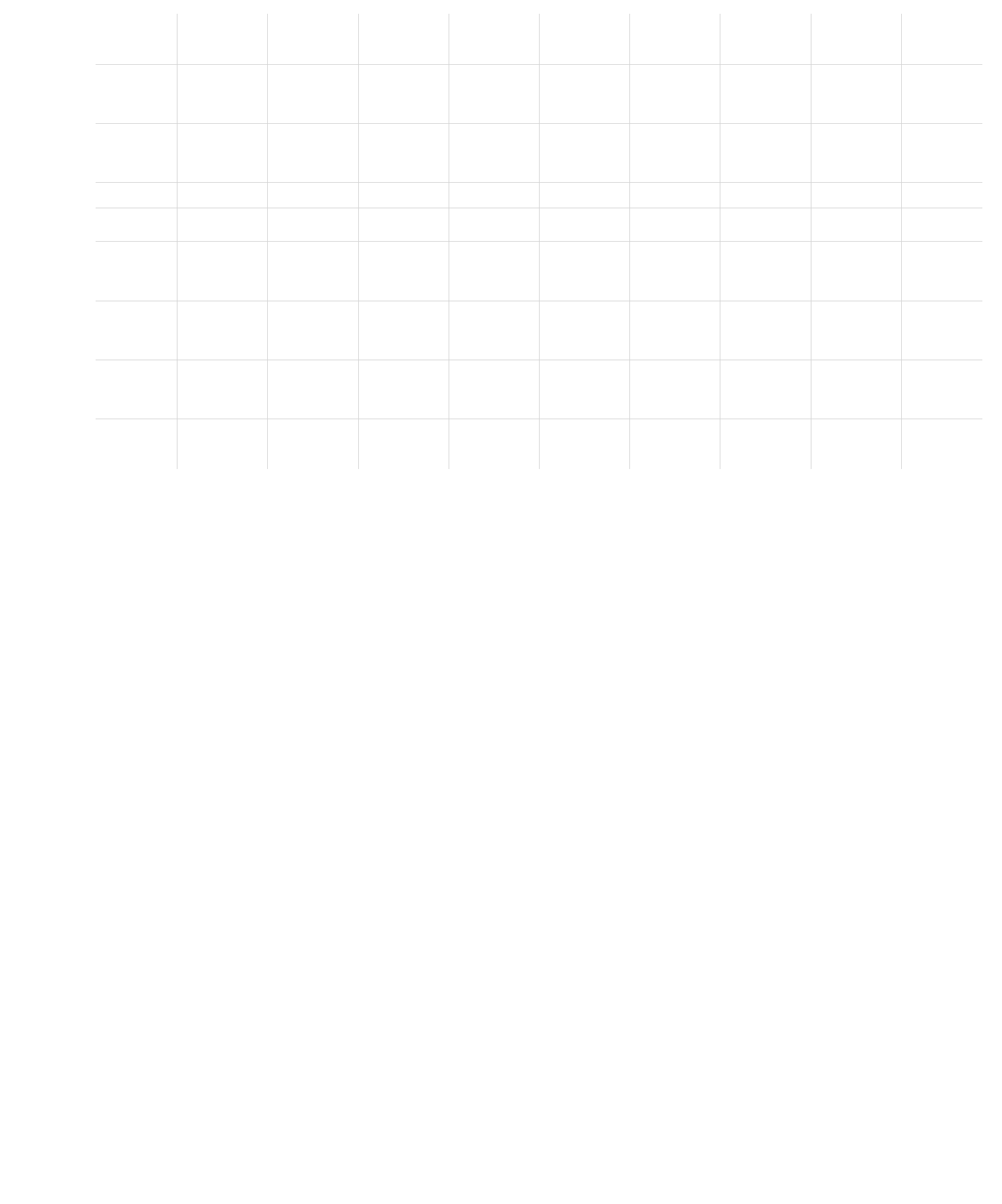
    \vspace{0.2cm}
    \caption{The tip displacements of an aluminum swinging rod computed for force frequencies of $0.88$ Hz and $4.90$ Hz.}
    \label{fig-alu-rod-u-tip}
\end{figure*}

Our last example is 
a long rod submerged in water, which simulates a sort of mooring lines or cables considered in aero-hydro-elastic simulations and offshore wind engineering. 
To this end, we consider a combination of 
the gravitational field, the surrounding water at rest, and a horizontal pulsating force applied at the free end of the swinging rod. 
We choose a sinusoidal pulsating force that is:
\begin{align}\label{eq:pulsF}
    \vect{F}\,(t) \,=\, A_F \, \sin \left( \frac{\omega_F}{2\,\pi} \, t\right) \, \vect{E}_1 \,,
\end{align}

\noindent
where $A_F$ and $\omega_F$ are the force amplitude and angular frequency, respectively. 
For this example, we choose an amplitude $A_f \,=\, 350.0$ kN, and different frequencies $(\omega_F/2\pi)$ ranging from $0.1$ Hz to $8$ Hz. 
The surrounding water is at rest, i.e. $\mat{V}_{\infty}=\mat{0}$ and $\mat{a}_{\infty} = \mat{0}$, 
and the mass density of the surrounding water is $\rho_f \,=\, 1000$ kg/m$^3$. 
In this example, we consider an aluminium rod with an initial director \reviewerII{$\mat{d}=(0,\;-1,\;0)$}, a length of $250$ m, a circular cross-section with a radius of $0.02$ m, Young's modulus $E\,=\,7\cdot10^{10}$ N/m$^2$, and mass density $\rho \,=\, 2700$ kg/m$^3$. 
We simulate the gravitational field with a direction of $(0,\;-1,\;0)$, which allows the rod to deform only in the $xy$-plane. 
We compute this example for $1000$ s and $2000$ s such that a steady state is observed during the last $100$ s, which is identified as the response becomes periodic in time.

Fig. \ref{fig-alu-rod-displ-wF} illustrates the mean value and the amplitude of the horizontal displacement $u_x$ at the free tip of the studied swinging rod after it reaches the steady state, as a function of the force frequencies $(\omega_F/2\pi)$. 
We observe that the mean value of the tip displacement (Fig. \ref{fig-alu-rod-displ-wF}a), which corresponds to the position, around which the rod oscillates, i.e. the static equilibrium position, 
jumps at force frequencies smaller than $2.0$ Hz and those larger than $5.0$ Hz. 
It means that the equilibrium position and configuration change abruptly when the force frequency changes, which 
indicates a nonlinear behavior of the rod in different frequency ranges. 
Focusing on the amplitude of the tip displacement in Fig. \ref{fig-alu-rod-displ-wF}b, we can see that 
after a steep increase at low frequencies, it decreases with increasing force frequency and slightly jumps downwards around $4.9$ Hz. 
This relation between the amplitude and the force frequency can be explained based on the linear theory. 
Consider a damped harmonic oscillator of mass $m$, spring stiffness $k$, and damping coefficient $b$, subjected to a sinusoidal pulsating force, $f_0\sin(\Omega t)$, of frequency $\Omega$ and amplitude $f_0$. 
It is well-known from classical vibration theory that the response amplitude of this linear system is:
\begin{equation}\label{eqn:linearoscillator}
    A = \frac{f_0}{\sqrt{m(\Omega^2-\omega_0^2)^2+b^2\Omega^2}}
\end{equation}

\noindent
where $\omega_0=\sqrt{k/m}$ is the natural frequency of the system. 
We can see that for the cases of $\Omega < \omega_0$, increasing $\Omega$ increases the amplitude $A$, and for the cases of $\Omega > \omega_0$, increasing $\Omega$ decreases $A$. 
Since the studied nonlinear rod can be considered as a damped (by considering the surrounding flow) distributed-parameter elastic pendulum subjected to a pulsating force at its free end, 
a similar relation between the amplitude and the force frequency is expected for both linear and the nonlinear cases. 
This supports our observation in Fig. \ref{fig-alu-rod-displ-wF}b, except for the jump at $4.9$ Hz, which  
associates with the highest positive peak of the mean value in Fig. \ref{fig-alu-rod-displ-wF}a.

To gain better insights into the rod behavior, in Fig. \ref{fig-alu-rod-snap-shots}, we plot some deformed configurations at different times for $0.88$ Hz and $4.9$ Hz, i.e. before and close to the frequency value where the amplitude jumps. 
We observe that for both two frequencies, the rod behavior is a combination of elastic vibrations and rotations as a rigid body that oscillates around a static equilibrium position. 
While this equilibrium position is along the vertical axis in the case of $0.88$ Hz (Fig. \ref{fig-alu-rod-snap-shots}a), 
it, as well as the equilibrium configuration, changes in the case of $4.9$ Hz (Fig. \ref{fig-alu-rod-snap-shots}b), since the lower part of the rod is bent to a horizontal segment, and the rod then oscillates around the new deformed position. 
This behavior together with the amplitude jump is also observed in the time history of the tip displacement, illustrated in Fig. \ref{fig-alu-rod-u-tip} for $0.88$ Hz and $4.9$ Hz. 
We can also see that both two displacement components oscillate over time, while their mean value jumps to a value around $100$ m in the case of $4.9$ Hz. 
This is further reflected in the relation between the pulsating force frequency and the amplitude of forces induced by the surrounding flow, illustrated in Fig. \ref{fig-alu_rod-fluid-force-wF}. 
In Fig. \ref{fig-alu_rod-fluid-force-wF}, we illustrate the amplitude $A_f^{\omega}$ of three components of the resulting force induced by the surrounding flow (added mass, normal drag, and tangential drag), integrated over the rod length once the rod reaches the steady state, that is: 
\begin{align}
    A_f^{\omega}(\omega_F)=\underset{\text{steady state}}{\max}\left[f_f^t(t)\right]-\underset{\text{steady state}}{\min}\left[f_f^t(t)\right] \nonumber \,,
\end{align}
where the integrated force $f_f^t(t)$ is:
\begin{align}
    f_f^t(t)=\int_0^L\abss{\mat{F}_f(s,t)}\,ds \,, \nonumber
\end{align}
as a function of the pulsating force frequency. 
We observe that the amplitude of the normal and tangential drag forces also jumps approximately at $4.9$ Hz. 
In particular, 
the normal drag drops, while the tangential drag jumps upwards, which is consistent with the fact that the lower part of the rod is bent to a horizontal segment, i.e. only the upper part is mainly affected by the norm drag, while the tangential drag is the dominating force in the lower part. 
Regarding the robustness of the applied discretization scheme, we did not obtain unstable computations and results containing high-frequency contents. 
We conclude that different nonlinear behaviors of the swinging rod can be represented and studied using the rod formulation \cite{gebhardt_2021_beam}, together with the isogeometric discretization scheme, improved using outlier removal, and with an energy-momentum preserving implicit time integration scheme. 
This has been shown to be a sufficiently robust approach for studying nonlinear structures such as swinging rods modeling mooring lines, which can be further investigated for complex behaviors such as parametric resonances and chaotic behavior in future works.

\section{Summary and conclusions}\label{sec-summary}

In this paper, we explored the application of the nonlinear formulation \cite{gebhardt_2021_beam} for rods that exhibit only axial and bending deformations, using isogeometric spatial discretizations. 
Our results illustrate different convergence rates for odd and even polynomial degrees, which is known from other isogeometric discretization methods, see e.g.\ \cite{Auricchio2013,Schillinger:13.1} and the references therein. 
They also show that the continuity generally does not affect the accuracy and convergence, except the convergence in $H^2$ semi-norm when using odd degrees, which decreases with decreasing continuity. 
We demonstrated computationally for studied dynamic benchmarks of two- and three-dimensional rods that isogeometric discretizations using B-splines with $C^1$ continuity or higher are less robust than the standard scheme using Hermite functions. 
Increasing the smoothness of cubic spline basis functions leads to a smaller phase, which may reduce the robustness and may relate to the approximation power of the corresponding discretizations with higher continuity in the case of nonlinear rods \cite{gebhardt_2021_beam}. 
We showed that robustness can be improved 
via a strong approach of outlier removal \cite{hiemstra_outlier_2021} without compromising the accuracy. 
Alternatively, reducing the time step or employing forces with damping characteristics leads to more robust computations. 
We have shown that the robustness is closely related to round-off errors due to floating-point arithmetic. 
In addition, we demonstrated computationally for an unconstrained rod subjected to out-of-plane vanishing forces that the configuration-dependent mass matrix does not behave as a regular perturbation and thus cannot be simplified to a constant matrix. 
Lastly, we applied our nonlinear transient formulation to a swinging rubber rod subjected to gravity, forces induced by a surrounding flow such as wind and water, and a pulsating force of different frequencies. 
Our results also show that the isogeometric discretization scheme is robust and reliable for such an analysis.

The results presented in this work open up several directions for future works. 
One open question is the accuracy and convergence behavior of the discretization scheme in nonlinear problems. 
Another question is the irregular behavior of the configuration-dependent mass term that, to our best knowledge, is not yet proved analytically. This is particularly interesting for choosing and developing an efficient and robust time integration scheme. 
It is also desirable to employ our nonlinear formulation with the isogeometric discretization scheme in highly nonlinear problems with complex loads and geometries.

\bmhead{Acknowledgments}

C.G.\ Gebhardt, B.A. Roccia and T.-H. Nguyen gratefully acknowledge the financial support from the European Research Council through the ERC Consolidator Grant “DATA-DRIVEN OFFSHORE” (Project ID 101083157).
 
D.\ Schillinger, R.R. Hiemstra and T.-H. Nguyen gratefully acknowledge financial support from the German Research Foundation (Deutsche Forschungsgemeinschaft) through the DFG Emmy Noether Grant SCH 1249/2-1 and the standard DFG grant SCH 1249/5-1.

\section*{Statements and Declarations}

The authors declare that they have no known competing financial interests or personal relationships that could have appeared to influence the work reported in this paper.

% \appendix
\begin{appendices}

\section{An elastic pendulum}\label{sec-appendix1}

\begin{figure}[ht!]
    \centering
    \def\svgwidth{0.7\columnwidth}
    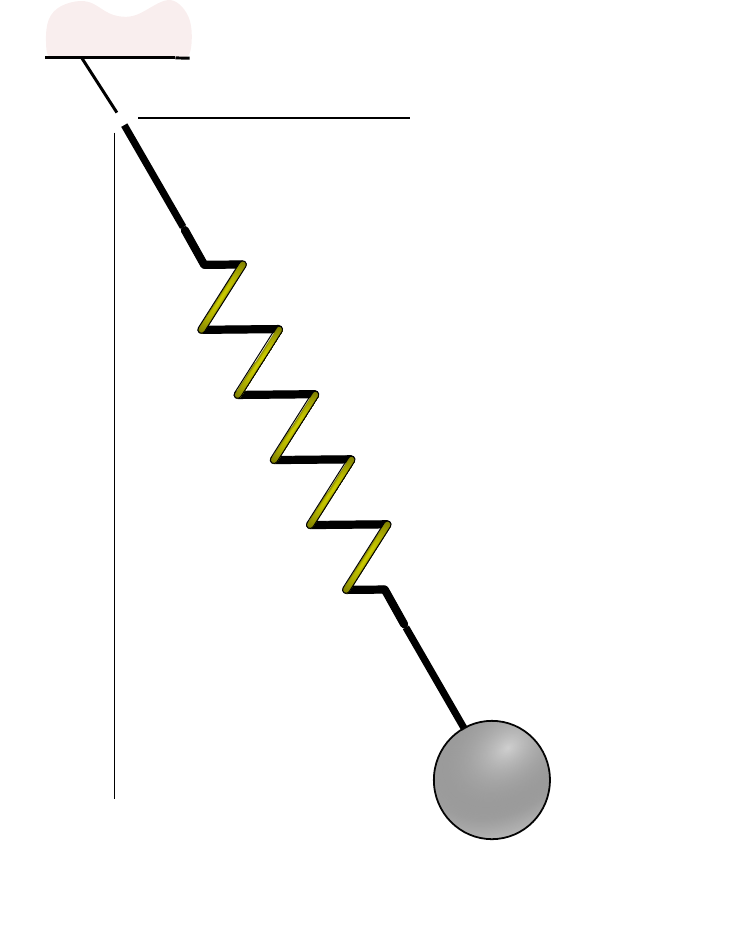
    \vspace{0.2cm}
    \caption{Schematic of an elastic pendulum with two degrees of freedom.} \label{fig-pendulum}
\end{figure}

\begin{figure*}[hbt!]
    \centering
    \def\svgwidth{0.95\textwidth}
    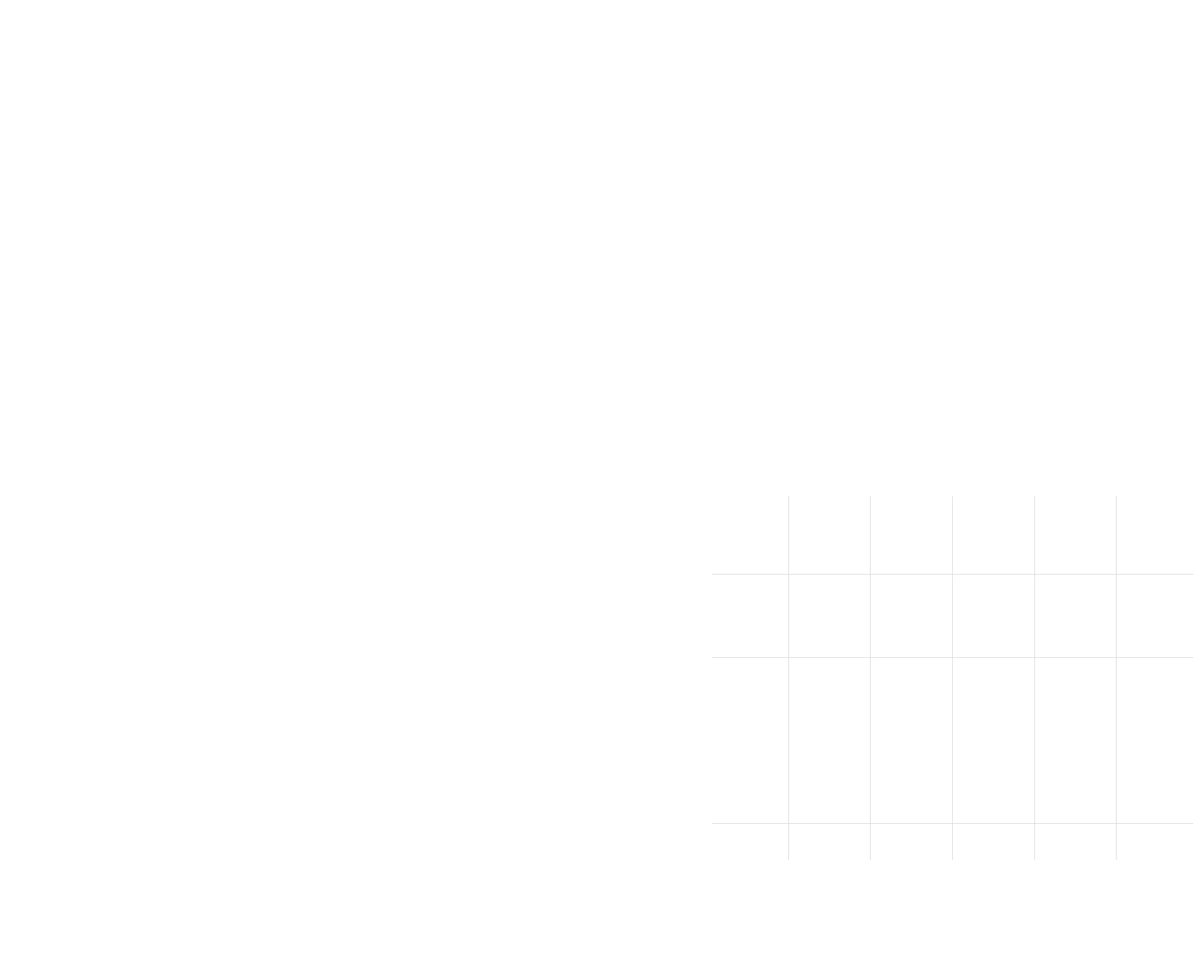
    \caption{Energy, the third component of angular momentum and the second precision quotient of the elastic pendulum in Fig. \ref{fig-pendulum} with the initial conditions \eqref{eq-pendulum-init1}.}
    \label{fig-pendulum-noF}
\end{figure*}

\begin{figure*}[hbt!]
    \centering
    \def\svgwidth{0.95\textwidth}
    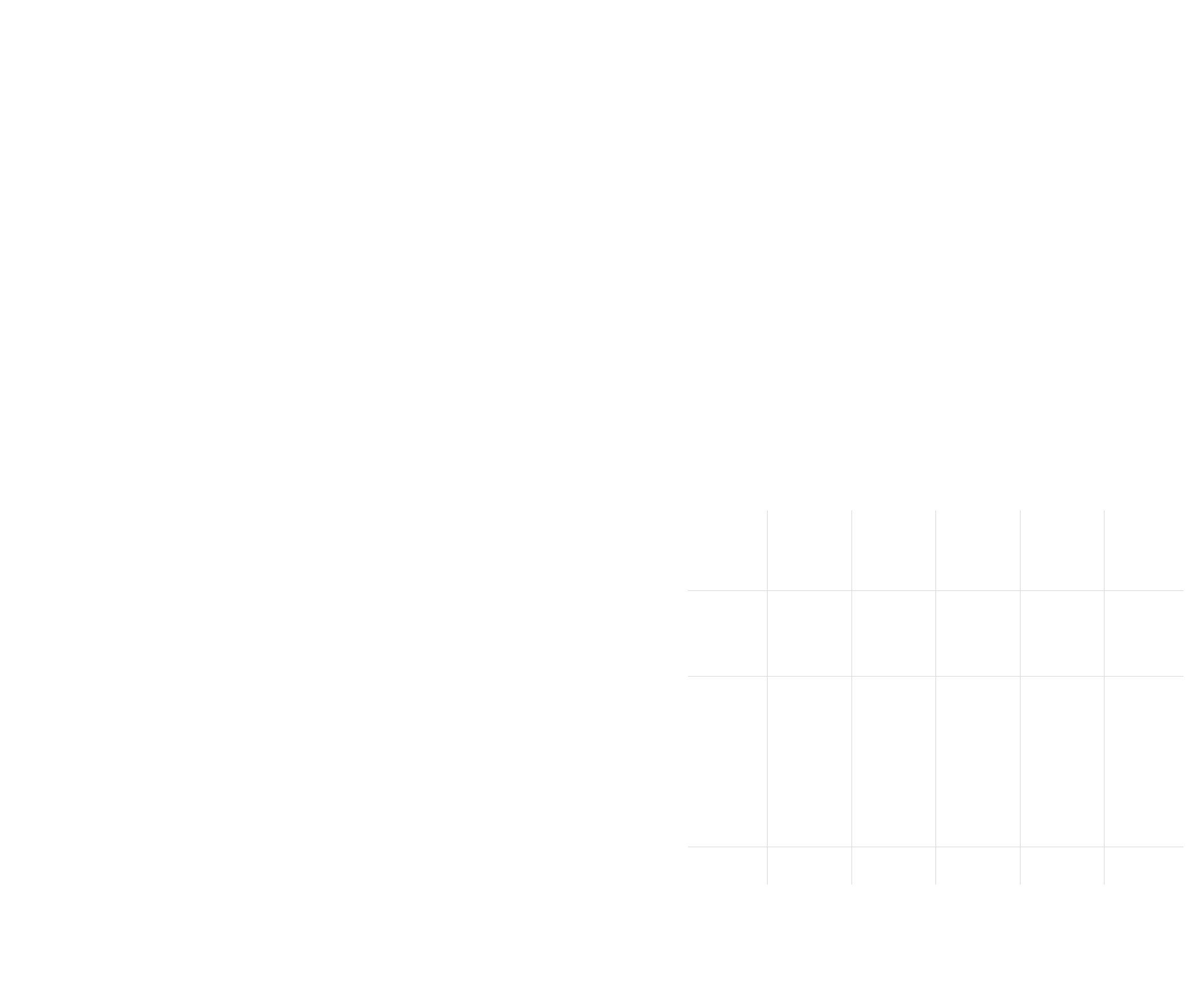
    \caption{Energy, the third component of angular momentum and the second precision quotient of the elastic pendulum in Fig. \ref{fig-pendulum}, subjected to gravity and a parabolic wind profile, with the initial conditions \eqref{eq-pendulum-init2}.}
    \label{fig-pendulum-F}
\end{figure*}

To benchmark our implementation regarding the time integration scheme and to verify that 
the chosen implicit time integration scheme achieves second-order accuracy, approximately preserves energy, and exactly preserves the linear and angular momentum \cite{gebhardt_2021_beam,gebhardt_implicit_2020}, 
discussed in Section \ref{sec-time-int}, 
we consider the dynamics of a 
two degrees-of-freedom elastic pendulum, illustrated in Fig. \ref{fig-pendulum}. 
Its configuration space is $\mathcal{D}_{\text{pen}}=\mathbb{R}\times SO(2)$, where $SO(2)$ is the special orthogonal group in two dimensions. 
For this pendulum, the two degrees of freedom are the angular rotation with respect to the vertical axis, $\theta(t)$, and the axial deformation of the spring, $\eta(t)$.  
Unit vectors $\mat{E}_1$ and $\mat{E}_2$ are the first two canonical Cartesian bases of $\mathbb{R}^3$, $L_0$, $k$, $m$, and $g$ are the spring's natural length, spring stiffness, mass, and gravitational acceleration, respectively. We consider a simulation time of $30$ s and a time step $\Delta\,t\,=\,0.005$ s.

\begin{table*}[hbt!]
	\centering
	{\footnotesize
	\begin{tabularx}{0.9\linewidth}{ >{\centering\arraybackslash\hsize=.18\hsize}X | >{\centering\arraybackslash\hsize=.2\hsize}X >{\centering\arraybackslash\hsize=.2\hsize}X >{\centering\arraybackslash\hsize=.2\hsize}X >{\centering\arraybackslash\hsize=.2\hsize}X }
	\toprule
	  	  		        & $t\,=\,7.5$ s 		 & $t\,=\,15$ s 			 & $t\,=\,22.5$ s  		 & $t\,=\,30$ s  	\\
	\midrule
	 Total energy  		& $26.824953231727577$ 		 & $26.824943982990575$ 			 & $26.824999851284034$  		 & $26.824976045230380$  	\\
	\midrule
	 Angular momentum  		& $-0.605000000000103$ 		 & $-0.605000000000105$ 			 & $-0.605000000000100$  		 & $-0.605000000000120$  	\\
	\bottomrule
\end{tabularx}}
\caption{Total energy and the angular momentum of an elastic pendulum illustrated in Fig. \ref{fig-pendulum-noF}.
}
\label{tab:energy-angMon-pendulum-noF}
\end{table*}

We first consider the case of no gravitational acceleration ($g\,=\,0$) and no external load with the following initial conditions:
\begin{subequations}\label{eq-pendulum-init1}
    \begin{align}
        & \theta(t\,=\,0) = 0\,, \qquad \eta(t\,=\,0) = 0.1\,, \\
        & \dot{\theta}(t\,=\,0) = -0.5\,, \qquad \dot{\eta}(t\,=\,0) = 0.25\,.
    \end{align}    
\end{subequations}
Fig. \ref{fig-pendulum-noF}a-b illustrates the time evolution of the energy and the third component of the angular momentum, $j_3$. 
We note 
that for the studied pendulum, the first and second components of the angular momentum are zero, and thus are not illustrated here. 
We observe that both the total energy and $j_3$ are virtually constant over the simulation time. 
Table \ref{tab:energy-angMon-pendulum-noF} shows their values at four snapshots. 
We can see that the total energy is approximately preserved, while the angular momentum is exactly preserved up to a tolerance of the machine accuracy. 
Fig. \ref{fig-pendulum-noF}c-d shows the second precision quotient of the response during the simulation time, 
which is computed to verify the implementation correctness of a numerical integration scheme and is defined as \cite{kreiss_ortiz_2014}: 
\begin{align}
    Q_{II} \,=\, Q_{II}(t) \,=\, \frac{\norm{ \mat{u}_{\Delta\,t} \,-\, \mat{u}_{\Delta\,t/2} }}{\norm{ \mat{u}_{\Delta\,t/2} \,-\, \mat{u}_{\Delta\,t/4} }} \,,
\end{align} 

\noindent
where $\mat{u}\,=\,\mat{u}(t)$ is a time-dependent variable, and its subscripts, $\Delta t$, $\Delta t/2$, ..., denote the time step  
employed to compute $\mat{u}$. 
For well-implemented time integration schemes, the precision quotient $Q_{II}$ is approximately $2^q$, where $q$ is the order of accuracy of the integration scheme \cite{kreiss_ortiz_2014}. For our implicit scheme, $q=2$. 
We note that our choice of the second precision quotient instead of the first precision quotient is due to the lack of an analytical solution for the studied elastic pendulum \cite{kreiss_ortiz_2014}. 
We observe in Fig. \ref{fig-pendulum-noF}c-d the expected value of $2^2 = 4$ for $Q_{II}$, which verifies the correctness of our implementation.

We then consider the load case consisting of the gravitational field and the following parabolic wind profile:
\begin{align}
    \hspace{-0.5cm}& \mat{v}_w (x_1\,t) \,=\, \nonumber \\
    \hspace{-0.5cm}& x_1^2 \, \left(\, 1.0 \,+\, 0.1\,\sin\left(\,\frac{1}{50}\,\frac{\sqrt{k\,/\,m}}{2\,\pi}\,t\right) \,\right) \, \mat{E}_2 \,.
\end{align}

\noindent
We choose the following initial conditions of the pendulum:
\begin{subequations}\label{eq-pendulum-init2}
    \begin{align}
        & \theta(t\,=\,0) = \pi/2\,, \qquad \eta(t\,=\,0) = 0.0\,, \\
        & \dot{\theta}(t\,=\,0) = 0.0\,, \qquad \dot{\eta}(t\,=\,0) = 0.0\,.
    \end{align}    
\end{subequations}

\noindent
We observe in Fig. \ref{fig-pendulum-F}a-b that the total energy and angular momentum are no longer preserved due to the presence of external forces, as expected. 
We can also see that after approximately $17$ s, the pendulum does not contain any kinetic energy, i.e. it approximately achieves a static equilibrium state. 
The second precision quotient $Q_{II}$, illustrated in Fig. \ref{fig-pendulum-F}c-d, also implies a second-order accurate time integration scheme in this case. 
We conclude that the chosen implicit time integration scheme approximately preserves the total energy, exactly preserves the angular momentum, and is second-order accurate.

\section{Simplified model of a force field induced by a surrounding flow}\label{sec-flow-force}

\begin{figure}[h!]
    \centering
    \def\svgwidth{0.65\columnwidth}
    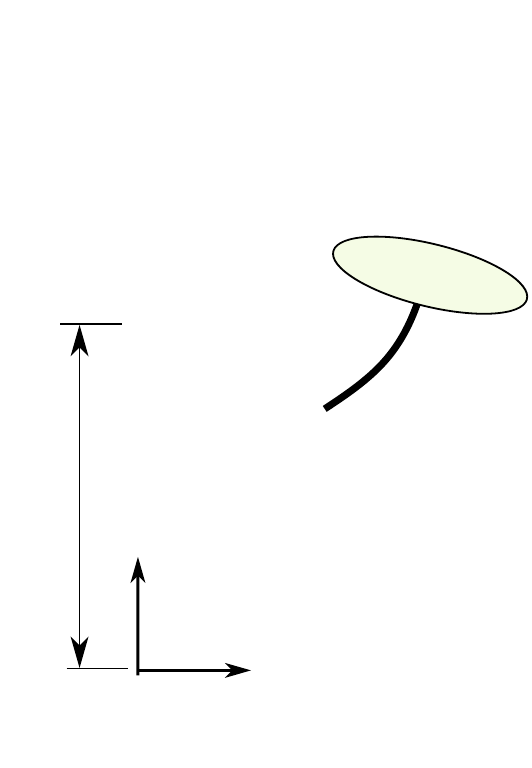
    \vspace{0.2cm}
    \caption{Schematic of a wind profile.}
    \label{fig.flow.1}
\end{figure}

For the numerical examples included in subsequent sections, we consider forces induced by a surrounding flow such as wind and water. Fig.~\ref{fig.flow.1} illustrates the schematic of an exemplary wind profile acting on a rod configuration. 
We consider a simplified model of such forces \cite{Huston1981} that consist of three counterparts: 
the resulting force due to the accelerated surrounding flow by the moving rod, which we refer to as the added mass force, 
the tangential drag force, 
and the normal drag force. 
To this end, 
the resulting force $\mat{F}_{f}$ per unit length at an arbitrary point of the rod is expressed as follows:
\begin{equation}\label{EqC.1}
    \mat{F}_f = C_1\mat{a}_N + C_2\abss{\mat{V}_N}\mat{V}_N + C_3\abss{\mat{V}_T}\mat{V}_T \,,
\end{equation}

\noindent
where $\mat{a}_N$ is the normal component of the relative flow acceleration with respect to the rod, 
$\mat{V}_N$ and $\mat{V}_T$ the normal and tangential component of the relative flow velocity, respectively,  
and the coefficients $C_1$, $C_2$ and $C_3$ are given by: 
\begin{equation}\label{EqC.2}
\begin{split}
    C_1 &= \frac{1}{4}\pi\,C_M\rho_f\,\diameter^2 \,, \\
    C_2 &= \frac{1}{2}C_N\rho_f\,\diameter \,, \\
    C_3 &= \frac{1}{2}C_T\rho_f\,\diameter \,, 
\end{split}
\end{equation}

\noindent
where $\rho_f$ is the mass density of the surrounding flow, and $\diameter$ is the diameter of the cylindrical cross-section of the rod. 
The coefficients $C_M$, $C_N$, and $C_T$ depend on the Reynolds number and are commonly determined experimentally \cite{Huston1981}. 

To determine the forces 
acting on a rod segment of length $ds$ due to the added mass, normal and tangential counterparts,  
it is necessary to describe the relative flow acceleration and velocity in terms of the ambient motion of the fluid and the rod. 
Considering the kinematic description of the rod given in the previous subsection, the quantities $\mat{a}_N$, $\mat{V}_N$, and $\mat{V}_T$ can be expressed as follows:
\begin{equation}\label{EqC.3}
\begin{split}
    \mat{a}_N &= \Pd\,\mat{a}, \\
    \mat{V}_N &= \Pd\mat{V},\quad\text{and} \\
    \mat{V}_T &= \left(\mat{d}\otimes\mat{d}\right)\mat{V},
\end{split}
\end{equation}

\noindent
where $\mat{V}=\mat{V}_{\infty}-\dot{\vect{\varphi}}$, and $\mat{a}=\mat{a}_{\infty}-\ddot{\vect{\varphi}}$. 
The magnitude and direction of the free-stream velocity $\mat{V}_{\infty}(z,t)$ is a function of the height $z$ above the ground level (or below the sea level if ocean structures are considered) and time $t$. 
The free-stream acceleration is the time derivative of the free-stream velocity, i.e., $\mat{a}_{\infty}(z,t)=\frac{\partial}{\partial t}\mat{V}_{\infty}(z,t)$. 
Inserting (\ref{EqC.3}) into (\ref{EqC.1}) and integrating along the rod, we obtain the resultant forces as follows: 
\begin{equation}\label{EqC.4}
\begin{split}
    \mat{F}_f &= C_1\,\int_{0}^{S}\Pd\,(\mat{a}_{\infty}-\ddot{\vect{\varphi}})\,ds \\
    & + C_2\int_{0}^{S}\abss{\Pd\,(\mat{V}_{\infty}-\dot{\vect{\varphi}})}\,\Pd\,(\mat{V}_{\infty}-\dot{\vect{\varphi}})\,ds  \nonumber \\
    & + C_3\,\int_{0}^{S}\abss{\left(\mat{d}\otimes\mat{d}\right)(\mat{V}_{\infty}-\dot{\vect{\varphi}})}\left(\mat{d}\otimes\mat{d}\right)(\mat{V}_{\infty}-\dot{\vect{\varphi}})\,ds. \nonumber
\end{split}
\end{equation}

We note that coefficients $C_i$, $i=1,2,3$, are assumed to be independent of the position along the rod.

\section{Linearization}\label{sec-appendix2}

%********************************************************************************************************************
%                                            G E N E R A L I T I E S
%********************************************************************************************************************

\subsection{Preliminaries}\label{Appendix_1}

We recall the semi-discrete formulation for the rod motions in Section \ref{sec-discretization}, that is:
\begin{equation}\label{eqA.1}
    \begin{split}
        \mat{g}_d = \int_0^S \, & \left(\, \mat{M} (\mat{\qhat}) \, \nabla_{\dot{\mat{\qhat}}} \, \dot{\mat{\qhat}} \,+\, \mat{B} (\mat{\qhat})^T \, \boldsymbol{\sigma}_h \right.  \\
        & \qquad \left. -\, \mat{\bspline}^T \, \mat{f}^{\text{ext}} \,\right) \, \mathrm{d}s \,=\, \mat{0}\,.        
    \end{split}
\end{equation}
Analogously, the variational formulation that describes the equilibrium of the rod is:
\begin{equation}\label{eqA.1a}
        \mat{g}_s = \int_0^S \, \left(\, \mat{B} (\mat{\qhat})^T \, \boldsymbol{\sigma}_h \,-\, \mat{\bspline}^T \, \mat{f}^{\text{ext}} \,\right) \, \mathrm{d}s \,=\, \mat{0}\,.
    \end{equation}

We employ a standard approach based on a Taylor expansion of (\ref{eqA.1a}) and (\ref{eqA.1}) to obtain the tangent stiffness matrix associated with $\mat{g}_s$ and $\mat{g}_d$, respectively. To this end, Taylor’s approximation for a vector function $\mat{g}(\mat{q},\dot{\mat{q}},\ddot{\mat{q}})$ is given by:
\begin{align}\label{eqA.2}
    \hspace{-0.5cm} &\mat{g}(\mat{q}+\Delta\mat{q},\dot{\mat{q}}+\Delta\dot{\mat{q}},\ddot{\mat{q}}+\Delta\ddot{\mat{q}}) = \\
    \hspace{-0.5cm} &\quad \mat{g}(\mat{q},\dot{\mat{q}},\ddot{\mat{q}}) + \text{D}\,\mat{g}(\mat{q},\dot{\mat{q}},\ddot{\mat{q}})\cdot\left(\Delta\mat{q},\Delta\dot{\mat{q}},\Delta\ddot{\mat{q}}\right) \nonumber \\
    \hspace{-0.5cm} &+ \text{D}^2\,\mat{g}(\mat{q},\dot{\mat{q}},\ddot{\mat{q}}):\left(\left(\Delta\mat{q},\Delta\dot{\mat{q}},\Delta\ddot{\mat{q}}\right)\otimes\left(\Delta\mat{q},\Delta\dot{\mat{q}},\Delta\ddot{\mat{q}}\right)\right) \nonumber \\ 
    \hspace{-0.5cm} & + O\left(\norm{\Delta\mat{q}}^3,\norm{\Delta\dot{\mat{q}}}^3,\norm{\Delta\ddot{\mat{q}}}^3,...,\norm{\Delta\dot{\mat{q}}}^2\norm{\Delta\ddot{\mat{q}}},...\right)\,, \nonumber
\end{align}

\noindent
where $\text{D}^i(\cdot)$, $i=1,2,...$, is a $(i+1)$-order tensor of type $(0,i)$, and $(:)$ denotes the double-contraction tensor operation. Assuming that the higher-order terms are negligible, we obtain:
\begin{align}\label{eqA.3}
    \hspace{-0.5cm}&\mat{g}(\mat{q}+\Delta\mat{q},\dot{\mat{q}}+\Delta\dot{\mat{q}},\ddot{\mat{q}}+\Delta\ddot{\mat{q}}) \approx \nonumber \\
    \hspace{-0.5cm}& \qquad \mat{g}(\mat{q},\dot{\mat{q}},\ddot{\mat{q}}) + \partial_{\mat{q}}\,\mat{g}(\mat{q},\dot{\mat{q}},\ddot{\mat{q}})\cdot\Delta\mat{q} \\ 
    \hspace{-0.5cm}&\qquad +\partial_{\dot{\mat{q}}}\,\mat{g}(\mat{q},\dot{\mat{q}},\ddot{\mat{q}})\cdot\Delta\dot{\mat{q}} + \partial_{\ddot{\mat{q}}}\,\mat{g}(\mat{q},\dot{\mat{q}},\ddot{\mat{q}})\cdot\Delta\ddot{\mat{q}}\,, \nonumber
\end{align}

\noindent
where $\partial_{\mat{q}}(\cdot)$, $\partial_{\dot{\mat{q}}}(\cdot)$, and $\partial_{\ddot{\mat{q}}}(\cdot)$ denote partial derivatives with respect to $\mat{q}$, $\dot{\mat{q}}$, and $\ddot{\mat{q}}$, respectively. Introducing (\ref{eqA.1a}) into (\ref{eqA.3}) yields:
\begin{align}\label{eqA.4}
    \hspace{-0.5cm}&\mat{g}_s(\mat{q}+\Delta\mat{q}) \approx \mat{g}_s(\mat{q}) + \\
    \hspace{-0.5cm}& \partial_{\mat{q}}\,\left[\int_0^S \left(\, \mat{B} (\mat{\qhat})^T \, \boldsymbol{\sigma}_h \,-\, \mat{\bspline}^T \, \mat{f}^{\text{ext}} \,\right) \, \mathrm{d}s\right]\cdot \Delta \mat{q} \approx  \nonumber \\
    \hspace{-0.5cm}&\mat{g}_s(\mat{q})+\left[\int_0^S \partial_{\mat{q}}\left(\mat{B} (\mat{\qhat})^T \, \boldsymbol{\sigma}_h - \mat{\bspline}^T \, \mat{f}^{\text{ext}}\right) \, \mathrm{d}s\right]\cdot\Delta\mat{q}\,, \nonumber
\end{align}

\noindent
and (\ref{eqA.1}) into (\ref{eqA.3}):
\begin{align}\label{eqA.5}
    &\mat{g}_d(\mat{q}+\Delta\mat{q},\dot{\mat{q}}+\Delta\dot{\mat{q}},\ddot{\mat{q}}+\Delta\ddot{\mat{q}}) \approx \mat{g}_d(\mat{q},\dot{\mat{q}},\ddot{\mat{q}}) + \nonumber \\
    &\qquad \partial_{\mat{q}}\,\left[...\right]\cdot\Delta\mat{q} + \partial_{\dot{\mat{q}}}\,\left[...\right]\cdot\Delta\dot{\mat{q}} + \partial_{\ddot{\mat{q}}}\,\left[...\right]\cdot\Delta\ddot{\mat{q}} \nonumber \\
    &\approx \mat{g}_d(\mat{q},\dot{\mat{q}},\ddot{\mat{q}}) + \left[\int_0^S\partial_{\mat{q}}\left(...\right)\mathrm{d}s\right]\cdot\Delta\mat{q} \\
    &+\left[\int_0^S\partial_{\dot{\mat{q}}}\left(...\right)\mathrm{d}s\right]\cdot\Delta\dot{\mat{q}} + \left[\int_0^S\partial_{\ddot{\mat{q}}}\left(...\right)\mathrm{d}s\right]\cdot\Delta\ddot{\mat{q}}\,. \nonumber
\end{align}

In tensor notation, derivatives with respect to a vector can be expressed as \footnote{We note that Einstein's summation is implied.} 
$\partial_{\mat{q}}(\cdot) = \partial_{\qhat^i}(\cdot)\otimes\mat{G}_i$,
where $\mat{q}=(q^1,q^2,...,q^{3\m})^T$, and $\{\mat{G}_1,...,\mat{G}_{3\m}\}$ is an orthonormal basis for $\mathbb{R}^{3\m}$. 
We consider the three following tangent matrices resulting 
from (\ref{eqA.4}) and (\ref{eqA.5}):
\begin{align}\label{eqA.6}
    \mat{K}_S &= \partial_{q^i}(\mat{B}(\mat{q})^T\mat{a})\vert_{\mat{a}=\boldsymbol{\sigma}_h}\otimes\mat{G}_i \nonumber \\
    & + \mat{B}(\mat{q})^T\partial_{q^i}(\boldsymbol{\sigma}_h)\otimes\mat{G}_i\,, \nonumber \\
    \mat{K}_F &= \partial_{q^i}(\mat{N}^T\mat{f}^{\text{ext}})\otimes\mat{G}_i + \partial_{\dot{q}^i}(\mat{N}^T\mat{f}^{\text{ext}})\otimes\mat{G}_i \nonumber \\
    & + \partial_{\ddot{q}^i}(\mat{N}^T\mat{f}^{\text{ext}})\otimes\mat{G}_i\,, \\
    \mat{K}_{M} &=\partial_{q^i}(\mat{M} (\mat{\qhat})\,\nabla_{\dot{\mat{\qhat}}}\dot{\mat{\qhat}})\otimes\mat{G}_i \nonumber\\
    & + \partial_{\dot{q}^i}(\mat{M} (\mat{\qhat})\,\nabla_{\dot{\mat{\qhat}}}\dot{\mat{\qhat}})\otimes\mat{G}_i \nonumber \\
    & + \partial_{\ddot{q}^i}(\mat{M} (\mat{\qhat})\,\nabla_{\dot{\mat{\qhat}}}\dot{\mat{\qhat}})\otimes\mat{G}_i\,, \nonumber
\end{align}

\noindent
where $\mat{a}\in\mathbb{R}^{3\m}$ is a constant vector, $\mat{K}_S$, $\mat{K}_F$ and $\mat{K}_M$ are the so-called \textit{static tangent stiffness matrix}, \textit{force tangent stiffness matrix} and \textit{mass tangent stiffness matrix}, respectively. 

%********************************************************************************************************************
%                               S T I F F N E S S   T A N G E N T   O P E R A T O R
%********************************************************************************************************************

\subsection{Static tangent stiffness matrix}\label{Appendix_2}

The static tangent stiffness matrix $\mat{K}_S$ consists of two terms: the first term resulting from the derivative of the operator $\mat{B}(\mat{q})$ (geometric stiffness), and the second arising from the derivative of vector $\boldsymbol{\sigma}_h$ (elastic stiffness), i.e., $\mat{K}_S=\mat{K}_S^g+\mat{K}_S^e$. 
We first recall 
the discrete stress and strain measures from Section \ref{sec-discretization}, which are:
\begin{equation}\label{eqA.2.1}
\begin{split}
    \boldsymbol{\sigma}_h = 
    \left\{\begin{matrix} \mat{n}_h \\ \mat{m}_h \end{matrix}\right\} & = 
        \begin{bmatrix} EA\,\boldsymbol{\epsilon}_h \\[0.3em] EI\,\boldsymbol{\kappa}_h \end{bmatrix} \\
    & = \underbrace{\begin{bmatrix} EA\,\mat{I}_3 & \mat{0}_3 \\ \mat{0}_3 & EI\,\mat{I}_3 \end{bmatrix}}_{\mat{C}(E,A,I)}\,\left\{\begin{matrix} \boldsymbol{\epsilon}_h \\ \boldsymbol{\kappa}_h  \end{matrix}\right\} \,.
\end{split}
\end{equation}

\noindent
Considering the discrete linearized strain operator $\mat{B}(\mat{q}) = \begin{bmatrix}
    \mat{B}_{11} & \mat{0} \\ \mat{B}_{21} & \mat{B}_{22}
\end{bmatrix}$ in (\ref{B-matrix}), 
we compute the derivative of the sub-operators $\mat{B}_{11}$, $\mat{B}_{21}$, and $\mat{B}_{22}$, as follows:  
\begin{equation}\label{eqA.2.2}
\begin{split}
    \partial_{q^i}&\left[\mat{B}_{11}^T(\mat{q})\mat{a}\right]\vert_{\mat{a}=\mat{n}_h} \\
    &= \partial_{q^i}\left.\left[\mat{N}^{\prime T}\left( \mat{I}-\abss{\phic_h^\prime}^{-1} \Pdh \right)\mat{a}\right]\right\vert_{\mat{a}=\mat{n}_h} \\
    &= \mat{N}^{\prime T}\mat{A}_1\,\partial_{q^i}(\phic_h^\prime)\,, \\
    \partial_{q^i}&\left[\mat{B}_{21}^T(\mat{q})\mat{a}\right]\vert_{\mat{a}=\mat{m}_h} \\
    &= \partial_{q^i}\left.\left[\mat{N}^{\prime T}\left( \abss{\phic_h^\prime}^{-2}\Hdh[\phic_h^{\prime\prime}]_{\times} \right)\mat{a}\right]\right\vert_{\mat{a}=\mat{m}_h} \\
    &= \mat{N}^{\prime T}\mat{A}_2\,\partial_{q^i}(\phic_h^\prime) + \mat{N}^{\prime T}\mat{A}_3\,\partial_{q^i}(\phic_h^{\prime\prime})\,, \\
    \partial_{q^i}&\left[\mat{B}_{22}^T(\mat{q})\mat{a}\right]\vert_{\mat{a}=\mat{m}_h} \\
    &= \partial_{q^i}\left.\left[\mat{N}^{\prime\prime T}\left( \abss{\phic_h^\prime}^{-1}[\mat{d}_h]_{\times} \right)\mat{a}\right]\right\vert_{\mat{a}=\mat{m}_h} \\
    &= \mat{N}^{\prime\prime T}\mat{A}_4\,\partial_{q^i}(\phic_h^{\prime})\,,
\end{split}
\end{equation}

\noindent
where:
\begin{equation}\label{eqA.2.3}
\begin{split}
    \mat{A}_1 &= \frac{1}{\abss{\phic_h^\prime}^{2}}\left[ 2\mat{d}_h\odot\mat{n}_h - 3(\mat{d}_h\cdot\mat{n}_h)\mat{d}_h\otimes\mat{d}_h \right. \\
    & \left. + \left(\mat{d}_h \cdot \mat{n}_h \right) \, \mat{I} \right]\,, \\ 
    \mat{A}_2 &= -\frac{2}{\abss{\phic_h^\prime}^{3}}\left[ 2\left([\phic_h^{\prime\prime}]_{\times}\mat{m}_h\odot\mat{d}_h\right) \right. \\
    &\left. + ([\phic_h^{\prime\prime}]_{\times}\mat{m}_h\cdot\mat{d}_h)(2\Hdh-\mat{I})\right]\,, \\
    \mat{A}_3 &= -\frac{1}{\abss{\phic_h^\prime}^{2}}\Hdh\,[\mat{m}_h]_{\times}\,, \\
    \mat{A}_4 &= \frac{1}{\abss{\phic_h^\prime}^{2}}[\mat{m}_h]_{\times}\,\Hdh\,.
\end{split}
\end{equation}

\noindent
Applying the spatial approximation of $\phic(s,t)$ introduced in (\ref{eq-discretize}), 
we obtain:
\begin{equation}\label{eqA.2.4}
\begin{split}
    \partial_{q^i}(\phic_h^{\prime}) &= \partial_{q^i}\left( \frac{d}{ds}(\mat{N}\mat{q}) \right) \\
    & = \mat{N}^{\prime} \partial_{q^i}\left(\mat{q}\right) = \mat{N}^{\prime}\mat{G}_i\,, \\
    \partial_{q^i}(\phic_h^{\prime\prime}) &= \mat{N}^{\prime\prime}\mat{G}_i\,.
\end{split}
\end{equation}

Introducing (\ref{eqA.2.4}) into (\ref{eqA.2.2}), 
we obtain the first counterpart of $\mat{K}_S$, 
$\mat{K}_S^g$, that is:
\begin{equation}\label{eqA.2.5}
\begin{split}
    \mat{K}_S^g &= \partial_{q^i}(\mat{B}(\mat{q})^T\mat{a})\vert_{\mat{a}=\boldsymbol{\sigma}_h}\otimes\mat{G}_i \\
    &= \left[\mat{N}^{\prime T}(\mat{A}_1+\mat{A}_2)\mat{N}^{\prime} + \mat{N}^{\prime T}\mat{A}_3\,\mat{N}^{\prime\prime} \right. \\
    & \qquad \left. + \mat{N}^{\prime\prime T}\mat{A}_4\mat{N}^{\prime}\right]\, \overbrace{\mat{G}_i \otimes \mat{G}_i}^{\mat{I}_{3\m}} \\
    &= \begin{bmatrix} \mat{N}^{\prime T} & \mat{N}^{\prime\prime T} \end{bmatrix}\,\begin{bmatrix} \mat{A}_1+\mat{A}_2 & \mat{A}_3 \\ \mat{A}_4 & \mat{0}_3\end{bmatrix}\,\begin{bmatrix} \mat{N}^{\prime} \\ \mat{N}^{\prime\prime} \end{bmatrix},
\end{split}
\end{equation}

\noindent
where $\mat{I}_{3\m}$ is the identity matrix of dimension $3\m\times 3\m$.

The second counterpart,  
$\mat{K}_S^e$, 
employing (\ref{eqA.2.1}), 
can be computed as follows:
\begin{equation}\label{eqA.2.6}
\begin{split}
    \mat{K}_S^e &= \mat{B}(\mat{q})^T\partial_{\mat{q}}(\boldsymbol{\sigma}_h) \\
    &= \mat{B}(\mat{q})^T\partial_{\mat{q}}\left( \mat{C}\left\{\begin{matrix} \boldsymbol{\epsilon}_h \\ \boldsymbol{\kappa}_h \end{matrix}\right\} \right)\,.
\end{split}
\end{equation}
We recall $(\boldsymbol{\delta\epsilon},\delta\boldsymbol{\kappa})^T=\mathcal{B}(\phic^{\prime},\phic^{\prime\prime})\, \delta\phic$. 
Thus, the derivative $\partial_{\mat{q}}(\boldsymbol{\epsilon}_h,\boldsymbol{\kappa}_h)^T$, directly related to the discretized strain operator $\mat{B}(\mat{q})$, is:
\begin{equation}\label{eqA.2.7}
\begin{split}
    \partial_{\mat{q}}\left(\left\{\begin{matrix} \boldsymbol{\epsilon}_h \\ \boldsymbol{\kappa}_h \end{matrix}\right\} \right)=\mat{B}(\mat{q})\,\underbrace{\partial_{\mat{q}}(\mat{q})}_{\mat{I}} \,.
\end{split}
\end{equation}
$\mat{K}_S^e$ then takes the following form:
\begin{equation}\label{eqA.2.8}
\begin{split}
    \mat{K}_S^e &=\mat{B}(\mat{q})^T\mat{C}\,\mat{B}(\mat{q})\,,
\end{split}
\end{equation}

\noindent
where the constitutive matrix $\mat{C}$ is assumed to be constant. 
The static tangent stiffness matrix $\mat{K}_S$ is then:
\begin{equation}\label{eqA.2.9}
\begin{split}
    \mat{K}_S &= \mat{K}_S^g + \mat{K}_S^e = \\
    &\begin{bmatrix} \mat{N}^{\prime T} & \mat{N}^{\prime\prime T} \end{bmatrix}\,\begin{bmatrix} \mat{A}_1+\mat{A}_2 & \mat{A}_3 \\ \mat{A}_4 & \mat{0}_3\end{bmatrix}\,\begin{bmatrix} \mat{N}^{\prime} \\ \mat{N}^{\prime\prime} \end{bmatrix} \\
    &+ \mat{B}(\mat{q})^T\mat{C}\,\mat{B}(\mat{q})\,.
\end{split}
\end{equation}

%********************************************************************************************************************
%                                 F O R C E   T A N G E N T   O P E R A T O R
%********************************************************************************************************************

\subsection{Force tangent stiffness matrix}\label{Appendix_3}

We now derive the force tangent stiffness matrix $\mat{K}_F$ that is required for our computations in this work. 
We consider two different types of forces: (\textit{i}) follower forces for two-dimensional study cases, and (\textit{ii}) forces induced by a surrounding flow described in Appendix \ref{sec-flow-force}.

\subsubsection{Follower forces in two-dimensional studies}

\begin{figure}[ht!]
    \centering
    \def\svgwidth{0.8\columnwidth}
    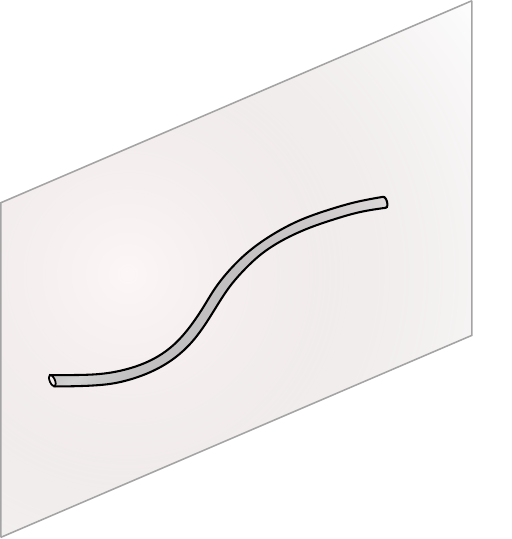
    \caption{Schematic for a two-dimensional follower force.}
    \label{figA.1}
\end{figure}

In two-dimensional cases, the motion of the rod is confined in a single plane at all times (see Fig. \ref{figA.1}). Thus, we can define a lumped follower force as follows: 
\begin{equation}\label{eqA.3.1}
    \mat{f}_f = f_0\,(\mat{E}_2\times\mat{d}_h)\,,
\end{equation}

\noindent
where $f_0$ is the magnitude of the force and is assumed to be constant, and $\mat{E}_2$ is the unit normal to the plane of motion. According to (\ref{eqA.6}), the tangent stiffness matrix $\mat{K}_F$ resulting from $\mat{f}_f$ is:
\begin{equation}\label{eqA.3.2}
\begin{split}
    \mat{K}_F &= \partial_{q^i}\left( \mat{N}^T f_0\,(\mat{E}_2\times\mat{d}_h)\right)\otimes\mat{G}_i \\
    &=\frac{f_0}{\abss{\phic_h^\prime}}\mat{N}^T \, [\mat{E}_2]_{\times}\,\Pdh\mat{N}^\prime\,\underbrace{\mat{G}_i\otimes\mat{G}_i}_{\mat{I}_{3\m}} \\
    &=\frac{f_0}{\abss{\phic_h^\prime}}\mat{N}^T \, [\mat{E}_2]_{\times}\,\Pdh\mat{N}^\prime \,.
    \end{split}
\end{equation}

We note that since 
the configuration space of Kirchhoff rods employed in this work requires only the specification of $\phic(s,t)$ and the director $\mat{d}$, a modeling of three-dimensional follower forces is generally not possible.

\subsubsection{Force induced by a surrounding flow}

As mentioned in Appendix \ref{sec-flow-force}, we consider a simplified model for a force induced by a surrounding flow, that consists of three counterparts: an added mass force, a tangential drag force, and a normal drag force. 
In this subsection, we derive the tangent stiffness matrix corresponding to each counterpart. 
To simplify the involved algebra for this derivation, we consider the resulting expression of the semi-discrete formulation after employing the implicit time integration scheme described in Section \ref{sec-time-int}. 
We note that the derivation of the tangent stiffness matrices can be alternatively performed with the equations of motion and/or the semi-discrete equations, and thus is independent of the numerical scheme applied.

The chosen implicit scheme in this work (see also Section \ref{sec-time-int}) evaluates 
$\mat{g}_d$ at the time instant $t_{n+\frac{1}{2}}$, and thus also the force term, i.e. $(\mat{N}^T\mat{f}^{\textrm{ext}})_{n+\frac{1}{2}}$. 
The term associated with the force is then:
\begin{equation}\label{eqA.3.3}
\begin{split}
    \int_0^{S}&\left[(\mat{N}^T\mat{F}_f)_{n+\frac{1}{2}}\right]\mathrm{d}s = C_1\int_0^{S}\left[(\mat{N}^T\Pdh\mat{a}_{h})_{n+\frac{1}{2}}\right]\mathrm{d}s \\ 
    &+ C_2\int_0^{S}\left[(\mat{N}^T\abss{\Pdh\mat{V}_{h}}\Pdh\mat{V}_{h})_{n+\frac{1}{2}}\right]\mathrm{d}s \\ 
    &+ C_3\int_0^{S}\left[(\mat{N}^T\abss{(\mat{d}_h\otimes\mat{d}_h)\mat{V}_{h}}(\mat{d}_h\otimes\mat{d}_h)\mat{V}_{h})_{n+\frac{1}{2}}\right]\mathrm{d}s\,, \nonumber
    \end{split}
\end{equation}

\noindent
where $\mat{a}_{h}=\mat{a}_{\infty}-\ddot{\phic}_h$ is the discrete relative acceleration of the surrounding flow, and $\mat{V}_{h}=\mat{V}_{\infty}-\dot{\phic}_h$ the discrete relative velocity.

%--------------------------------------------------------------------------------------------------------------------------
%                                               A D D E D   M A S S

\vspace{0.5cm}
\underline{\textbf{The added mass force}}, denoted as $\mat{F}^{am}$, is a function of the free-stream, rod acceleration, and rod configuration. 
Employing the midpoint rule and a standard Taylor expansion up to first order, we obtain the following approximation for $\mat{F}^{am}$ at the time instance $t_{n+\frac{1}{2}}$:
\begin{equation}\label{eqA.3.4}
\begin{split}
    &\mat{F}^{am}_{n+\frac{1}{2}}(\mat{q}_{n+1}+\Delta\mat{q},\dot{\mat{q}}_{n+1}+\Delta\dot{\mat{q}}) \\
    & \qquad \approx \mat{F}^{am}_{n+\frac{1}{2}}(\mat{q}_{n+1},\dot{\mat{q}}_{n+1}) \\
    & \qquad + \partial_{\mat{q}_{n+1}}\mat{F}^{am}_{n+\frac{1}{2}}(\mat{q}_{n+1},\dot{\mat{q}}_{n+1})\cdot\Delta\mat{q} \\
    & \qquad + \partial_{\dot{\mat{q}}_{n+1}}\mat{F}^{am}_{n+\frac{1}{2}}(\mat{q}_{n+1},\dot{\mat{q}}_{n+1})\cdot\Delta\dot{\mat{q}}\,,
    \end{split}
\end{equation}

\noindent
where:
\begin{align}\label{eqA.3.5}
   &\mat{F}^{am}_{n+\frac{1}{2}}(\mat{q}_{n},\mat{q}_{n+1},\dot{\mat{q}}_{n},\dot{\mat{q}}_{n+1}) \\
   &= C_1\int_0^S\mat{N}^T\left\{\frac{1}{2}\left[(\Pdh\mat{a}_{\infty})_n+(\Pdh\mat{a}_{\infty})_{n+1}\right]\right\}\mathrm{d}s \nonumber\\
   &- C_1\int_0^S\mat{N}^T\left\{\frac{1}{\Delta t}\left[(\Pdh\dot{\phic}_h)_{n+1}-(\Pdh\dot{\phic}_h)_{n}\right]\right\}\mathrm{d}s \,.\nonumber
\end{align}
Considering (\ref{eqA.3.4}) and (\ref{eqA.3.5}), the corresponding tangent stiffness matrix per unit of length is:
\begin{equation}\label{eqA.3.6}
\begin{split}
   \mat{K}_F^{am} = C_1\mat{N}^T&\left[\frac{1}{2}\underbrace{\partial_{q_{n+1}^i}(\Pdh\mat{a}_{\infty})_{n+1}}_{\mat{f}_1} \right. \\
   & - \frac{1}{\Delta t}\underbrace{\partial_{q_{n+1}^i}(\Pdh\dot{\phic}_h)_{n+1}}_{\mat{f}_2} \\
   & \left. - \frac{1}{\Delta t}\underbrace{\partial_{\dot{q}_{n+1}^i}(\Pdh\dot{\phic}_h)_{n+1}}_{\mat{f}_3}\right]\otimes\mat{G}_i \,.
    \end{split}
\end{equation}

We assume that both the magnitude and direction of the force vary with the altitude (or depth for marine applications). 
For the sake of simplicity, we consider $\mat{a}_{\infty}$ as a function of the vertical coordinate $z$ and time $t$, i.e., $\mat{a}_{\infty}(z,t)$, where $z$ depends on the current configuration of the rod, that is $z=\phic_h\cdot\mat{E}_3$. 
We then obtain:
\begin{align}\label{eqA.3.7}
   \mat{f}_1 &= \left\{\underbrace{-\frac{1}{\abss{\phic_h^\prime}}\left[ (\mat{a}_{\infty}\cdot\mat{d}_h)\Hdh + \mat{d}_h\otimes\mat{a}_{\infty}\right]_{n+1}}_{\mat{K}_1}\mat{N}^{\prime} \right. \nonumber \\
   & \quad \left. + \underbrace{\left[ \Pdh(\partial_z\mat{a}_{\infty}\otimes\mat{E}_3) \right]_{n+1}}_{\mat{K}_2}\mat{N}\right\}\,\mat{G}_i \,, \\
   \mat{f}_2 &= \underbrace{-\frac{1}{\abss{\phic_h^\prime}}\left[ (\dot{\phic}_h\cdot\mat{d}_h)\Hdh + \mat{d}_h\otimes\dot{\phic}_h \right]_{n+1}}_{\mat{K}_3}\mat{N}^{\prime}\,\mat{G}_i \,, \nonumber\\
   \mat{f}_3 &= \underbrace{\left[ \Pdh\mat{N} \right]_{n+1}}_{\mat{K}_4}\,\mat{G}_i \,. \nonumber
\end{align}
Introducing (\ref{eqA.3.7}) into (\ref{eqA.3.6}), and recalling that $\mat{G}_i\otimes\mat{G}_i = \mat{I}_{3\m}$ and $\Delta\dot{\mat{q}}=\frac{2}{h}\Delta\mat{q}$, we obtain:
\begin{equation}\label{eqA.3.8}
\begin{split}
   \mat{K}_F^{am} = C_1\mat{N}^T & \left[\frac{1}{2}\mat{K}_1\mat{N}^{\prime} + \frac{1}{2}\mat{K}_2\mat{N} \right. \\
   & \left. - \frac{1}{\Delta t}\mat{K}_3\mat{N}^{\prime} - \frac{2}{\Delta t^2}\mat{K}_4\mat{N}\right] \,.
    \end{split}
\end{equation}

%--------------------------------------------------------------------------------------------------------------------------
%                                      N O R M A L   D R A G   F O R C E

\vspace{0.5cm}
\underline{\textbf{The normal drag force}}, denoted as $\mat{F}^{cn}$, is a function of the free-stream, rod velocity, and rod configuration. 
The normal drag force of the surrounding flow, evaluated at $t_{n+\frac{1}{2}}$, is:
\begin{equation}\label{eqA.3.9}
\begin{split}
    & \mat{F}^{cn}_{n+\frac{1}{2}}(\mat{q}_{n},\mat{q}_{n+1},\dot{\mat{q}}_{n},\dot{\mat{q}}_{n+1}) \\
    & = C_2\int_0^S\mat{N}^T\left\{\frac{1}{2}\left[(\abss{\Pdh\mat{V}_h}\Pdh\mat{V}_h)_n \right. \right. \\
    & \qquad \qquad \qquad \quad \left. \left. +(\abss{\Pdh\mat{V}_h}\Pdh\mat{V}_h)_{n+1}\right]\right\}\mathrm{d}s \,. \nonumber
    \end{split}
\end{equation}
Analogously to the tangent stiffness matrix corresponding to the added mass force derived above,
the tangent stiffness matrix per unit of length associated with the normal drag force is:
\begin{equation}\label{eqA.3.10}
\begin{split}
   \mat{K}_F^{cn} &= C_2\mat{N}^T\left[\frac{1}{2}\mat{K}_1\mat{N} - \frac{1}{2}\mat{K}_2\mat{N}^{\prime} - \frac{1}{\Delta t}\mat{K}_3\mat{N}\right]\,,
    \end{split}
\end{equation}

\noindent
where:
\begin{equation}\label{eqA.3.11}
\begin{split}
   \mat{K}_1 &= \abss{\Pdh\mat{V}_h}(2\mat{I}-\mat{P}_{\text{u}})\Pdh(\partial_z\mat{V}_{\infty}\otimes\mat{E}_3)\,, \\
   \mat{K}_2 &= \frac{\abss{\Pdh\mat{V}_h}}{\abss{\phic_h^\prime}}(2\mat{I}-\mat{P}_{\text{u}})\left[(\mat{V}_h\cdot\mat{d}_h)\Hdh + \mat{d}_h\otimes\mat{V}_h\right]\,, \\
   \mat{K}_3 &= \abss{\Pdh\mat{V}_h}(2\mat{I}-\mat{P}_{\text{u}})\Pdh, \\
   \mat{P}_{\text{u}} &= \mat{I} - \frac{1}{\abss{\Pdh\mat{V}_h}^2}(\Pdh\mat{V}_h)\otimes(\Pdh\mat{V}_h)\,. \nonumber
    \end{split}
\end{equation}

%--------------------------------------------------------------------------------------------------------------------------
%                                      T A N G E N T I A L   D R A G   F O R C E

\vspace{0.5cm}
\underline{\textbf{The tangential drag force}}, denoted as $\mat{F}^{ct}$, is also a function of the free-stream, rod velocity, and rod configuration. 
The tangential drag force of the surrounding flow, evaluated at $t_{n+\frac{1}{2}}$, is:
\begin{equation}\label{eqA.3.12}
\begin{split}
   &\mat{F}^{ct}_{n+\frac{1}{2}}(\mat{q}_{n},\mat{q}_{n+1},\dot{\mat{q}}_{n},\dot{\mat{q}}_{n+1}) \\
   &= C_3\int_0^S\mat{N}^T\left\{\frac{1}{2}\left[\left(\abss{(\mat{d}_h\otimes\mat{d}_h)\mat{V}_h}(\mat{d}_h\otimes\mat{d}_h)\mat{V}_h\right)_n \right.\right. \\ 
   & + \left.\left.(\abss{(\mat{d}_h\otimes\mat{d}_h)\mat{V}_h}(\mat{d}_h\otimes\mat{d}_h)\mat{V}_h)_{n+1}\right]\right\}\mathrm{d}s\,. \nonumber
    \end{split}
\end{equation}
Analogously, 
the tangent stiffness matrix per unit of length associated with the tangential drag force is:
\begin{equation}\label{eqA.3.13}
\begin{split}
   \mat{K}_F^{ct} &= C_3\mat{N}^T\left[\frac{1}{2}\mat{K}_1\mat{N} + \frac{1}{2}\mat{K}_2\mat{N}^{\prime} - \frac{1}{\Delta t}\mat{K}_3\mat{N}\right]\,, \nonumber
    \end{split}
\end{equation}

\noindent
where:
\begin{equation}\label{eqA.3.14}
\begin{split}
   \mat{K}_1 &= \abss{(\mat{d}_h\otimes\mat{d}_h)\mat{V}_h}\left[2\mat{I}-\mat{P}_{\text{v}}\right](\mat{d}_h\otimes\mat{d}_h)(\partial_z\mat{V}_{\infty}\otimes\mat{E}_3)\,, \\
   \mat{K}_2 &= \frac{\abss{(\mat{d}_h\otimes\mat{d}_h)\mat{V}_h}}{\abss{\phic_h^\prime}}\left(2\mat{I}-\mat{P}_{\text{v}}\right)\left[(\mat{V}_h\cdot\mat{d}_h)\Hdh + \mat{d}_h\otimes\mat{V}_h\right]\,, \\
   \mat{K}_3 &= \abss{(\mat{d}_h\otimes\mat{d}_h)\mat{V}_h}(2\mat{I}-\mat{P}_{\text{v}})\left[\mat{d}_h\otimes\mat{d}_h\right]\,, \\
   \mat{P}_{\text{v}} &= \mat{I} - \frac{1}{\abss{(\mat{d}_h\otimes\mat{d}_h)\mat{V}_h}^2}\left[(\mat{d}_h\otimes\mat{d}_h)\mat{V}_h\right]\otimes\left[(\mat{d}_h\otimes\mat{d}_h)\mat{V}_h\right]\,. \nonumber
    \end{split}
\end{equation}

%********************************************************************************************************************
%                                 M A S S   T A N G E N T   O P E R A T O R
%********************************************************************************************************************

\subsection{Mass tangent stiffness matrix}\label{Appendix_4}

Lastly, we derive the mass tangent stiffness matrix, $\mat{K}_M$, corresponding to the inertial contribution $\mat{M}(\mat{q})\nabla_{\dot{\mat{\qhat}}}\dot{\mat{q}}$ of the mass operator. 
Similarly to the tangent stiffness matrix corresponding to the force induced by a surrounding flow in the previous subsection, 
we consider the resulting expression of the semi-discrete formulation after employing the implicit time integration scheme described in Section \ref{sec-time-int}. 
Considering (\ref{eqs}) together with the midpoint rule formulas, we obtain the following approximation for the inertia term of the mass operator, evaluated at $t=n+\frac{1}{2}$, which is: 
\begin{align}\label{eqA.4.1}
   &\mat{f}^{I}_{n+\frac{1}{2}}(\mat{q}_{n},\mat{q}_{n+1},\dot{\mat{q}}_{n},\dot{\mat{q}}_{n+1}) \\
   &= \int_0^S\left\{\frac{1}{\Delta t}\left[\left(\mat{M}\dot{\mat{q}}\right)_{n+1} - \left(\mat{M}\dot{\mat{q}}\right)_n\right]\right. \nonumber \\
   & + \frac{1}{2}\mat{N}^{\prime T}\left[2\,I_{\rho} \left(\frac{1}{\abss{\phic_h^\prime}^3}[\Pdh\odot(\dot{\phic}_h^\prime\otimes\mat{d}_h)]\dot{\phic}_h^\prime\right)_n \right. \nonumber \\
   &\qquad \left. \left. + 2\,I_{\rho}\left(\frac{1}{\abss{\phic_h^\prime}^3}[\Pdh\odot(\dot{\phic}_h^\prime\otimes\mat{d}_h)]\dot{\phic}_h^\prime\right)_{n+1}\right]\right\}\mathrm{d}s \,. \nonumber
\end{align}
Employing 
the standard Taylor expansion up to first order for this term leads to:
\begin{equation}\label{eqA.4.2}
\begin{split}
   \mat{f}^{I}_{n+\frac{1}{2}}& (\mat{q}_{n+1}+\Delta\mat{q},\dot{\mat{q}}_{n+1}+\Delta\dot{\mat{q}}) \\
   &\approx \mat{f}^{I}_{n+\frac{1}{2}}(\mat{q}_{n+1},\dot{\mat{q}}_{n+1}) \\
   &+ \partial_{\mat{q}_{n+1}}\mat{F}^{I}_{n+\frac{1}{2}}(\mat{q}_{n+1},\dot{\mat{q}}_{n+1})\cdot\Delta\mat{q} \\
   &+ \partial_{\dot{\mat{q}}_{n+1}}\mat{F}^{I}_{n+\frac{1}{2}}(\mat{q}_{n+1},\dot{\mat{q}}_{n+1})\cdot\Delta\dot{\mat{q}} \,.
    \end{split}
\end{equation}
Introducing (\ref{eqA.4.1}) into (\ref{eqA.4.2}),   
we obtain the following counterparts associated with the inertia contribution of the mass operator:
\begin{equation}\label{eqA.4.3}
\begin{split}
    &\mat{K}_1 = \partial_{q^i_{n+1}}(\mat{M}(\mat{q})\dot{\mat{q}})_{n+1}\otimes\mat{G}_i \\
   &= -\frac{I_{\rho}}{\abss{\phic_h^\prime}^3}\mat{N}^{\prime T}\left\{ (\mat{d}_h\cdot\dot{\phic}_h^{\prime})(2\Hdh-\mat{I}) \right. \\
   & \left. \, + \mat{d}_h\otimes\dot{\phic}_h^{\prime} + 2\,\dot{\phic}_h^{\prime}\otimes\mat{d}_h \right\}\mat{N}^{\prime} \,, \\
   &\mat{K}_2 = \partial_{\dot{q}^i_{n+1}}(\mat{M}(\mat{q})\dot{\mat{q}})_{n+1}\otimes\mat{G}_i = \mat{N}^T\mat{M}(\mat{q}_{n+1})\mat{N} \,, \\
   &\mat{K}_3 = 2\,I_{\rho}\mat{N}^{\prime T} \partial_{q^i_{n+1}}\left(\frac{[\Pdh\odot(\dot{\phic}_h^\prime\otimes\mat{d}_h)]\dot{\phic}_h^\prime}{\abss{\phic_h^\prime}^3}\right)_{n+1}\otimes\mat{G}_i \\
   &= \frac{-I_{\rho}\,\mat{N}^{\prime\, T}}{\abss{\phic_h^\prime}^4} \left\{ 2(\dot{\phic}_h^\prime\cdot\mat{d}_h) \left[4 \dot{\phic}_h^\prime\odot\mat{d}_h + (\dot{\phic}_h^\prime\cdot\mat{d}_h)[3\Hdh-2\mat{I}]\right] \right. \\
   &- \left. \dot{\phic}_h^\prime\otimes\dot{\phic}_h^\prime + (\dot{\phic}_h^\prime\cdot\dot{\phic}_h^\prime)[\mat{I} - 2\Hdh] \right\}\,\mat{N}^{\prime} \,, \\
   &\mat{K}_4 = 2\,I_{\rho}\mat{N}^{\prime T} \partial_{\dot{q}^i_{n+1}}\left(\frac{[\Pdh\odot(\dot{\phic}_h^\prime\otimes\mat{d}_h)]\dot{\phic}_h^\prime}{\abss{\phic_h^\prime}^3}\right)_{n+1}\otimes\mat{G}_i \\
   &= 2\frac{I_{\rho}}{\abss{\phic_h^\prime}^3}\mat{N}^{\prime T} \left\{ (\dot{\phic}_h^\prime\cdot\mat{d}_h)\Hdh + \mat{d}_h\otimes\dot{\phic}_h^\prime \right. \\
   & \qquad \qquad \qquad \left. + 2\Pdh\odot(\dot{\phic}_h^\prime\otimes\mat{d}_h) \right\}\,\mat{N}^{\prime} \,. \nonumber
\end{split}
\end{equation}

Recalling that $\Delta\dot{\mat{q}}=\frac{2}{h}\Delta\mat{q}$, we then obtain the tangent stiffness matrix per unit of length associated with the inertia term of the mass operator as follows:
\begin{equation}\label{eqA.4.4}
\begin{split}
   \mat{K}_M &=\left[\frac{1}{\Delta t}\mat{K}_1 + \frac{2}{\Delta t^2}\mat{K}_2 + \frac{1}{2}\mat{K}_3 + \frac{1}{\Delta t}\mat{K}_4\right] \,. \nonumber
    \end{split}
\end{equation}

\end{appendices}

\bibliography{refs}

\end{document}